\documentclass{jpsj3}
\usepackage{type1cm}
\usepackage{tabularx}

\title{
Twofold and Fourfold Symmetric 
Anisotropic Magnetoresistance Effect 
in A Model with Crystal Field
}
\author{
Satoshi Kokado$^1$\thanks{E-mail address: 
kokado.satoshi@shizuoka.ac.jp
} 
and Masakiyo Tsunoda$^2$
}
\inst{
$^1$Department of Electronics and Materials Science, 
Graduate School of Integrated Science and Technology, 
Shizuoka University, Hamamatsu 432-8561, Japan\\
$^2$Department of Electronic Engineering, 
Graduate School of Engineering, Tohoku University, 
Sendai 980-8579, Japan
}

\abst{
We theoretically study the twofold and fourfold symmetric 
anisotropic magnetoresistance (AMR) effects of ferromagnets. 
We here use the two-current model for 
a system consisting of a conduction state 
and localized d states. 
The localized d states are obtained from 
a Hamiltonian with a spin--orbit interaction, 
an exchange field, and a crystal field. 
From the model, 
we first derive general expressions for 
the coefficient of the twofold symmetric term ($C_2$) 
and 
that of the fourfold symmetric term ($C_4$) 
in the AMR ratio. 
In the case of a strong ferromagnet, 
the dominant term in $C_2$ is proportional to 
the difference in the partial densities of states (PDOSs) 
at the Fermi energy ($E_{\mbox{\tiny F}}$) 
between the $d\varepsilon$ and $d\gamma$ states, 
and 
that in $C_4$ is proportional to 
the difference in the PDOSs at $E_{\mbox{\tiny F}}$ 
among the $d\varepsilon$ states. 
Using the dominant terms, 
we next analyze the experimental results for Fe$_4$N, 
in which $|C_2|$ and $|C_4|$ increase with decreasing temperature. 
The experimental results can be reproduced 
by assuming that 
the tetragonal distortion increases with decreasing temperature. 
}

\begin{document}
\maketitle

\section{Introduction}
\label{intro}
The anisotropic magnetoresistance (AMR) effect 
is a phenomenon 
in which the electrical resistivity 
depends on the relative angle 
between the magnetization 
(${\mbox{\boldmath $M$}}$) 
direction and the electric current 
(${\mbox{\boldmath $I$}}$) 
direction (see Fig. \ref{sample}).
\cite{Thomson,Campbell1,Potter,McGuire,Malozemoff1,Malozemoff2,Miyazaki} 
The AMR effect has been studied extensively 
both experimentally and theoretically 
since 1857, 
when it was discovered by W. Thomson.~\cite{Thomson} 
The AMR ratio, which is the efficiency of the effect, 
is generally defined by
\begin{eqnarray}
\label{AMR}
&&\frac{\Delta \rho (\phi)}{\rho} = 
\frac{ \rho (\phi) - \rho_\perp}{\rho_\perp}, 
\end{eqnarray}
with 
$\rho_\perp$=$\rho (\pi/2)$. 
Here, 
$\phi$ is the relative angle between 
the thermal average of 
the spin $\langle {\mbox{\boldmath $S$}} \rangle$ 
($\propto$$-{\mbox{\boldmath $M$}}$) 
and ${\mbox{\boldmath $I$}}$, 
and $\rho(\phi)$ is the resistivity at $\phi$. 

Experimentally, 
$\Delta \rho(0)/\rho$ has been measured for various ferromagnets 
such as Fe,\cite{McGuire} Co,\cite{McGuire} Ni,\cite{McGuire} 
Ni-based alloys,\cite{Campbell1} 
and half-metallic ferromagnets.
\cite{half-metal,Yang1,Sakuraba1,Sakuraba2,Yang2,Ziese,Ueda,Du} 
In addition, 
the AMR ratios of 
many ferromagnets 
have been observed to be 
\begin{eqnarray}
\label{AMR_c}
&&\frac{\Delta \rho (\phi)}{\rho} = 
c_0 + c_2 \cos 2\phi, 
\end{eqnarray}
where $c_2$ is the coefficient of the twofold symmetric term 
and $c_0$ is chosen to be 
$c_2$ so as to satisfy $\Delta \rho(\pi/2)/\rho$=0.
\cite{Tsunoda1,Yang1,Yang2,Nishiwaki,Rowan}


Theoretically, 
expressions for $\Delta \rho(0)/\rho$ 
have often been derived by using electric transport theory based on 
the two-current model with 
$s$--$d$ scattering. 
\cite{Campbell1,Potter,Malozemoff1,Malozemoff2,Kokado1,Kokado2} 
The $s$--$d$ scattering means that 
the conduction electron (denoted as $s$) is scattered by impurities 
into the localized d states (denoted as $d$) with 
the exchange field and the spin--orbit interaction. 
As a representative study, 
Campbell, Fert, and Jaoul (CFJ)\cite{Campbell1} derived an expression for 
$\Delta \rho(0)/\rho$ for strong ferromagnets\cite{SW_FM} 
such as Ni-based alloys 
[see Eq. (\ref{C_2_CFJ})]. 
On the basis of the CFJ model,\cite{Campbell1} 
Malozemoff obtained an expression for $\Delta \rho(0)/\rho$ for 
weak ferromagnets\cite{SW_FM} 
as well as strong ferromagnets.\cite{Malozemoff1,Malozemoff2} 
In addition, we extended the CFJ model\cite{Campbell1} 
and the Malozemoff model \cite{Malozemoff1,Malozemoff2}
to a more general model, 
which could systematically explain 
the experimental results of 
$\Delta \rho(0)/\rho$ for various ferromagnets 
including half-metallic ferromagnets.\cite{Kokado1} 

We also derived 
the analytic expression for $\Delta \rho (\phi)/\rho$ 
given by Eq. (\ref{AMR_c}).\cite{Kokado2}  
%
We then showed that 
the twofold symmetric feature of Eq. (\ref{AMR_c}) 
could be intuitively explained 
by considering the d states distorted by the spin--orbit interaction. 
Note here that 
the crystal field of the d states 
was not taken into account 
in the derivation of Eq. (\ref{AMR_c}). 
The d states were indexed by $M$=0, $\pm 1$, and $\pm 2$, 
with $M$ being the magnetic quantum number of the 3d states. 
Furthermore, 
the partial density of states (PDOS) of each d state 
at the Fermi energy ($E_{\mbox{\tiny F}}$) 
was assumed to be constant 
regardless of the d state. 

Recently, 
the AMR ratios of 
several ferromagnets\cite{Tsunoda2,Ito1,Kabara,Ito2,ZRLi,Gorkom,Ramos,Rushforth,PLi,Liu} 
have been experimentally observed to be
\begin{eqnarray}
\label{AMR_c1}
&&\frac{\Delta \rho (\phi)}{\rho} = 
C_0 + C_2 \cos 2\phi + C_4 \cos 4\phi. 
\end{eqnarray}
Here, 
$C_2$ ($C_4$) is the coefficient of the twofold (fourfold) symmetric term, 
and $C_0$ is chosen to be 
$C_2-C_4$ so as to satisfy $\Delta \rho(\pi/2)/\rho$=0. 
For example, 
$|C_2|$ and $|C_4|$ for Fe$_4$N 
increase with decreasing temperature $T$ 
as shown later in Fig. \ref{best}.\cite{Tsunoda2,Ito1,Kabara,Ito2,ZRLi} 
The coefficients $C_2$ and $C_4$ were measured to be 
$C_2$=$-$0.0343 and $C_4$=0.00556 at $T$=4 K.\cite{Tsunoda2} 

The set of $C_0$, $C_2$, and $C_4$, however, 
has seldom been derived 
within the framework of transport theory 
and has often been represented by 
phenomenological expressions.\cite{Doring,Bozorth,Ramos,Gorkom,Rushforth} 
We anticipate that 
expressions for $C_0$, $C_2$, and $C_4$ 
obtained by transport theory 
will play an important role in 
the analysis and understanding of the AMR effect. 
We also predict that 
the fourfold symmetric term in Eq. (\ref{AMR_c1}) may appear 
under 
the crystal field of the d states, 
which was neglected in the previous models \cite{Kokado1} 
[i.e., Eq. (\ref{AMR_c})]. 

In this paper, we obtained $C_2$ and $C_4$ 
by extending our model \cite{Kokado1,Kokado2} to one with a crystal field. 
We first performed a numerical calculation of $C_2$ and $C_4$ 
for a strong ferromagnet 
using the d states, which were obtained by applying 
the exact diagonalization method (EDM) to 
a Hamiltonian of the d states with a crystal field. 
The result revealed that 
$C_4$ appears under a crystal field of tetragonal symmetry, 
whereas 
it vanishes under a crystal field of cubic symmetry. 
We next derived general expressions for the resistivity, $C_2$, and $C_4$ 
for ferromagnets with the tetragonal field 
using the d states, which were obtained 
by applying first- and second-order perturbation theory (PT) to 
the Hamiltonian. 
From the expressions, 
we obtained expressions for $C_2$ and $C_4$ 
for the strong ferromagnet with the tetragonal field. 
The result showed that 
$C_2 \cos 2\phi$ is related to the real part of the probability amplitudes 
of the specific hybridized states 
and $C_4 \cos 4\phi$ is related to 
the probabilities of the specific hybridized states. 
In addition, 
we performed a simple analysis of the experimental results of 
$C_2$ and $C_4$ for Fe$_4$N 
using the dominant terms in $C_2$ and $C_4$ obtained by PT. 
The experimental results could be reproduced 
by assuming that 
the tetragonal distortion increases with decreasing $T$. 

The present paper is organized as follows: 
In Sec. \ref{theory}, 
we obtain wave functions of the localized d states 
by applying first- and second-order PT to 
the Hamiltonian of the localized d states. 
Using the wave functions, 
we derive general expressions for the resistivity, 
$C_2$, and $C_4$ for ferromagnets. 
In Sec. \ref{appl_strong}, 
we obtain 
expressions for $C_2$ and $C_4$ 
for a strong ferromagnet 
from the above-mentioned $C_2$ and $C_4$. 
In addition, we perform the numerical calculation of $C_2$ and $C_4$ 
using the d states, which are obtained by applying 
the EDM to the Hamiltonian. 
We then compare 
$C_2$ and $C_4$ obtained by PT 
and the respective values obtained by the EDM. 
In Sec. \ref{appl_fe4n}, 
we analyze the experimental results of $C_2$ and $C_4$ 
for Fe$_4$N. 
The conclusion is presented in Sec. \ref{conclusion}. 
In Appendix \ref{H_matrix}, 
we show the matrix of the Hamiltonian. 
In Appendix \ref{zero-order}, 
we give the zero-order states of the d states, 
which are obtained 
by performing the unitary transformation on the perturbation term. 
In Appendix \ref{selection}, 
we describe the overlap integrals of the $s$--$d$ scattering rate. 
In Appendix \ref{sd_scatter}, we give 
an expression for the $s$--$d$ scattering rate. 
Section \ref{relation} shows 
that the present $\Delta \rho(0)/\rho$ (=$2 C_2$) 
coincides with 
our previous model\cite{Kokado1,Kokado2} 
and the CFJ model\cite{Campbell1} 
under appropriate conditions.

\section{Theory}
\label{theory}
\begin{figure}[ht]
\begin{center}
\includegraphics[width=0.55\linewidth]{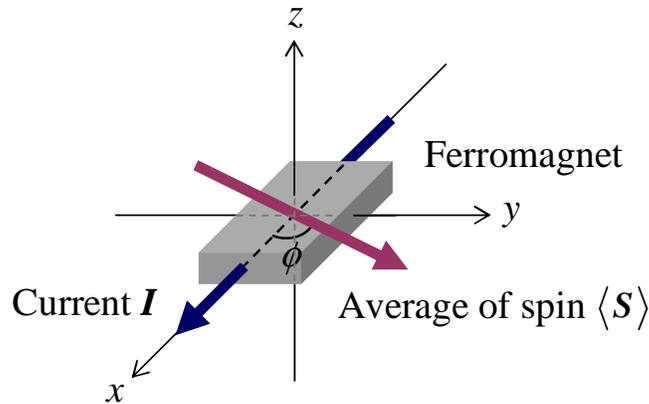}
\caption{
(Color online) 
Sketch of the sample geometry. 
The current ${\mbox{\boldmath $I$}}$ flows in the $x$ direction, 
the thermal average of the spin 
$\langle {\mbox{\boldmath $S$}} \rangle$ 
($\propto$$-{\mbox{\boldmath $M$}}$) lies in the $xy$ plane, 
and $\phi$ is the relative angle between 
the ${\mbox{\boldmath $I$}}$ direction 
and the $\langle {\mbox{\boldmath $S$}} \rangle$ direction. 
}
\label{sample}
\end{center}
\end{figure}

In this section, 
we obtain general expressions for the resistivity, $C_2$, and $C_4$ 
in a model in which 
${\mbox{\boldmath $I$}}$ flows in the $x$ direction 
and 
$\langle {\mbox{\boldmath $S$}} \rangle$ ($\propto$$-{\mbox{\boldmath $M$}}$) 
lies in the $xy$ plane (see Fig. \ref{sample}). 
We here use 
the two-current model with $s$--$d$ scattering 
in which the conduction electron is scattered into the localized d states 
by nonmagnetic impurities.
\cite{Campbell1,Potter,Malozemoff1,Malozemoff2,Kokado1,Kokado2}  
The d states are obtained 
by applying PT to a Hamiltonian of the d states. 
We also explain the numerical calculation method 
for $C_2$ and $C_4$, 
in which 
the d states are obtained 
by applying the EDM to the Hamiltonian. 


\subsection{Hamiltonian}



We first present the Hamiltonian ${\cal H}$ of the localized d states 
of a single atom \cite{Kokado1,comment_H} in a ferromagnet with 
a spin--orbit interaction, an exchange field, 
and a crystal field of tetragonal symmetry. 
This crystal field represents 
the case that distortion in the $z$ direction is added to 
the crystal field of cubic symmetry.\cite{tetra} 
Note that 
$C_4$ appears under a crystal field of tetragonal symmetry, 
whereas 
it vanishes under a crystal field of cubic symmetry, 
as will be described in Sec. \ref{c2c4_edm_r}. 

The Hamiltonian ${\cal H}$ is expressed as
\begin{eqnarray}
\label{Hamiltonian}
&&{\cal H} = {\cal H}_0 + V, \\
\label{Hamiltonian0}
&&{\cal H}_0 = {\cal H}_{\rm cubic} 
- {\mbox{\boldmath $H$}} \cdot {\mbox{\boldmath $S$}}, \\
\label{V}
&&V = V_{\rm so} + V_{\rm tetra}, 
\end{eqnarray}
with
\begin{eqnarray}
&&
{\cal H}_{\rm cubic}
=
\sum_{\sigma=\pm}
\Big[ 
E_\varepsilon 
(
|xy, \chi_\sigma (\phi) \rangle \langle xy, \chi_\sigma (\phi)| 
+ |yz, \chi_\sigma (\phi) \rangle \langle yz, \chi_\sigma (\phi)| 
+ |xz, \chi_\sigma (\phi) \rangle 
\langle xz, \chi_\sigma (\phi)| 
) \nonumber \\
&& \hspace*{1.5cm}+ 
E_\gamma 
(
|x^2-y^2, \chi_\sigma (\phi)\rangle \langle x^2-y^2, \chi_\sigma (\phi)| 
+ 
|3z^2-r^2, \chi_\sigma (\phi) \rangle \langle 3z^2-r^2, \chi_\sigma (\phi)| 
) \Big], \\
&&V_{\rm so} = \lambda {\mbox{\boldmath $L$}} \cdot {\mbox{\boldmath $S$}}, \\
&&
V_{\rm tetra}
=\sum_{\sigma=\pm}
\Big[ 
\delta_{\varepsilon} ( 
|xz, \chi_\sigma (\phi) \rangle \langle xz, \chi_\sigma (\phi) | 
+ |yz, \chi_\sigma (\phi) \rangle \langle yz, \chi_\sigma (\phi) | ) 
\nonumber \\
&&\hspace*{1.2cm} + \delta_{\gamma} 
|3z^2 -r^2, \chi_\sigma (\phi) \rangle \langle 3z^2-r^2, \chi_\sigma (\phi) | 
\Big], 
\end{eqnarray}
and
\begin{eqnarray}
&&{\mbox{\boldmath $S$}}=(S_x, S_y, S_z), \\
&&{\mbox{\boldmath $L$}}=(L_x, L_y, L_z), \\
&&{\mbox{\boldmath $H$}}=H (\cos \phi, \sin \phi, 0), 
\end{eqnarray}
where $H > 0$. 
Here, 
${\mbox{\boldmath $S$}}$ is the spin angular momentum 
and 
${\mbox{\boldmath $L$}}$ is the orbital angular momentum. 
The spin quantum number $S$ 
and the azimuthal quantum number $L$ 
are chosen to be 
$S$=1/2 and $L$=2.\cite{Kokado1}

\begin{figure}[ht]
\begin{center}
\includegraphics[width=0.39\linewidth]{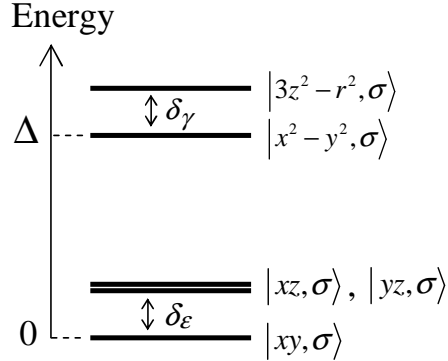}
\caption{
Energy levels of the 3d states in the crystal field of tetragonal symmetry. 
The energy levels are measured from $E_\varepsilon$. 
}
\label{energy}
\end{center}
\end{figure}

The above terms are explained as follows: 
The term 
${\cal H}_{\rm cubic}$ represents the crystal field of cubic symmetry. 
The term $-{\mbox{\boldmath $H$}} \cdot {\mbox{\boldmath $S$}}$ 
is the Zeeman interaction due to the exchange field of the ferromagnet 
${\mbox{\boldmath $H$}}$, 
where 
${\mbox{\boldmath $H$}}$$\propto$$-{\mbox{\boldmath $M$}}$ 
and ${\mbox{\boldmath $H$}}$$\propto$$\langle {\mbox{\boldmath $S$}} \rangle$. 
The term 
$V_{\rm so}$ is the spin--orbit interaction, 
where $\lambda$ is the spin--orbit coupling constant. 
The term $V_{\rm tetra}$ is an additional term 
to reproduce the crystal field of tetragonal symmetry. 
The state $|i,\chi_\sigma (\phi) \rangle$ is expressed by 
$|i,\chi_\sigma (\phi) \rangle$=$|i \rangle |\chi_\sigma (\phi) \rangle$. 
The state $|i \rangle$ is the orbital state, 
defined by 
$|xy \rangle$=$xyf(r)$, 
$|yz \rangle$=$yzf(r)$, 
$|xz \rangle$=$xzf(r)$, 
$|x^2-y^2 \rangle$=$\frac{1}{2}(x^2-y^2)f(r)$, and 
$|3z^3-r^2 \rangle$=$\frac{1}{2\sqrt{3}}(3z^2-r^2)f(r)$, 
with $f(r)$ being the radial part of the 3d orbital, 
where $r$=$\sqrt{x^2 + y^2 + z^2}$. 
The states 
$|xy \rangle$, $|yz \rangle$, and $|xz \rangle$ 
are referred to as $d\varepsilon$ orbitals 
and 
$|x^2-y^2 \rangle$ and $|3z^2-r^2 \rangle$ are 
referred to as $d\gamma$ orbitals. 
The quantity 
$E_\varepsilon$ is the energy level of $|xy \rangle$ and 
$E_\gamma$ is that of $|x^2-y^2 \rangle$. 
The quantity 
$\Delta$ is defined as $\Delta$=$E_\gamma - E_\varepsilon$, 
$\delta_{\varepsilon}$ is the energy difference 
between $|xz \rangle$ (or $|yz \rangle$) and $|xy \rangle$, 
and $\delta_{\gamma}$ is that 
between $|3z^2-r^2 \rangle$ and $|x^2-y^2 \rangle$ 
(see Fig. \ref{energy}). 
The state 
$|\chi_\sigma (\phi) \rangle$ ($\sigma$=$+$, $-$) is the spin state, i.e., 
\begin{eqnarray}
\label{+spin}
&&|\chi_+ (\phi) \rangle = \frac{1}{\sqrt{2}} ( e^{-i\phi} |\uparrow \rangle + |\downarrow \rangle ), \\
\label{-spin}
&&|\chi_- (\phi) \rangle = \frac{1}{\sqrt{2}} ( -e^{-i\phi} |\uparrow \rangle + |\downarrow \rangle ), 
\end{eqnarray}
which are eigenstates 
of $- {\mbox{\boldmath $H$}} \cdot {\mbox{\boldmath $S$}}$. 
Here, $|\chi_+ (\phi) \rangle$ ($|\chi_- (\phi) \rangle$) 
denotes the up spin state (down spin state) 
for the case that 
the quantization axis is chosen along the direction of 
$\langle {\mbox{\boldmath $S$}} \rangle$. 
The state 
$|\uparrow \rangle$ ($|\downarrow \rangle$) represents 
the up spin state (down spin state) 
for the case that 
the quantization axis is chosen along the $z$ axis. 

Regarding the parameters, 
we assume 
the relations 
$\Delta/H \ll 1$, 
$|\lambda|/\Delta \ll 1$, 
$\delta_\varepsilon/\Delta \ll 1$, and $\delta_\gamma/\Delta \ll 1$, 
bearing a typical ferromagnet in mind. 
In particular, 
$H$, $\Delta$, and $|\lambda|$ are roughly set to 
$H$$\sim$1 eV, $\Delta$$\sim$0.1 eV, 
and 
$|\lambda|$$\sim$0.01 eV.\cite{parameter,Yosida} 

\subsection{Wave functions of localized d states}
\label{wf}




To obtain the wave functions of the d states, 
we apply first- and second-order PT to ${\cal H}$ 
of Eq. (\ref{Hamiltonian}). 
Here, ${\cal H}_0$ of Eq. (\ref{Hamiltonian0}) 
is the unperturbed term, 
while $V$ of Eq. (\ref{V}) 
is 
the perturbed term. 
When the matrix of ${\cal H}$ is represented in the basis set 
$|xy,\chi_\sigma (\phi) \rangle$, 
$|yz,\chi_\sigma (\phi) \rangle$, 
$|xz,\chi_\sigma (\phi) \rangle$, 
$|x^2-y^2,\chi_\sigma (\phi) \rangle$, 
and $|3z^2-r^2,\chi_\sigma (\phi) \rangle$, 
the unperturbed system is degenerate 
(see Table \ref{matrix1} in Appendix \ref{H_matrix}). 
We therefore use PT 
for the case that the unperturbed system is degenerate.
\cite{Sakurai,Motizuki} 
First, the unitary transformation is performed 
for the subspace with the basis set 
$|xy,\chi_\sigma (\phi) \rangle$, $|yz,\chi_\sigma (\phi) \rangle$, 
and $|xz,\chi_\sigma (\phi) \rangle$ 
as mentioned in Appendix \ref{zero-order}. 
As a result, we obtain the zero-order states as 
$|\xi_+,\chi_+ (\phi) \rangle$, 
$|\delta_\varepsilon,\chi_+ (\phi) \rangle$, 
$|\xi_-,\chi_+ (\phi) \rangle$, 
$|\xi_+,\chi_- (\phi) \rangle$, 
$|\delta_\varepsilon,\chi_- (\phi) \rangle$, 
and 
$|\xi_-,\chi_- (\phi) \rangle$. 
Here, $\xi_\pm$ and $\delta_\varepsilon$ represent 
the eigenvalues of $V$ in the above subspace, 
where $\xi_\pm$ is given by Eq. (\ref{xi_pm}). 
The respective zero-order states are expressed as 
\begin{eqnarray}
\label{+,+}
&&|\xi_+,\chi_+ (\phi)\rangle = A \Big[ (\delta_\varepsilon - \sqrt{\delta_\varepsilon^2 + \lambda^2} ) |xy, \chi_+ (\phi) \rangle 
+ i \lambda \sin \phi |yz, \chi_+ (\phi) \rangle 
- i\lambda \cos \phi |xz, \chi_+ (\phi) \rangle  \Big], \nonumber \\\\
\label{d,+}
&&|\delta_\varepsilon,\chi_+ (\phi) \rangle = \cos \phi |yz,\chi_+ (\phi) \rangle + \sin \phi |xz, \chi_+ (\phi) \rangle, \\
\label{-,+}
&&|\xi_-,\chi_+ (\phi) \rangle = B \Big[ (\delta_\varepsilon + \sqrt{\delta_\varepsilon^2 + \lambda^2} ) |xy, \chi_+ (\phi) \rangle 
+ i \lambda \sin \phi |yz, \chi_+ (\phi) \rangle 
- i \lambda \cos \phi |xz, \chi_+ (\phi) \rangle  \Big], 
\nonumber \\\\
%
%
\label{+,-}
&&|\xi_+,\chi_- (\phi)\rangle = A \Big[ (\delta_\varepsilon - \sqrt{\delta_\varepsilon^2 + \lambda^2} ) |xy, \chi_- (\phi) \rangle 
- i \lambda \sin \phi |yz, \chi_- (\phi) \rangle 
+ i\lambda \cos \phi |xz, \chi_- (\phi) \rangle  \Big], 
\nonumber \\\\
\label{d,-}
&&|\delta_\varepsilon,\chi_- (\phi)\rangle = \cos \phi |yz,\chi_- (\phi) \rangle + \sin \phi |xz, \chi_- (\phi) \rangle, \\
\label{-,-}
&&|\xi_-,\chi_- (\phi) \rangle = B \Big[ (\delta_\varepsilon + \sqrt{\delta_\varepsilon^2 + \lambda^2} ) |xy, \chi_- (\phi) \rangle 
- i \lambda \sin \phi |yz, \chi_- (\phi) \rangle 
+ i \lambda \cos \phi |xz, \chi_- (\phi) \rangle  \Big], \nonumber \\
%
%
\end{eqnarray}
with
\begin{eqnarray}
\label{AAA}
&&A = (
2 \delta_\varepsilon^2 + 2 \lambda^2 - 2 \delta_\varepsilon \sqrt{\delta_\varepsilon^2 + \lambda^2}
)^{-1/2}, \\
\label{BBB}
&&B=(
2 \delta_\varepsilon^2 + 2 \lambda^2 + 2 \delta_\varepsilon \sqrt{\delta_\varepsilon^2 + \lambda^2}
)^{-1/2}. 
\end{eqnarray}
Next, using the basis set 
$|\xi_+,\chi_\pm (\phi)\rangle$, 
$|\delta_\varepsilon,\chi_\pm (\phi) \rangle$, 
$|\xi_-,\chi_\pm (\phi) \rangle$, 
$|x^2-y^2, \chi_\pm (\phi) \rangle$, 
and 
$|3z^2-r^2, \chi_\pm (\phi) \rangle$, 
we construct the matrix of ${\cal H}$ of Eq. (\ref{Hamiltonian}) 
as shown in Table \ref{matrix2}. 
In the construction, 
we perform, for example, the following operations: 
\begin{eqnarray}
\label{ope_1}
\lambda (L_xS_x + L_yS_y) |\xi_\pm,\chi_- (\phi) \rangle &=&
\frac{1}{\sqrt{2\delta_\varepsilon^2 + 2 \lambda^2 \mp 2\delta_\varepsilon \sqrt{\delta_\varepsilon^2 + \lambda^2}}} 
\nonumber \\
&&
\times \Bigg\{ 
\frac{\sqrt{3}\lambda^2}{2}|3z^2-r^2 \rangle 
(- i \cos 2\phi |\chi_+ (\phi)\rangle  - \sin 2\phi |\chi_- (\phi) \rangle ) \nonumber \\
&&
+i \frac{\lambda^2}{2} |x^2-y^2,\chi_+ (\phi) \rangle -\frac{1}{2} \lambda^2 |xy,\chi_- (\phi) \rangle \nonumber \\
&&
+ i \frac{\lambda}{2}(\delta_\varepsilon \mp \sqrt{\delta_\varepsilon^2 + \lambda^2}) \nonumber \\
&& \times 
[ |xz \rangle (i \sin \phi|\chi_+ (\phi) \rangle -\cos \phi |\chi_- (\phi) \rangle ) \nonumber \\
&&
+|yz\rangle (i \cos \phi |\chi_+ (\phi) \rangle +\sin \phi |\chi_- (\phi) \rangle ) 
] 
\Bigg\}, 
\\
\label{ope_2}
\lambda (L_xS_x + L_yS_y) |\delta_\varepsilon, \chi_- (\phi) \rangle &=&
i \frac{\sqrt{3}\lambda}{2} |3z^2-r^2 \rangle 
(i \sin 2\phi |\chi_+ (\phi) \rangle - \cos 2\phi |\chi_- (\phi) \rangle ) \nonumber \\
&&- i \frac{\lambda}{2} |x^2-y^2,\chi_- (\phi) \rangle 
+ \frac{\lambda}{2} |xy,\chi_+ (\phi) \rangle. 
\end{eqnarray}
Equations (\ref{ope_1}) and (\ref{ope_2}) 
play an important role in $C_2$ and $C_4$ 
as described in the $\phi$ dependence of the wave functions 
in this section. 

\begin{table}
\caption{
Matrix representation of ${\cal H}$ of Eq. (\ref{Hamiltonian}) 
in the basis set 
$|\xi_+,\chi_+ (\phi) \rangle$, 
$|\delta_\varepsilon,\chi_+ (\phi) \rangle$, 
$|\xi_-,\chi_+ (\phi) \rangle$, 
$|\xi_+,\chi_- (\phi) \rangle$, 
$|\delta_\varepsilon,\chi_- (\phi) \rangle$, 
$|\xi_-,\chi_- (\phi) \rangle$, 
$|x^2-y^2, \chi_+ (\phi) \rangle$, 
$|x^2-y^2, \chi_- (\phi) \rangle$, 
$|3z^2-r^2, \chi_+ (\phi) \rangle$, 
and $|3z^2-r^2, \chi_- (\phi) \rangle$. 
Here, $\xi_\pm$ is defined as 
$\xi_\pm$=$(\delta_\varepsilon \pm \sqrt{\delta_\varepsilon^2 + \lambda^2})/2$ 
[see Eq. (\ref{xi_pm})]. 
In addition, $A$ and $B$ are given by Eqs. (\ref{AAA}) and (\ref{BBB}), 
respectively. 
In this table, 
$(\phi)$ in $\chi_\sigma (\phi)$ is omitted due to limited space. 
}
{\tiny 
\hspace*{-2.3cm}
\begin{tabularx}{20cm}{XXXXXXXXXXX}
\hline 
 & $|\xi_+,\chi_+  \rangle$ & $|\delta_\varepsilon,\chi_+  \rangle$ & $|\xi_-,\chi_+  \rangle$ & $|\xi_+,\chi_-  \rangle$ & $|\delta_\varepsilon,\chi_-  \rangle$ & $|\xi_-,\chi_-  \rangle$ & $|x^2-y^2$, $\chi_+  \rangle$ & $|x^2-y^2$, $\chi_-  \rangle$ & $|3z^2-r^2$, $\chi_+  \rangle$ & $|3z^2-r^2$, $\chi_-  \rangle$ \\
\hline 
$\langle \xi_+,\chi_+ |$ & $- \frac{H}{2}+\xi_+$ & 0 & 0 & 0 & $\frac{\lambda A}{2} $ & 0 & 0 & $- i \lambda A$ & $- \frac{\sqrt{3}\lambda^2 A}{2} $ & $i \frac{\sqrt{3}\lambda^2 A}{2}$ \\
 &  &  &  &  & $\times ( 2\xi_- - \lambda )$ &  &  & $\times ( 2\xi_- + \frac{\lambda}{2} )$ & $\times \sin 2\phi$ & $\times \cos 2\phi$ \\
\\
$\langle\delta_\varepsilon,\chi_+  |$ & 0 & $- \frac{H}{2}+\delta_\varepsilon$ & 0 & $\frac{\lambda A}{2} $ & 0 & $\frac{\lambda B}{2}$ & $-i \frac{\lambda}{2}$ & 0 & $- i \frac{\sqrt{3}\lambda}{2}$ & $\frac{\sqrt{3}\lambda}{2}$ \\
 &  &  &  & $\times ( -2\xi_- + \lambda )$ &  & $\times  ( -2\xi_- + \lambda )$ &  &  & $\times \cos 2\phi$ & $\times \sin 2\phi$ \\
\\
$\langle \xi_-,\chi_+  |$ & 0 & 0 & $- \frac{H}{2}+\xi_-$ & 0 & $\frac{\lambda B}{2} $ & 0 & 0 & $- i \lambda B$ & $- \frac{\sqrt{3}\lambda^2 B}{2}$ & $i \frac{\sqrt{3}\lambda^2 B}{2}$ \\
 &  &  &  &  & $\times ( 2\xi_+ - \lambda )$ &  &  & $\times \left( 2\xi_+ + \frac{\lambda}{2} \right)$ & $\times \sin 2\phi$ & $\times \cos 2\phi$ \\
\\
$\langle \xi_+,\chi_- |$ & 0 & $\frac{\lambda A}{2} $ & 0 & $\frac{H}{2}+\xi_+$ & 0 & 0 & $-i \lambda A $ & 0 & $i \frac{\sqrt{3}\lambda^2 A}{2}$ & $- \frac{\sqrt{3}\lambda^2 A}{2}$ \\
 &  & $\times ( -2\xi_- + \lambda )$ &  &  &  &  & $\times \left( 2\xi_- + \frac{\lambda}{2} \right)$ &  & $\times \cos 2\phi$ & $\times \sin 2\phi$ \\
\\
$\langle\delta_\varepsilon,\chi_-|$ & $\frac{\lambda A}{2} $ & 0 & $\frac{\lambda B}{2}$ & 0 & $\frac{H}{2}+\delta_\varepsilon$ & 0 & 0 & $i \frac{\lambda}{2}$ & $- \frac{\sqrt{3}\lambda}{2}$ & $i \frac{\sqrt{3}\lambda}{2}$ \\
 & $\times ( 2\xi_- - \lambda )$ &  & $\times ( 2\xi_+ - \lambda )$ &  &  &  &  &  & $\times \sin 2\phi$ & $\times \cos 2\phi$ \\
\\
$\langle \xi_-,\chi_- |$ & 0 & $\frac{\lambda B}{2} $ & 0 & 0 & 0 & $\frac{H}{2}+\xi_-$ & $-i \lambda B$ & 0 & $i \frac{\sqrt{3}\lambda^2 B}{2}$ & $- \frac{\sqrt{3}\lambda^2 B}{2}$ \\
 & & $\times ( -2\xi_- + \lambda )$ & & & &  & $\times \left( 2\xi_+ + \frac{\lambda}{2} \right)$ &  & $\times \cos 2\phi$ & $\times \sin 2\phi$ \\
\\
$\langle x^2-y^2$, $\chi_+ |$ & 0 & $i \frac{\lambda}{2}$ & 0 & $i \lambda A$ & 0 & $i \lambda B$ & $- \frac{H}{2}+\Delta$ & 0 & 0 & 0 \\
 & & & & $\times \left( 2\xi_- + \frac{\lambda}{2} \right)$ & & $\times \left( 2\xi_+ + \frac{\lambda}{2} \right)$ & & & & \\
\\
$\langle x^2-y^2$, $\chi_- |$ & $i \lambda A$ & 0 &  $i \lambda B$ & 0 & $-i \frac{\lambda}{2}$ & 0 & 0 & $\frac{H}{2}+\Delta$ & 0 & 0 \\
 & $\times \left( 2\xi_- + \frac{\lambda}{2} \right)$ & & $\times \left( 2\xi_+ + \frac{\lambda}{2} \right)$ & & & & & & & \\
\\
$\langle 3z^2-r^2$, $\chi_+ |$ & $- \frac{\sqrt{3}\lambda^2 A}{2}$ & $i \frac{\sqrt{3}\lambda}{2}$ & $- \frac{\sqrt{3}\lambda^2 B}{2}$ & $-i \frac{\sqrt{3}\lambda^2 A}{2}$ & $- \frac{\sqrt{3}\lambda}{2}$ & $-i \frac{\sqrt{3}\lambda^2 B}{2}$ & 0 & 0 & $- \frac{H}{2} + \Delta + \delta_\gamma$ & 0 \\
 & $\times \sin 2\phi$ & $\times \cos 2\phi$ & $\times \sin 2\phi$ & $\times \cos 2\phi$ & $\times \sin 2\phi$ & $\times \cos 2\phi$ &  &  &  &  \\
\\
$\langle 3z^2-r^2$, $\chi_- |$ & $-i \frac{\sqrt{3} \lambda^2 A}{2}$ & $\frac{\sqrt{3}\lambda}{2}$ & $-i \frac{\sqrt{3}\lambda^2 B}{2}$ & $- \frac{\sqrt{3}\lambda^2 A}{2}$ & $-i \frac{\sqrt{3}\lambda}{2}$ & $- \frac{\sqrt{3} \lambda^2 B}{2}$ & 0 & 0 & 0 & $\frac{H}{2} + \Delta + \delta_\gamma$ \\
 & $\times \cos 2\phi$ & $\times \sin 2\phi$ & $\times \cos 2\phi$ & $\times \sin 2\phi$ & $\times \cos 2\phi$ & $\times \sin 2\phi$ &  &  &  &  \\
\hline 
\end{tabularx}
}
\label{matrix2}
\end{table} 

Applying the usual first- and second-order PT to 
${\cal H}$ in Table \ref{matrix2}, 
we obtain 
$|i,\chi_\varsigma (\phi) )$, 
where $i$ ($\varsigma$) 
denotes the orbital index 
(spin index) of the dominant state in $|i,\chi_\varsigma (\phi))$. 
The d state of the up spin $|i,\chi_+(\phi) )$ 
is expressed as
\begin{eqnarray}
\label{|+,+)}
|\xi_+,\chi_+ (\phi)) &=& c_{\xi_+,+} 
( |x^2-y^2,\chi_- (\phi) \rangle 
+ w^{\xi_+,+}_{3z^2-r^2,+}  \sin 2\phi|3z^2-r^2,\chi_+ (\phi) \rangle 
\nonumber \\
&&+ 
w^{\xi_+,+}_{3z^2-r^2,-}  \cos 2\phi |3z^2 -r^2,\chi_- (\phi) \rangle ) 
\ldots~, \\
\label{|delta_e,+)}
|\delta_\varepsilon,\chi_+ (\phi)) &=& c_{\delta_\varepsilon,+} 
( 
|x^2-y^2,\chi_+ (\phi) \rangle 
+ w^{\delta_\varepsilon,+}_{3z^2-r^2,+}  \cos 2\phi|3z^2-r^2,\chi_+ (\phi) \rangle \nonumber \\
&&+ w^{\delta_\varepsilon,+}_{3z^2-r^2,-} \sin 2\phi 
|3z^2 -r^2,\chi_- (\phi) \rangle ) 
\ldots~, \\
\label{|-,+)}
|\xi_-,\chi_+ (\phi)) &=& c_{\xi_-,+} ( |x^2-y^2,\chi_- (\phi) \rangle  
+w^{\xi_-,+}_{3z^2-r^2,+} 
\sin 2\phi|3z^2-r^2,\chi_+ (\phi) \rangle \nonumber \\
&&+ w^{\xi_-,+}_{3z^2-r^2,-}  \cos 2\phi 
|3z^2 -r^2,\chi_- (\phi) \rangle ) 
\ldots~, \\
\label{|x2,+)}
|x^2-y^2,\chi_+ (\phi)) &=&c_{x^2-y^2,+} ( |x^2-y^2,\chi_+ (\phi) \rangle 
+ w^{x^2-y^2,+}_{3z^2-r^2,+}  \cos 2\phi|3z^2-r^2,\chi_+ (\phi) \rangle ) 
\ldots~, \nonumber \\\\
\label{|3z2,+)}
|3z^2-r^2,\chi_+ (\phi)) &=& c_{3z^2-r^2,+} ( |3z^2-r^2,\chi_+ (\phi) \rangle 
+ w^{3z^2-r^2,+}_{x^2-y^2,+}  \cos 2\phi|x^2-y^2,\chi_+ (\phi) \rangle ) 
\ldots~, \nonumber \\
\end{eqnarray}
and the d state of the down spin $|i,\chi_-(\phi))$ 
is expressed as
\begin{eqnarray}
\label{|+,-)}
|\xi_+,\chi_- (\phi) ) &=& c_{\xi_+,-} ( |x^2-y^2,\chi_+ (\phi) \rangle 
+ w^{\xi_+,-}_{3z^2-r^2,+} 
\cos 2\phi|3z^2-r^2,\chi_+ (\phi) \rangle \nonumber \\
&&+ 
w^{\xi_+,-}_{3z^2-r^2,-}  \sin 2\phi |3z^2 -r^2,\chi_- (\phi) \rangle ) 
\ldots~, \\
\label{|delta_e,-)}
|\delta_\varepsilon,\chi_- (\phi)) &=& c_{\delta_\varepsilon,-} 
( 
|x^2-y^2,\chi_- (\phi) \rangle 
+ w^{\delta_\varepsilon,-}_{3z^2-r^2,+} \sin 2\phi 
|3z^2 -r^2,\chi_+ (\phi) \rangle \nonumber \\
&&
+ w^{\delta_\varepsilon,-}_{3z^2-r^2,-}  \cos 2\phi|3z^2-r^2,\chi_- (\phi) \rangle 
) 
\ldots~, \\
\label{|-,-)}
|\xi_-,\chi_- (\phi)) &=& c_{\xi_-,-} ( |x^2-y^2,\chi_+ (\phi) \rangle 
+w^{\xi_-,-}_{3z^2-r^2,+} 
\cos 2\phi|3z^2-r^2,\chi_+ (\phi) \rangle \nonumber \\
&&+ w^{\xi_-,-}_{3z^2-r^2,-}  \sin 2\phi 
|3z^2 -r^2,\chi_- (\phi) \rangle ) 
\ldots~, \\
\label{|x2,-)}
|x^2-y^2,\chi_- (\phi)) &=& c_{x^2-y^2,-} ( |x^2-y^2,\chi_- (\phi)\rangle 
+ w^{x^2-y^2,-}_{3z^2-r^2,-}  \cos 2\phi|3z^2-r^2,\chi_- (\phi) \rangle ) 
\ldots~, \nonumber \\\\
\label{|3z2,-)}
|3z^2-r^2,\chi_- (\phi))&=&c_{3z^2-r^2,-} ( |3z^2-r^2,\chi_- (\phi)\rangle 
+ w^{3z^2-r^2,-}_{x^2-y^2,-}  \cos 2\phi|x^2-y^2,\chi_- (\phi) \rangle ) 
\ldots~. \nonumber \\
\end{eqnarray}
In the right-hand side of Eqs. (\ref{|+,+)})$-$(\ref{|3z2,-)}), 
we specify only 
the $|3z^2-r^2,\chi_\sigma (\phi) \rangle$ 
and $|x^2-y^2,\chi_\sigma (\phi) \rangle$ terms 
because 
these states contribute to the present transport in which 
${\mbox{\boldmath $I$}}$ flows in the $x$ direction 
(see Appendix \ref{selection}). 
The dominant states in 
$|\xi_+,\chi_\sigma (\phi) )$, 
$|\delta_\varepsilon,\chi_\sigma (\phi) )$, 
$|\xi_-,\chi_\sigma (\phi))$, 
$|x^2-y^2,\chi_\sigma (\phi))$, 
and $|3z^2-r^2,\chi_\sigma (\phi))$ 
are respectively written as 
$|\xi_+,\chi_\sigma (\phi) \rangle$, 
$|\delta_\varepsilon,\chi_\sigma (\phi) \rangle$, 
$|\xi_-,\chi_\sigma (\phi) \rangle$, 
$|x^2-y^2,\chi_\sigma (\phi) \rangle$, 
and $|3z^2-r^2,\chi_\sigma (\phi) \rangle$, 
although they are not shown in Eqs. (\ref{|+,+)})$-$(\ref{|-,+)}) 
and (\ref{|+,-)})$-$(\ref{|-,-)}). 
The other states, except for the dominant state 
in each $|i,\chi_\varsigma (\phi))$, 
represent the slightly hybridized states 
due to the spin--orbit interaction. 
The quantity 
$w^{i,\varsigma}_{j,\sigma} \cos 2\phi$ or 
$w^{i,\varsigma}_{j,\sigma} \sin 2\phi$ 
represents the probability amplitude of $|j,\chi_\sigma (\phi) \rangle$ 
normalized by $c_{i,\varsigma}$. 
Here, $w^{i,\varsigma}_{j,\sigma}$ is the coefficient of 
the $\cos 2\phi$ or $\sin 2\phi$ term normalized 
by $c_{i,\varsigma}$, 
while 
$c_{i,\varsigma}$ 
is the coefficient of 
the constant term, which does not depend on $\phi$. 
Such $w^{i,\varsigma}_{j,\sigma}\cos 2\phi$ 
and 
$w^{i,\varsigma}_{j,\sigma}\sin 2\phi$ 
generate 
the twofold and fourfold symmetric terms of $\Delta \rho(\phi)/\rho$ 
as described in Sec. \ref{sec_resistivity}. 

\begin{figure*}
\begin{center}
\includegraphics[width=0.76\linewidth]{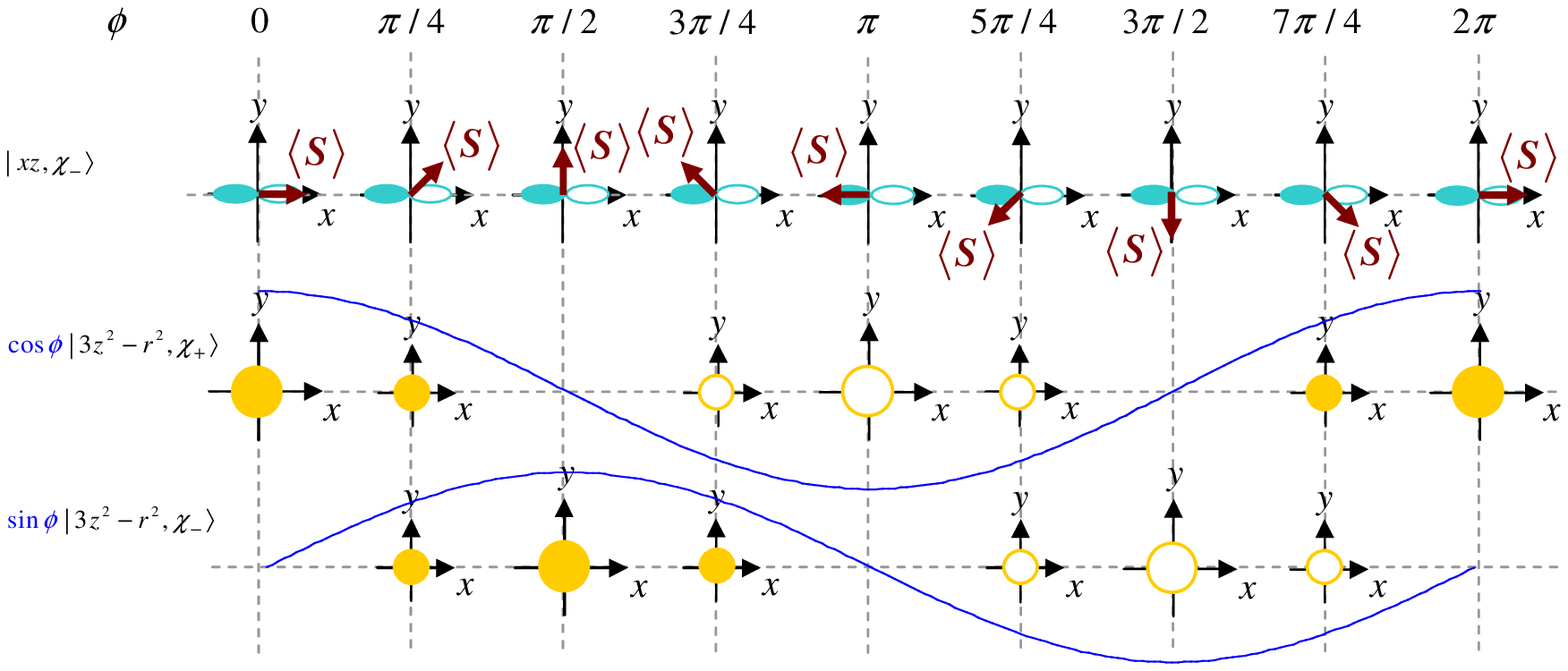}\\
\hspace*{-15cm}(a)\\
\includegraphics[width=0.76\linewidth]{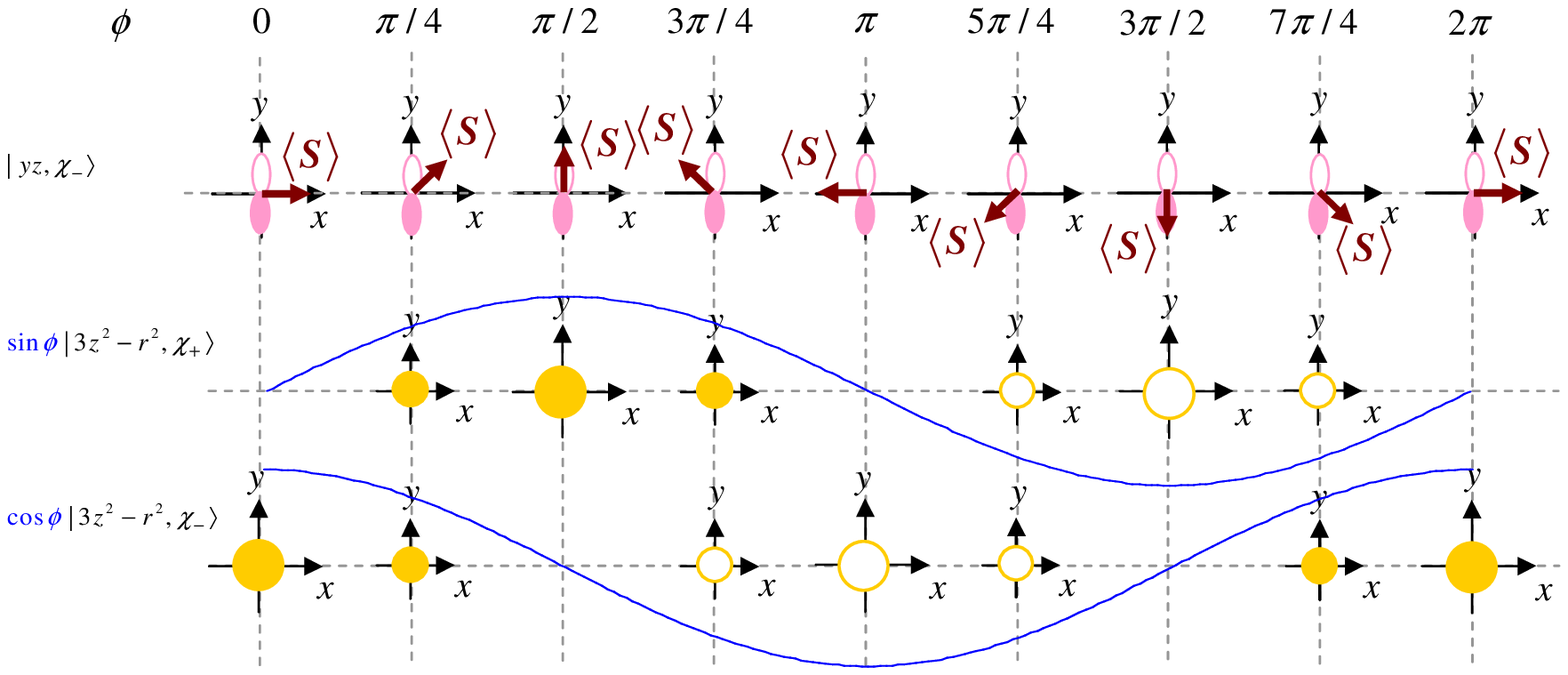}\\
\hspace*{-15cm}(b)
\caption{
(Color online) 
(a) Schematic illustration of 
the $\phi$ dependences of 
$|xz,\chi_- (\phi) \rangle$, 
$\cos \phi |3z^2-r^2,\chi_+ (\phi) \rangle$, 
and 
$\sin \phi |3z^2-r^2, \chi_- (\phi) \rangle$ 
in $\lambda L_yS_y |xz,\chi_- (\phi) \rangle$ 
of Eq. (\ref{LySy}). 
The upper part shows 
the top view (looking down along the $z$ axis) 
of $|xz,\chi_- (\phi) \rangle$. 
The middle part shows 
$\cos \phi |3z^2 - r^2,\chi_+ (\phi) \rangle$ in the $xy$ plane. 
The lower part shows 
$\sin \phi|3z^2 - r^2,\chi_- (\phi) \rangle$ in the $xy$ plane. 
Here, 
$|xz, \chi_- (\phi)$ is shown 
by the sky-blue or sky-blue-bordered orbital 
and 
$|3z^2-r^2, \chi_\pm (\phi) \rangle$ 
is shown by the yellow or yellow-bordered orbital. 
The blue curve in the middle part is $\cos \phi$ 
and that in the lower part is $\sin \phi$. 
Note that the $\phi$ dependent coefficients of 
$|3z^2-r^2, \chi_\pm (\phi) \rangle$ 
are given by only $\cos \phi$ and $\sin \phi$; 
that is, 
the prefactor of $\cos \phi$ or $\sin \phi$ is ignored. 
In addition, 
the color-filled orbitals (white orbitals with a colored border) 
express regions with a negative sign (positive sign) 
in the wave function, 
where the $\phi$ dependent coefficients 
are taken into consideration 
in regard to $|3z^2-r^2, \chi_\pm (\phi) \rangle$. 
(b) Schematic illustration of 
the $\phi$ dependences of 
$|yz,\chi_- (\phi) \rangle$, 
$\sin \phi|3z^2-r^2,\chi_+ (\phi) \rangle$, 
and 
$\cos \phi|3z^2-r^2,\chi_- (\phi) \rangle$ 
in 
$\lambda L_xS_x |yz,\chi_- (\phi) \rangle$ 
of Eq. (\ref{LxSx}). 
The upper part shows 
the top view (looking down along the $z$ axis) 
of $|yz,\chi_- (\phi) \rangle$. 
The middle part shows 
$\sin \phi |3z^2 - r^2,\chi_+ (\phi) \rangle$ in the $xy$ plane. 
The lower part shows 
$\cos \phi|3z^2 - r^2,\chi_- (\phi) \rangle$ in the $xy$ plane. 
Here, $|yz, \chi_- (\phi) \rangle$ 
is shown by the pink or pink-bordered orbital. 
The blue curve in the middle part is $\sin \phi$ 
and that in the lower part is $\cos \phi$. 
The other notation is the same as in (a). 
}
\label{wf_xzyz}
\end{center}
\end{figure*}

\begin{figure*}
\begin{center}
\includegraphics[width=0.7\linewidth]{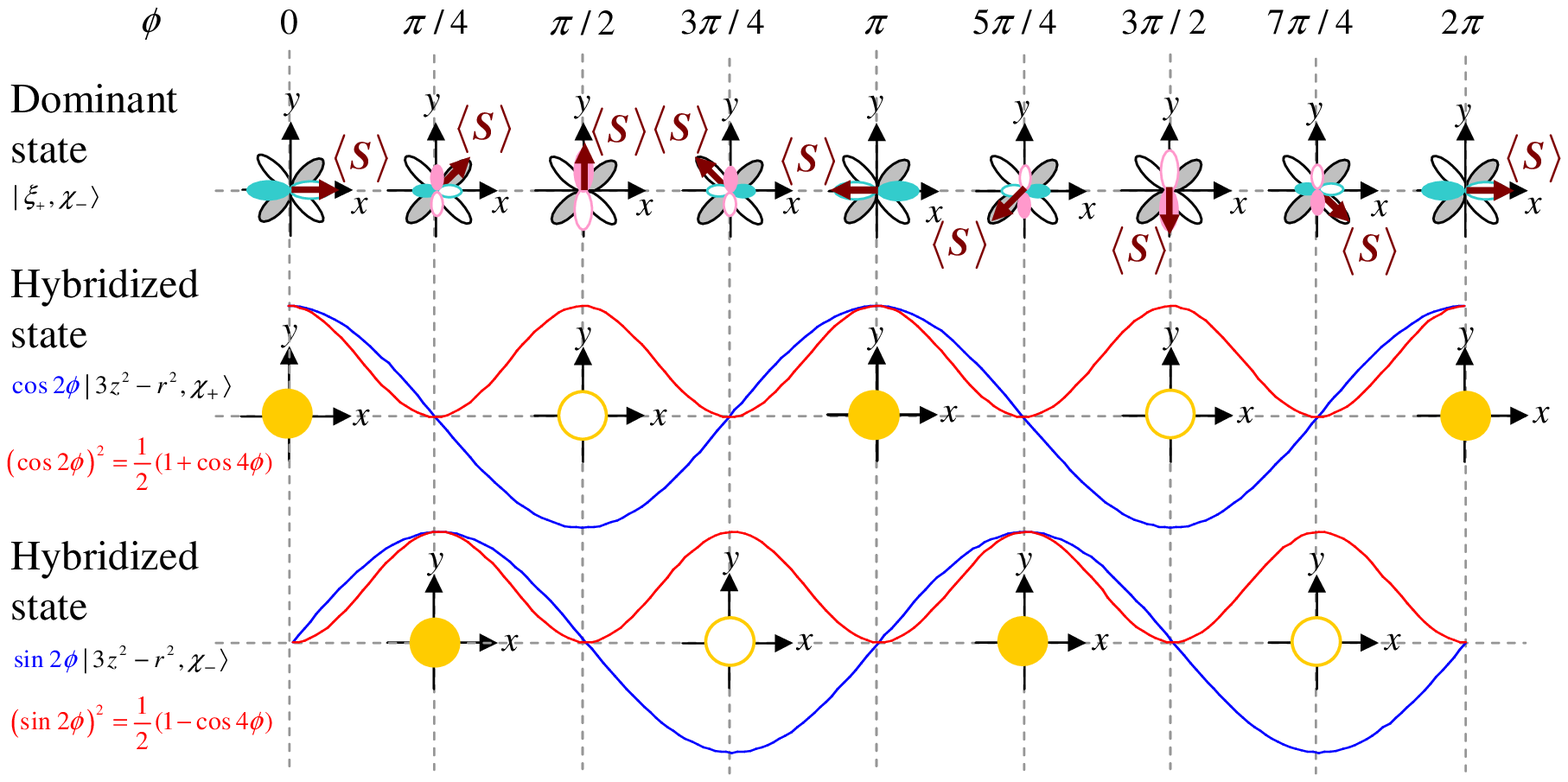}\\[-0.8cm]
\hspace*{-15cm}(a)\\
\vspace{0.2cm}
\includegraphics[width=0.7\linewidth]{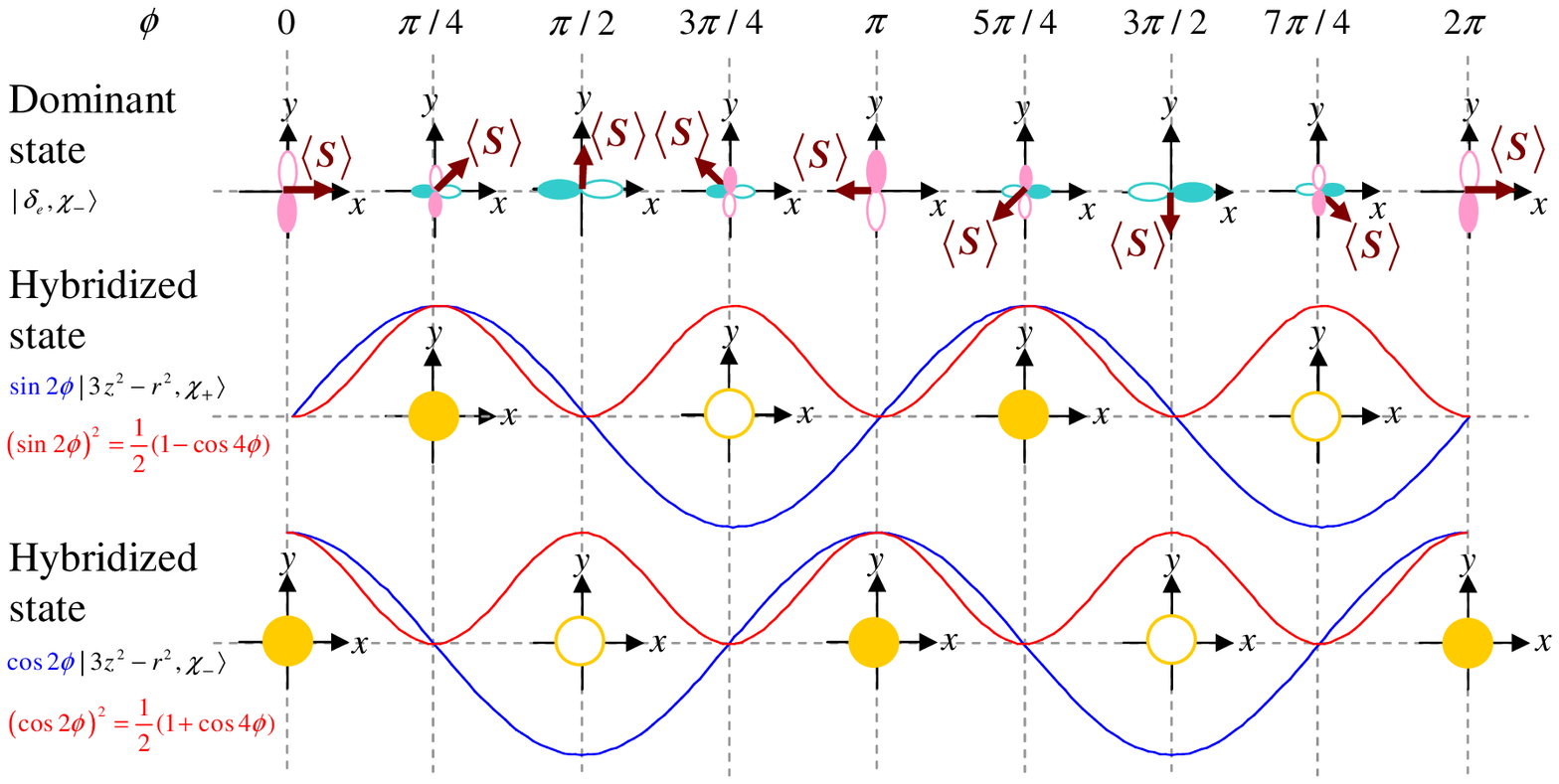}\\[-0.8cm]
\hspace*{-15cm}(b)\\
\vspace{0.2cm}
\includegraphics[width=0.7\linewidth]{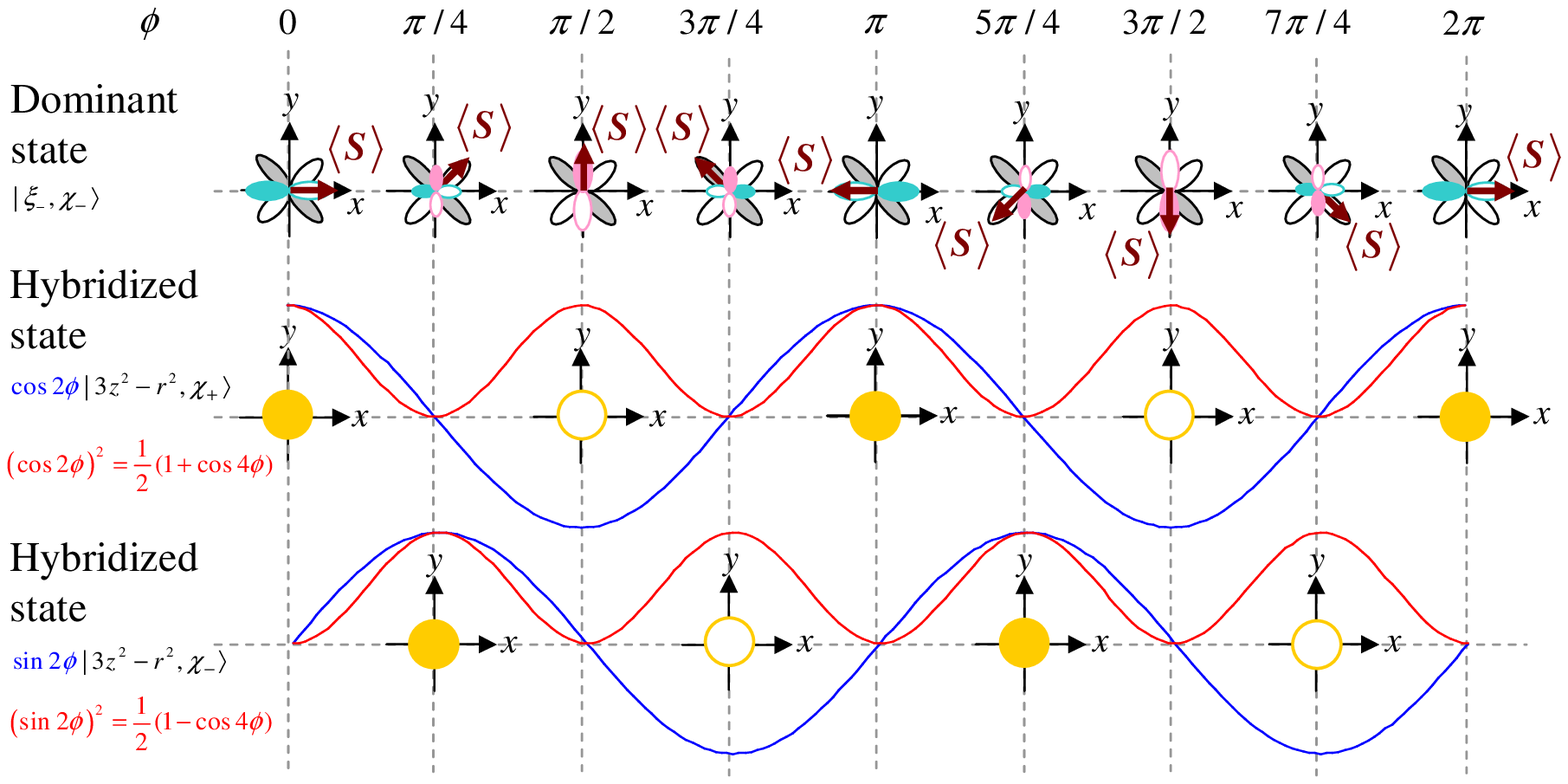}\\[-0.8cm]
\hspace*{-15cm}(c)\\
\caption{\footnotesize 
(Color online) 
Schematic illustration of 
the $\phi$ dependences of 
the dominant states in $d\varepsilon$ states 
and the hybridized states 
in Eqs. (\ref{|+,-)})$-$(\ref{|-,-)}). 
The dominant states in (a), (b), and (c) 
are 
$|\xi_+,\chi_- (\phi) \rangle$ of Eq. (\ref{+,-}), 
$|\delta_\varepsilon,\chi_- (\phi) \rangle$ of Eq. (\ref{d,-}), and 
$|\xi_-,\chi_- (\phi) \rangle$ of Eq. (\ref{-,-}), respectively. 
The hybridized states are 
represented by expressions 
with a probability amplitude 
of $\cos 2\phi$ or $\sin 2\phi$, i.e., 
$\cos 2\phi |3z^2-r^2,\chi_\pm (\phi) \rangle$ 
and 
$\sin 2\phi |3z^2-r^2,\chi_\pm (\phi) \rangle$, 
where 
the prefactor of $\cos 2\phi$ or $\sin 2\phi$ is ignored. 
In each panel, 
the upper part shows 
the top view (looking down along the $z$ axis) 
of the dominant state. 
In (a) and (c), 
the middle part shows 
$\cos 2\phi |3z^2 - r^2,\chi_+ (\phi) \rangle$ in the $xy$ plane, 
and the lower part shows 
$\sin 2\phi|3z^2 - r^2,\chi_- (\phi) \rangle$ in the $xy$ plane. 
In (b), 
the middle part shows 
$\sin 2\phi|3z^2 - r^2,\chi_- (\phi) \rangle$ 
in the $xy$ plane, 
and the lower part shows 
$\cos 2\phi |3z^2 - r^2,\chi_+ (\phi) \rangle$ 
in the $xy$ plane. 
The state $|yz,\chi_- (\phi) \rangle$ is shown 
by the pink or pink-bordered orbital, 
$|xz,\chi_- (\phi) \rangle$ is shown 
by the sky-blue or sky-blue-bordered orbital, 
and 
$|xy,\chi_- (\phi) \rangle$ is shown 
by the gray or gray-bordered orbital. 
The state $|3z^2-r^2,\chi_- (\phi) \rangle$ is represented 
by the yellow or yellow-bordered orbital. 
The color-filled orbitals (white orbitals with a colored border) 
express regions with a negative sign (positive sign) 
in the wave function 
including the probability amplitude. 
In the middle part of (a) and (c) 
and the lower part of (b), 
the blue and red curves 
are $\cos 2\phi$ and $\cos^2 2\phi$ [=$(1+\cos 4\phi)/2$], respectively. 
In the lower parts of (a) and (c) 
and the middle part of (b), 
the blue and red curves 
are $\sin 2\phi$ and $\sin^2 2\phi$ [=$(1-\cos 4\phi)/2$], respectively. 
}
\label{wf_epsilon}
\end{center}
\end{figure*}

\begin{figure*}
\begin{center}
\includegraphics[width=0.7\linewidth]{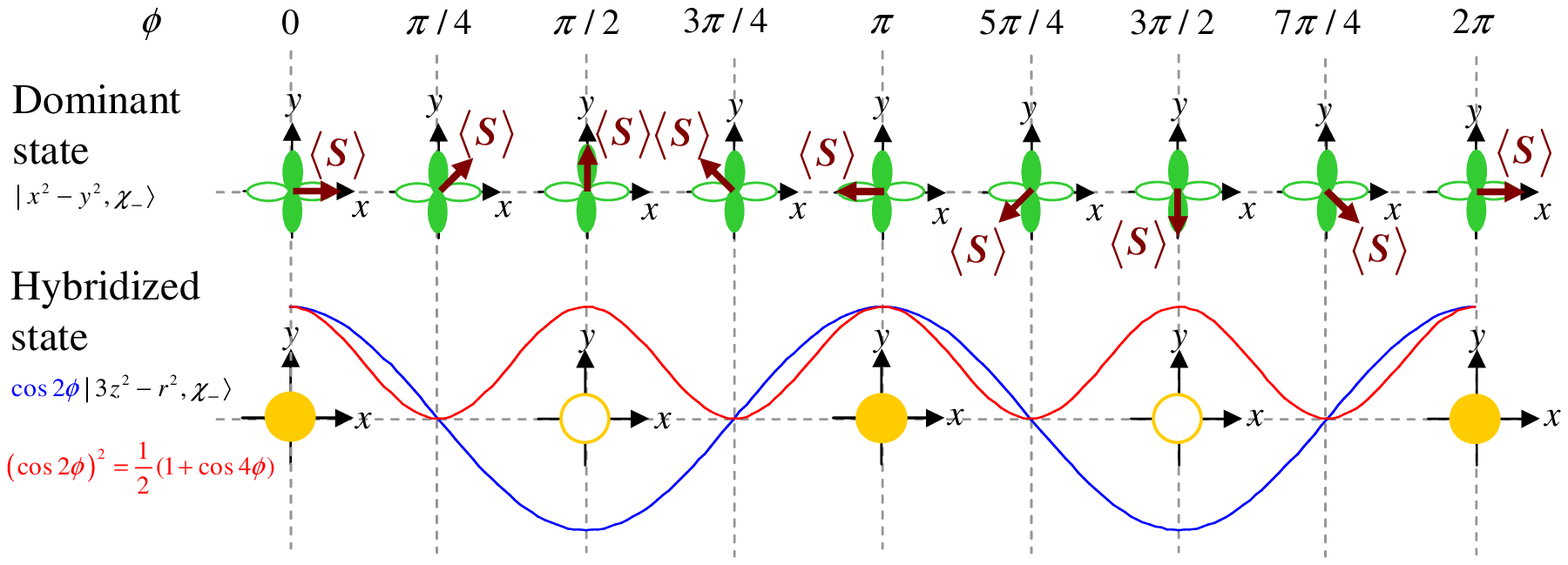}\\[-0.8cm]
\hspace*{-15cm}(a)\\
\vspace{0.2cm}
\includegraphics[width=0.7\linewidth]{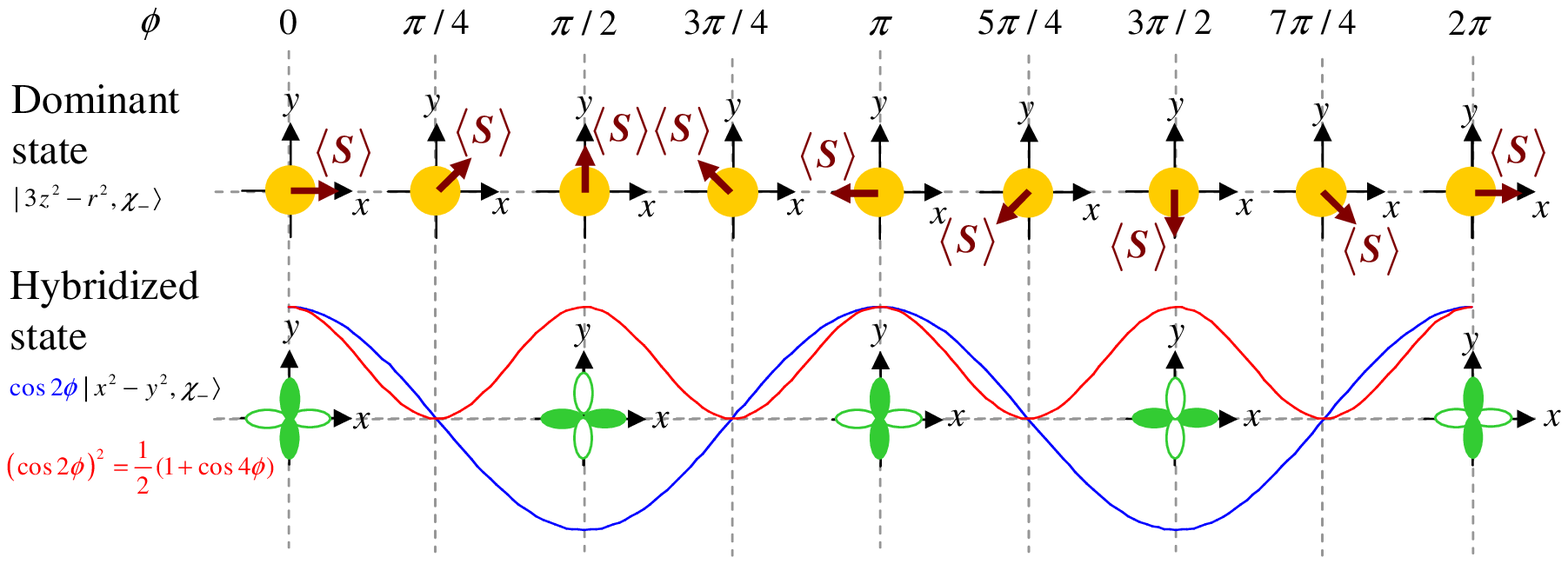}\\[-0.8cm]
\hspace*{-15cm}(b)
\caption{
(Color online) 
Schematic illustration of 
the $\phi$ dependences of 
the dominant states in $d\gamma$ states 
and the hybridized states in Eqs. (\ref{|x2,-)}) and (\ref{|3z2,-)}). 
The dominant states in (a) and (b) 
are 
$|x^2-y^2,\chi_- (\phi) \rangle$ and $|3z^2-y^2,\chi_- (\phi) \rangle$, 
respectively. 
The hybridized states are represented by expressions 
with a probability amplitude of $\cos 2\phi$, i.e., 
$\cos 2\phi |3z^2-r^2,\chi_- (\phi) \rangle$ 
and 
$\cos 2\phi |x^2-y^2,\chi_- (\phi) \rangle$, 
where 
the prefactor of $\cos 2\phi$ is ignored. 
In each panel, 
the upper part shows 
the top view (looking down along the $z$ axis) 
of the dominant state. 
In (a), 
the lower part shows 
$\cos 2\phi |3z^2 - r^2,\chi_- (\phi) \rangle$ in the $xy$ plane. 
In (b), 
the lower part shows 
$\sin 2\phi|x^2 - y^2,\chi_- (\phi) \rangle$ in the $xy$ plane. 
The state $|3z^2-r^2,\chi_- (\phi) \rangle$ is shown 
by the yellow or yellow-bordered orbital. 
The state $|x^2-y^2,\chi_- (\phi) \rangle$ is shown 
by the green or green-bordered orbital. 
The color-filled orbitals (white orbitals with a colored border) 
express regions with a negative sign (positive sign) 
in the wave function, 
where 
the probability amplitude 
is taken into consideration 
in regard to 
$|3z^2-r^2,\chi_- (\phi) \rangle$ 
and 
$|x^2-y^2,\chi_- (\phi) \rangle$. 
In the lower parts of (a) and (b), 
the blue and red curves 
are 
$\cos 2\phi$ and $\cos^2 2\phi$ [=$(1 + \cos 4\phi)/2$], respectively. 
}
\label{wf_gamma}
\end{center}
\end{figure*}

\subsection{Origin of $\cos 2\phi$ and $\sin 2\phi$ terms 
in d states}
We explain the origin of 
the $\cos 2\phi$ and $\sin 2\phi$ terms in 
Eqs. (\ref{|+,+)})$-$(\ref{|3z2,-)}). 
In the $d\varepsilon$ states, 
the $\cos 2\phi$ and $\sin 2\phi$ terms appear 
through $d\varepsilon - d\gamma$ hybridization. 
In the $d\gamma$ states, 
they appear 
owing to 
$d\gamma - d\varepsilon - d\gamma'$ hybridization, 
in which 
the $d\gamma$ states are hybridized to the $d\gamma'$ states 
via the $d\varepsilon$ states. 
These hybridizations are due to 
the specific matrix elements in Table \ref{matrix2}, i.e., 
$\langle 3z^2-r^2, \chi_{\sigma'} (\phi) | {\cal H} | \xi_\pm, \chi_\sigma (\phi) \rangle$ 
and 
$\langle 3z^2-r^2, \chi_{\sigma'} (\phi) | {\cal H} | \delta_\varepsilon, \chi_\sigma (\phi) \rangle$, 
with $\sigma$=$+$, $-$ and $\sigma'$=$+$, $-$. 
We now focus on 
$\langle 3z^2-r^2, \chi_{\sigma'} (\phi) | {\cal H} | \xi_\pm, \chi_- (\phi) \rangle$ 
and 
$\langle 3z^2-r^2, \chi_{\sigma'} (\phi) | {\cal H} | \delta_\varepsilon, \chi_- (\phi) \rangle$, 
where 
$\langle 3z^2-r^2, \chi_{\sigma'} (\phi) | {\cal H} | \xi_\pm, \chi_+ (\phi) \rangle$ 
and 
$\langle 3z^2-r^2, \chi_{\sigma'} (\phi) | {\cal H} | \delta_\varepsilon, \chi_+ (\phi) \rangle$ 
can also be discussed in a similar way. 
These matrix elements originate from 
only the $\cos 2\phi$ and $\sin 2\phi$ terms in 
Eqs. (\ref{ope_1}) and (\ref{ope_2}). 
Here, $\cos 2\phi$ and $\sin 2\phi$ 
are formed by the multiplication of 
the following coefficients: 
\begin{enumerate}
\item[(i)] 
The $\phi$ dependent coefficients of 
$|3z^2-r^2, \chi_- (\phi) \rangle$ in 
$\lambda L_yS_y |xz,\chi_- (\phi) \rangle$ of Eq. (\ref{LySy}) 
and $\lambda L_xS_x |yz,\chi_- (\phi) \rangle$ of  Eq. (\ref{LxSx}).

Note that 
the $\phi$ dependent coefficients of 
$|x^2-y^2, \chi_- (\phi) \rangle$ in Eqs. (\ref{LySy}) and (\ref{LxSx}) 
are not responsible for 
the $\cos 2\phi$ and $\sin 2\phi$ terms 
in Eqs. (\ref{ope_1}) and (\ref{ope_2}). 
\item[(ii)] 
The $\phi$ dependent coefficients of 
$|xz, \chi_- (\phi) \rangle$ 
and 
$|yz, \chi_- (\phi) \rangle$ in 
Eqs. (\ref{+,-})$-$(\ref{-,-}). 
\end{enumerate}
We discuss (i). 
We first emphasize that the operations 
that generate $|3z^2-r^2,\chi_\sigma (\phi) \rangle$ 
are 
$\lambda L_yS_y |xz,\chi_- (\phi) \rangle$ of Eq. (\ref{LySy}) 
and $\lambda L_xS_x |yz,\chi_- (\phi) \rangle$ of  Eq. (\ref{LxSx}). 
In Fig. \ref{wf_xzyz}(a), 
we show the $\phi$ dependences of 
$\cos \phi |3z^2-r^2,\chi_+ (\phi) \rangle$ and 
$\sin \phi |3z^2-r^2,\chi_- (\phi) \rangle$ 
in $\lambda L_yS_y |xz,\chi_- (\phi) \rangle$ of Eq. (\ref{LySy}). 
Here, 
the coefficients of $|3z^2-r^2,\chi_\pm (\phi) \rangle$ 
are given by only $\cos \phi$ and $\sin \phi$; 
that is, 
the prefactor of $\cos \phi$ or $\sin \phi$ is ignored for simplicity. 
When $\phi$=0, 
the coefficient of $|3z^2-r^2,\chi_+ (0) \rangle$ is finite, 
whereas that of $|3z^2-r^2,\chi_- (0) \rangle$ is zero. 
In brief, 
since 
the spin direction of $|\chi_-(0) \rangle$ 
in $|xz,\chi_- (0) \rangle$ is the $x$ direction, 
$S_y |\chi_- (0) \rangle$ 
becomes 
$S_y |\chi_-(0) \rangle$
=$-\frac{i}{2}|\chi_+(0) \rangle$. 
Namely, 
the spin is reversed by 
the operation of $S_y$. 
In contrast, 
when $\phi$=$\pi/2$, 
the coefficient of $|3z^2-r^2,\chi_- (\pi/2) \rangle$ is finite, 
whereas that of $|3z^2-r^2,\chi_+ (\pi/2) \rangle$ is zero. 
In short, 
since the spin direction of $|\chi_-(\pi/2) \rangle$ 
in $|xz,\chi_- (\pi/2) \rangle$ is the $y$ direction, 
$S_y |\chi_-(\pi/2) \rangle$ becomes 
$S_y | \chi_-(\pi/2) \rangle$
=$-\frac{1}{2}| \chi_-(\pi/2)\rangle$. 
Namely, the spin is conserved 
under the operation of $S_y$. 
In a similar way, 
we can consider 
the $\phi$ dependence of 
the coefficient of 
$|3z^2-r^2,\chi_\pm (\phi) \rangle$ 
in 
$\lambda L_xS_x |yz,\chi_- (\phi) \rangle$ of  Eq. (\ref{LxSx}) 
[also see Fig. \ref{wf_xzyz}(b)].

\subsection{Illustration of d states}
In Figs. \ref{wf_epsilon} and \ref{wf_gamma}, 
we show schematic illustrations of 
the $\phi$ dependences of 
the dominant states and hybridized states 
in Eqs. (\ref{|+,-)})$-$(\ref{|3z2,-)}). 
The dominant states are 
$|\xi_+,\chi_- (\phi) \rangle$ of Eq. (\ref{+,-}), 
$|\delta_\varepsilon,\chi_- (\phi)\rangle$ of Eq. (\ref{d,-}), 
$|\xi_-,\chi_- (\phi)\rangle$ of Eq. (\ref{-,-}), 
$|x^2-y^2, \chi_- (\phi) \rangle$, and 
$|3z^2-r^2, \chi_- (\phi) \rangle$. 
The hybridized states are represented by expressions 
with a probability amplitude 
of $\cos 2\phi$ or $\sin 2\phi$, i.e., 
$\cos 2\phi |3z^2-r^2, \chi_\pm (\phi) \rangle$, 
$\sin 2\phi |3z^2-r^2, \chi_\pm (\phi) \rangle$, and 
$\cos 2\phi |x^2-y^2, \chi_- (\phi) \rangle$, 
where 
the prefactor of $\cos 2\phi$ or $\sin 2\phi$ is ignored for simplicity. 
Each probability is also given by 
$\cos^2 2\phi$ [=$(1+\cos 4\phi)/2$] 
or $\sin^2 2\phi$ [=$(1-\cos 4\phi)/2$]. 
Such $\phi$ dependences 
originate from 
the $\phi$ dependent coefficients of 
$|3z^2-r^2,\chi_\sigma (\phi) \rangle$ 
in 
$\lambda(L_x S_x + L_yS_y) |\xi_\pm,\chi_- (\phi) \rangle$ 
of Eq. (\ref{ope_1}) 
and 
$\lambda (L_xS_x + L_yS_y) |\delta_\varepsilon,\chi_- (\phi) \rangle$ of  Eq. (\ref{ope_2}). 
These operations are commented on as follows: 
\begin{enumerate}
\item[(i)] 
$\lambda(L_x S_x + L_yS_y) |\xi_\pm,\chi_- (\phi) \rangle$ of Eq. (\ref{ope_1}) \\
When $\phi$=0, 
only $\lambda L_y S_y | xz, \chi_-(0) \rangle$ 
in this operation 
generates 
$|3z^2-r^2,\chi_+ (0) \rangle$ 
[see the case of $\phi$=0 in Figs. \ref{wf_epsilon}(a) 
and \ref{wf_epsilon}(c)]. 
This feature comes from 
the case of $\phi$=0 in Fig. \ref{wf_xzyz}(a). 
When $\phi$=$\pi/2$, 
only $\lambda L_x S_x | yz, \chi_-(\pi/2) \rangle$ 
in this operation generates 
$|3z^2-r^2,\chi_+ (\pi/2) \rangle$ 
[see the $\phi$=$\pi$/2 case in Figs. \ref{wf_epsilon}(a) 
and \ref{wf_epsilon}(c)]. 
This feature is due to 
the case of $\phi$=$\pi/2$ in Fig. \ref{wf_xzyz}(b). 
\item[(ii)] $\lambda(L_x S_x + L_yS_y) |\delta_\varepsilon,\chi_- (\phi) \rangle$ of  Eq. (\ref{ope_2})\\
When $\phi$=0, 
only $\lambda L_x S_x | yz, \chi_-(0) \rangle$ 
in this operation 
generates $|3z^2-r^2,\chi_- (0) \rangle$ 
[see the case of $\phi$=0 in Fig. \ref{wf_epsilon}(b)]. 
This feature stems from 
the case of $\phi$=0 in Fig. \ref{wf_xzyz}(b). 
When $\phi$=$\pi$/2, 
only $\lambda L_y S_y | xz, \chi_-(\pi/2) \rangle$ 
in this operation 
generates 
$|3z^2-r^2,\chi_- (\pi/2) \rangle$ 
[see the case of $\phi$=$\pi$/2 in Fig. \ref{wf_epsilon}(b)]. 
This feature is due to 
the case of $\phi$=$\pi$/2 in Fig. \ref{wf_xzyz}(a). 
\end{enumerate}
Here, $| xy, \chi_- (\phi) \rangle$ in 
$|\xi_\pm,\chi_- (\phi) \rangle$ 
and $|\delta_\varepsilon,\chi_- (\phi) \rangle$
is not responsible for 
the $\phi$ dependent coefficients 
of $|3z^2-r^2,\chi_- (\phi) \rangle$ 
as found from the fact that 
$\langle 3z^2-r^2, \chi_{\sigma'} (\phi) | {\cal H} | xy, \chi_\sigma (\phi) \rangle$ 
does not depend on $\phi$ (see Table \ref{matrix1}).

\subsection{General expression for resistivity}
\label{sec_resistivity}
Using 
$|i,\chi_\varsigma (\phi))$ of Eqs. (\ref{|+,+)})$-$(\ref{|3z2,-)}), 
we can obtain a general expression for $\rho(\phi)$. 
The resistivity 
$\rho (\phi)$ is first described 
by the two-current model,\cite{Campbell1} i.e.,
\begin{eqnarray}
\label{rho_phi}
&&\rho (\phi) 
= \frac{ \rho_+ (\phi) \rho_- (\phi)}
{\rho_+ (\phi) + \rho_- (\phi)}. 
\end{eqnarray}
The quantity 
$\rho_\sigma (\phi)$ is the resistivity of the $\sigma$ spin at $\phi$ 
with $\sigma$=$+$, $-$, 
where 
$\sigma$=$+$ ($-$) denotes the up spin (down spin) 
for the case in which 
the quantization axis is chosen along the direction of 
$\langle {\mbox{\boldmath $S$}} \rangle$. 
The resistivity $\rho_\sigma (\phi)$ is written as 
\begin{eqnarray}
\label{rho_sigma}
\rho_\sigma (\phi)=\frac{m_{\sigma}^*}{n_{\sigma} e^2 \tau_{\sigma}(\phi)}, 
\end{eqnarray}
where $e$ is the electric charge and 
$n_\sigma$ ($m^*_\sigma$) is the number density (effective mass)
of the electrons in the conduction band 
of the $\sigma$ spin.\cite{Ibach,Grosso} 
The conduction band consists of 
the s, p, and conductive d states.\cite{Kokado1} 
In addition, $1/\tau_{\sigma}(\phi)$ is 
the scattering rate of the conduction electron of the $\sigma$ spin, 
expressed as
\begin{eqnarray}
\label{tau_inv}
&&\frac{1}{\tau_{\sigma}(\phi)} 
= \frac{1}{\tau_{s,\sigma}} + 
\sum_{i} \sum_{\varsigma=+,-}
\frac{1}{\tau_{s,\sigma \to d_i,\varsigma} (\phi)}, 
\end{eqnarray}
with
\begin{eqnarray}
\label{tau_sd_inv}
\frac{1}{\tau_{s,\sigma \to d_i,\varsigma}(\phi)} = 
\frac{2 \pi}{\hbar} n_{\rm imp} N_{\rm n} {V_{\rm imp}(R_{\rm n})}^2 
\left| (i,\chi_\varsigma (\phi)|e^{ik_\sigma x},\chi_\sigma (\phi) \rangle \right|^2 
D_{i,\varsigma}^{(d)}, 
\end{eqnarray}
where $i$=$\xi_+$, $\delta_\varepsilon$, $\xi_-$, 
$x^2-y^2$, and $3z^2-r^2$. 
Here, 
$1/\tau_{s,\sigma}$ is the $s$--$s$ scattering rate, 
which is 
proportional to 
the PDOS of the conduction state of the $\sigma$ spin 
at $E_{\mbox{\tiny F}}$, $D_\sigma^{(s)}$.\cite{Kokado1} 
The $s$--$s$ scattering means that 
the conduction electron of the $\sigma$ spin 
is scattered into the conduction state of the $\sigma$ spin 
by nonmagnetic impurities or phonons. 
The quantity $1/\tau_{s,\sigma \to d_i,\varsigma}(\phi)$ 
is the $s$--$d$ scattering rate.\cite{Kokado1,Kokado2} 
The $s$--$d$ scattering represents 
the scattering of 
the conduction electron of the $\sigma$ spin 
into 
the $\sigma$ spin state in the localized d state of $i$ and $\varsigma$ 
by nonmagnetic impurities. 
The quantities $i$ 
and $\varsigma$ 
respectively denote the orbital and spin indexes 
of the dominant state in $|i,\chi_\varsigma (\phi))$. 
The localized d states 
$|i,\chi_\varsigma (\phi))$ are 
given by Eqs. (\ref {|+,+)})$-$(\ref{|3z2,-)}) 
obtained from ${\cal H}$ of Eq. (\ref{Hamiltonian}). 
The quantity $D_{i,\varsigma}^{(d)}$ represents 
the PDOS of 
the wave function of the tight-binding model 
for the d state of the $i$ orbital and $\varsigma$ spin 
at $E_{\mbox{\tiny F}}$ 
as was described in Ref. \citen{Kokado1}.\cite{comment_H} 
The conduction state of the $\sigma$ spin 
$|e^{ik_\sigma x},\chi_\sigma (\phi) \rangle $ is represented by 
the plane wave, i.e., 
$|e^{ik_\sigma x},\chi_\sigma (\phi) \rangle$=
$(1/\sqrt{\Omega}) e^{ik_\sigma x} |\chi_\sigma (\phi) \rangle$, 
where $k_\sigma$ is the Fermi wavevector of the $\sigma$ spin 
in the $x$ direction 
(i.e., the ${\mbox{\boldmath $I$}}$ direction) 
and $\Omega$ is the volume of the system. 
The quantitiy 
$V_{\rm imp}(R_{\rm n})$ is 
the scattering potential at $R_{\rm n}$ 
due to a single impurity, 
where $R_{\rm n}$ is the distance between 
the impurity and the nearest-neighbor host atom.\cite{Kokado1} 
The quantity 
$N_{\rm n}$ is the number of nearest-neighbor host atoms 
around a single impurity,\cite{Kokado1} 
$n_{\rm imp}$ is the number density of impurities, 
and 
$\hbar$ is the Planck constant $h$ divided by 2$\pi$.




On the basis of 
$\left| (i,\chi_\varsigma (\phi)|e^{ik_\sigma x},\chi_\sigma (\phi) \rangle \right|^2$ described in Appendix \ref{selection}, 
we obtain 
$\sum_i 1/\tau_{s,\sigma \to d_i,-}(\phi)$ in Eq. (\ref{tau_inv}) 
up to the second order of 
$\lambda/H$, $\lambda/\Delta$, 
$\lambda/(H \pm \Delta)$, 
$\delta_t/H$, $\delta_t/\Delta$, or 
$\delta_t/(H \pm \Delta)$, 
with $t$=$\varepsilon$ or $\gamma$. 
Details are given in Appendix \ref{sd_scatter}. 


Using these results, 
we obtain $\rho_\sigma (\phi)$ of Eq. (\ref{rho_sigma}) as
\begin{eqnarray}
\label{rho_s_phi}
\rho_\sigma (\phi)= \rho_{0,\sigma} 
+ \rho_{2,\sigma} \cos 2\phi
+ \rho_{4,\sigma} \cos 4\phi,
\end{eqnarray}
where $\rho_{0,\sigma}$ is the constant term, 
which is independent of $\phi$, 
$\rho_{2,\sigma}$ is the coefficient of the $\cos 2\phi$ term, 
and $\rho_{4,\sigma}$ is that of the $\cos 4\phi$ term. 
These quantities are specified by 
\begin{eqnarray}
&&\rho_{0,\sigma} = \rho_{0,\sigma}^{(0)} + \rho_{0,\sigma}^{(2)},  \\
\label{rho_r2_sigma}
&&\rho_{2,\sigma} = \rho_{2,\sigma}^{(1)} + \rho_{2,\sigma}^{(2)},  \\
\label{rho_r4_sigma}
&&\rho_{4,\sigma} = \rho_{4,\sigma}^{(2)}, 
\end{eqnarray}
where $v$ of $\rho_{u,\sigma}^{(v)}$ ($u$=0, 2, 4 and $v$=0, 1, 2) 
denotes the order of 
$\lambda/H$, $\lambda/\Delta$, 
$\lambda/(H \pm \Delta)$, 
$\delta_t/H$, $\delta_t/\Delta$, or 
$\delta_t/(H \pm \Delta)$, 
with $t$=$\varepsilon$ or $\gamma$. 
The quantities $\rho_{u,\sigma}^{(v)}$ are obtained as
\begin{eqnarray}
%
\label{rho_0_pm^(0)}
&&\rho_{0,\pm}^{(0)} =\rho_{s,\pm} 
+ \frac{3}{4} \rho_{s,\pm \to x^2-y^2,\pm} + \frac{1}{4} \rho_{s,\pm \to 3z^2-r^2,\pm}, \\
&&\rho_{0,\pm}^{(2)} =
\frac{3}{32} \left( \frac{\lambda}{\Delta} \right)^2 
(\lambda^2 A^2 \rho_{s,\pm \to \xi_+,\pm} + 3 \rho_{s,\pm \to \delta_\varepsilon,\pm} + \lambda^2 B^2 \rho_{s,\pm \to \xi_-,\pm}) \nonumber \\
&&\hspace*{1cm}+ \frac{3}{4} \left( \frac{\lambda}{H-\Delta} \right)^2 
\left[ A^2 \left( d_-^2 + \frac{\lambda^2}{8} \right) 
\rho_{s,\pm \to \xi_+,\mp}
+\frac{1}{8} \rho_{s,\pm \to \delta_\varepsilon,\mp}
+B^2 \left( d_+^2 + \frac{\lambda^2}{8} \right) \rho_{s,\pm \to \xi_-,\mp} \right] \nonumber \\
&&\hspace*{1cm}+ \left\{
\frac{3}{4} \left[ - \frac{1}{4}\left( \frac{\lambda}{\Delta} \right)^2 
- \left(\frac{\lambda}{\Delta - H} \right)^2 (A^2 d_-^2 + B^2 d_+^2 ) \right]
\right. \nonumber \\
&&\left.\hspace*{1cm}
+ \frac{3}{128} \left( \frac{\lambda}{\delta_\gamma} \right)^2 
\left( - \frac{\lambda}{\Delta} 
+ \frac{\lambda}{\Delta \mp H} \right)^2 \right\} \rho_{s,\pm \to x^2-y^2,\pm} \nonumber \\
&&\hspace*{1cm}
+\left\{-\frac{3}{16} 
\left[ \left( \frac{\lambda}{\Delta} \right)^2 
+ \left( \frac{\lambda}{\Delta \mp H} \right)^2 \right]
+ \frac{9}{128} 
\left( \frac{\lambda}{\delta_\gamma} \right)^2 
\left( \frac{\lambda}{\Delta} - \frac{\lambda}{\Delta \mp H} \right)^2 
\right\} \rho_{s,\pm \to 3z^2-r^2,\pm}, \nonumber \\\\
\label{rho_2_pm^(1)}
&&\rho_{2,\pm}^{(1)} =
\frac{3}{8} \frac{\lambda}{\delta_\gamma} \left( \frac{\lambda}{\Delta}
- \frac{\lambda}{\Delta \mp H} \right) (\rho_{s,\pm \to x^2-y^2,\pm} - \rho_{s,\pm \to 3z^2-r^2,\pm}), \\
&&\rho_{2,\pm}^{(2)} =
\frac{3}{4} \left[ -\frac{1}{2} \left( \frac{\lambda}{\Delta} \right)^2 
\rho_{s,\pm \to \delta_\varepsilon,\pm} + \left( \frac{\lambda}{H \mp \Delta}\right)^2
( \lambda A^2 d_- \rho_{s,\pm \to \xi_+,\mp} + \lambda B^2 d_+ \rho_{s,\pm \to \xi_-,\mp} ) \right] \nonumber \\
&&\hspace*{1cm}+\frac{\lambda}{\delta_\gamma} \Bigg\{
\frac{3}{4} \left[ \frac{1}{2} \frac{\lambda \delta_\varepsilon}{\Delta^2} 
- \left( \frac{\lambda}{\Delta \mp H} \right)^2 
(\xi_+ A^2 d_- + \xi_- B^2 d_+) \right]   \nonumber \\
&&\hspace*{1cm} -\frac{3}{16} \frac{\lambda^2}{\Delta(\Delta \mp H)}
(\lambda A^2 l_+ + \lambda B^2 l_- -2 A^2 d_- l_+ - 2B^2 d_+ l_-) 
\Bigg\} \rho_{s,\pm \to x^2-y^2,\pm} \nonumber \\
&&\hspace*{1cm}+\frac{\lambda}{\delta_\gamma} 
\Bigg\{
-\frac{3}{4} \left\{ -\frac{1}{2} \frac{\lambda (\delta_\gamma-\delta_\varepsilon)}{\Delta^2} 
+ \left( \frac{\lambda}{\Delta \mp H} \right)^2 
\left[ A^2 (\delta_\gamma - \xi_+) d_- + B^2 (\delta_\gamma - \xi_-) d_+ \right]\right\} \nonumber \\
&& \hspace*{1cm} -\frac{3}{16} \frac{\lambda^2}{\Delta(\Delta \mp H)}
(- \lambda A^2 l_+ - \lambda B^2 l_- + 2 A^2 l_+ d_- + 2 B^2 l_- d_+) 
\Bigg\} 
\rho_{s,\pm \to 3z^2-r^2,\pm}, \\
\label{rho_4_pm^(2)}
&&\rho_{4,\pm}^{(2)} =
\frac{3}{32} \left( \frac{\lambda}{\Delta} \right)^2 
(-\lambda^2 A^2 \rho_{s,\pm \to \xi_+,\pm} + \rho_{s,\pm \to \delta_\varepsilon,\pm} 
- \lambda^2 B^2 \rho_{s,\pm \to \xi_-,\pm} ) \nonumber \\
&& \hspace*{1cm}+\frac{3}{32} \left( \frac{\lambda}{H \mp \Delta} \right)^2
(\lambda^2 A^2 \rho_{s,\pm \to \xi_+,\mp} - \rho_{s,\pm \to \delta_\varepsilon,\mp} + 
\lambda^2 B^2 \rho_{s,\pm \to \xi_-,\mp} ) \nonumber \\
&&\hspace*{1cm}+\frac{3}{128} \left( \frac{\lambda}{\delta_\gamma} \right)^2
\left( - \frac{\lambda}{\Delta} + \frac{\lambda}{\Delta\mp H} \right)^2 
\rho_{s,\pm \to x^2-y^2,\pm} \nonumber \\
&&\hspace*{1cm}+ \frac{9}{128} \left( \frac{\lambda}{\delta_\gamma} \right)^2 
\left( \frac{\lambda}{\Delta} - \frac{\lambda}{\Delta\mp H} \right)^2 
\rho_{s,\pm \to 3z^2-r^2,\pm},
\end{eqnarray}
with
\begin{eqnarray}
&&\rho_{s,\sigma}= \frac{m_\sigma^*}{n_\sigma e^2 \tau_{s,\sigma}}, \\
\label{rho_i_pm}
&&\rho_{s \sigma \to d_i,\varsigma}
= \frac{m_\sigma^*}{n_\sigma e^2 \tau_{s,\sigma \to d_i,\varsigma}}, \\
&&d_\pm = \delta_\varepsilon + \lambda/2 \pm \sqrt{\delta_\varepsilon^2 + \lambda^2}, \\
&&l_\pm = \lambda - \delta_\varepsilon \pm \sqrt{\delta_\varepsilon^2 + \lambda^2}, 
\end{eqnarray}
where $A$, $B$, and $\xi_\pm$ are respectively given 
by Eqs. (\ref{AAA}), (\ref{BBB}), and (\ref{xi_pm}) 
and $\lambda^2 (A^2 + B^2)$=1 and 
$\lambda (A^2 d_- + B^2 d_+ )$=1/2 are used. 
Here, $\rho_{s,\sigma}$ is the $s$--$s$ resistivity 
and 
$\rho_{s \sigma \to d_i,\varsigma}$ is the $s$--$d$ resistivity. 
The $s$--$d$ scattering rate $1/\tau_{s,\sigma \to d_i,\varsigma}$ 
is defined by
\begin{eqnarray}
\label{1/tau_i_pm}
&&\frac{1}{\tau_{s,\sigma \to d_i,\varsigma}}=
\frac{2\pi}{\hbar} n_{\rm imp}N_{\rm n} {V_{\rm imp}(R_{\rm n})}^2  
\left|\langle 3z^2 -r^2,\chi_\sigma (\phi) | e^{ik_\sigma z}, \chi_\sigma (\phi) \rangle \right|^2 
D_{i,\varsigma}^{(d)} \nonumber \\
&&\hspace*{1.5cm}=
\frac{2\pi}{\hbar} 
n_{\rm imp}N_{\rm n}\frac{1}{3}v_\sigma^2 D_{i, \varsigma}^{(d)},
\end{eqnarray}
with 
\begin{eqnarray}
\label{v_sigma}
v_{\sigma}=V_{\rm imp}(R_{\rm n}) g_\sigma, 
\end{eqnarray}
where 
$g_\sigma$ is given by Eq. (\ref{g_sigma}). 
The overlap integral 
$\langle 3z^2 -r^2,\chi_\sigma (\phi) | 
e^{ik_\sigma z}, \chi_\sigma (\phi) \rangle$ in Eq. (\ref{1/tau_i_pm}) 
can be calculated using Eq. (\ref{integral}). 
Note that Eq. (\ref{1/tau_i_pm}) 
has been introduced 
to investigate the relation 
between the present result and 
the previous results\cite{Campbell1,Kokado1} 
(see Appendix \ref{relation}). 
Equation (\ref{1/tau_i_pm}) was used 
in the previous models.\cite{Campbell1,Kokado1} 

On the basis of Eqs. (\ref{rho_sigma})$-$(\ref{rho_s_phi}) 
and (\ref{selection1})$-$(\ref{type2}) 
and Appendix \ref{sd_scatter}, 
the features of $\rho_{2,\sigma}\cos 2\phi$ 
and $\rho_{4,\sigma}\cos 4\phi$ in Eq. (\ref{rho_s_phi}) 
are described as follows: 
\begin{enumerate}
\item[(i)] The resistivity $\rho_{2,\sigma} \cos 2\phi$ 
[see Eqs. (\ref{rho_s_phi}) and (\ref{rho_r2_sigma})] 
is related to the real part of the probability amplitudes 
of $|3z^2 -r^2, \chi_\sigma (\phi) \rangle$ and $|x^2 -y^2, \chi_\sigma (\phi) \rangle$, 
which are given by 
${\rm Re}[w_{3z^2-r^2,\sigma}^{i,\varsigma}]\cos 2\phi$ 
and 
${\rm Re}[w_{x^2-y^2,\sigma}^{i,\varsigma}]\cos 2\phi$, respectively 
[see Eqs. (\ref{X_{2phi,up}}) and (\ref{X_{2phi,dw}})]. 
The quantity $\cos 2\phi$ in the probability amplitude 
is shown in Figs. \ref{wf_epsilon} and \ref{wf_gamma}. 
\item[(ii)] 
The resistivity $\rho_{4,\sigma} \cos 4\phi$ 
[see Eqs. (\ref{rho_s_phi}) and (\ref{rho_r4_sigma})] 
is related to the probabilities of 
$|3z^2 -r^2, \chi_\sigma (\phi) \rangle$ and $|x^2 -y^2, \chi_\sigma (\phi) \rangle$, 
which are given by 
$|w_{3z^2-r^2,\sigma}^{i,\varsigma}|^2 ( 1 \pm \cos 4\phi)/2$ 
and 
$|w_{x^2-y^2,\sigma}^{i,\varsigma}|^2 ( 1+\cos 4\phi)/2$, respectively 
[see Eqs. (\ref{X_{4phi,up}}) and (\ref{X_{4phi,dw}})]. 
The quantity $(1 \pm \cos 4\phi)/2$ in the probability 
is shown in Figs. \ref{wf_epsilon} and \ref{wf_gamma}. 
\end{enumerate}

\subsection{General expressions for C$_2$ and C$_4$}
\label{general_c2c4}

Using Eqs. (\ref{AMR}), (\ref{rho_phi}), 
and (\ref{rho_s_phi})$-$(\ref{rho_r4_sigma}), 
we obtain a general expression for $\Delta \rho (\phi)/\rho$ 
up to the second order of 
$\lambda/H$, $\lambda/\Delta$, 
$\lambda/(H \pm \Delta)$, 
$\delta_t/H$, $\delta_t/\Delta$, or 
$\delta_t/(H \pm \Delta)$, 
with $t$=$\varepsilon$ or $\gamma$. 
The AMR ratio $\Delta \rho (\phi)/\rho$ is explicitly expressed by 
the form 
$\Delta \rho (\phi)/\rho$=$C_0 + C_2 \cos 2\phi + C_4 \cos 4\phi$, 
where $C_0$=$C_2 - C_4$. 
The coefficients $C_2$ and $C_4$ are written as
\begin{eqnarray}
\label{C_2}
&&C_2=-\frac{\rho_{2,+}^{(1)} + \rho_{2,-}^{(1)}}
{( \rho_{0,+}^{(0)} + \rho_{0,-}^{(0)} )^2}
\left( \frac{\rho_{0,+}^{(0)}}{\rho_{0,-}^{(0)}}
\rho_{2,-}^{(1)}
+ \frac{\rho_{0,-}^{(0)}}{\rho_{0,+}^{(0)}}
\rho_{2,+}^{(1)}\right)
+ 
\frac{1}
{\rho_{0,+}^{(0)} + \rho_{0,-}^{(0)}}
\left( \frac{\rho_{0,+}^{(0)}}{\rho_{0,-}^{(0)}}
\rho_{2,-}^{(2)}
+ \frac{\rho_{0,-}^{(0)}}{\rho_{0,+}^{(0)}}
\rho_{2,+}^{(2)}\right) \nonumber \\
&&\hspace*{0.9cm}+
\frac{1}
{\rho_{0,+}^{(0)} + \rho_{0,-}^{(0)}}
\left( \frac{\rho_{2,-}^{(1)}}{\rho_{0,-}^{(0)}}
+ \frac{\rho_{2,+}^{(1)}}{\rho_{0,+}^{(0)}} \right)
\left( \frac{\rho_{0,+}^{(0)}}{\rho_{0,-}^{(0)}}
\rho_{2,-}^{(1)}
+ \frac{\rho_{0,-}^{(0)}}{\rho_{0,+}^{(0)}}
\rho_{2,+}^{(1)}\right) \nonumber \\
&&\hspace*{0.9cm} + 
\frac{1}
{\rho_{0,+}^{(0)} + \rho_{0,-}^{(0)}}
\left( \frac{\rho_{0,+}^{(0)}}{\rho_{0,-}^{(0)}}
\rho_{2,-}^{(1)}
+ \frac{\rho_{0,-}^{(0)}}{\rho_{0,+}^{(0)}}
\rho_{2,+}^{(1)}\right),
\\
\label{C_4}
&&C_4=
\frac{\rho_{0,+}^{(0)} \rho_{4,-}^{(2)}}
{\rho_{0,-}^{(0)} ( \rho_{0,+}^{(0)} + \rho_{0,-}^{(0)} )}
+
\frac{\rho_{0,-}^{(0)} \rho_{4,+}^{(2)}}
{\rho_{0,+}^{(0)} ( \rho_{0,+}^{(0)} + \rho_{0,-}^{(0)} )}
+
\frac{1}{2}
\frac{\rho_{2,+}^{(1)} \rho_{2,-}^{(1)}}
{\rho_{0,+}^{(0)} \rho_{0,-}^{(0)}} \nonumber \\
&&\hspace*{0.9cm}
-\frac{1}{2}\frac{\rho_{2,+}^{(1)} + \rho_{2,-}^{(1)}}
{( \rho_{0,+}^{(0)} + \rho_{0,-}^{(0)} )^2}
\left( \frac{\rho_{0,+}^{(0)}}{\rho_{0,-}^{(0)}}
\rho_{2,-}^{(1)}
+ \frac{\rho_{0,-}^{(0)}}{\rho_{0,+}^{(0)}}
\rho_{2,+}^{(1)}\right), 
\end{eqnarray}
where $\rho_{u,\sigma}^{(v)}$ is given by 
Eqs. (\ref{rho_0_pm^(0)})$-$(\ref{rho_4_pm^(2)}). 
Using Eqs. (\ref{C_2}), (\ref{C_4}), 
and (\ref{rho_0_pm^(0)})$-$(\ref{rho_4_pm^(2)}), 
we can investigate $C_2$ and $C_4$ for various ferromagnets. 
Also note that 
$\Delta \rho(0)/\rho$ (=$2C_2$) of the present model 
coincides with 
that of our previous model\cite{Kokado1} and 
that of the CFJ model\cite{Campbell1} 
under appropriate conditions (see Appendix \ref{relation}). 

\subsection{Calculation method of $C_2$ and $C_4$ 
by exact diagonalization method}
\label{edm}
As a different approach from PT, 
we perform a numerical calculation of 
$C_2$ and $C_4$ using the d states, 
which are obtained by applying the EDM 
to ${\cal H}$ in Table \ref{matrix2}. 
The first purpose of this approach is 
to find the crystal field that leads to $C_4$$\ne$0. 
The second purpose is 
to check the validity of the results obtained by PT 
(see Sec. \ref{appl_strong}). 
The calculation in the EDM 
is as follows: 
\begin{enumerate}
\item[(i)] We numerically obtain 
$\left|i,\chi_\varsigma (\phi) \right)$ in Eq. (\ref{tau_sd_inv}) 
by applying the EDM to ${\cal H}$ in Table \ref{matrix2}. 
\item[(ii)] 
Utilizing the obtained $\left|i,\chi_\varsigma (\phi) \right)$ and 
Table \ref{tab1}, 
we numerically calculate 
$\left| ( i,\chi_\varsigma (\phi)|e^{ik_\sigma x},\chi_\sigma (\phi) \rangle \right|^2$ 
of Eq. (\ref{selection1}). 
\item[(iii)] 
Using the calculated 
$\left| (i,\chi_\varsigma (\phi)|e^{ik_\sigma x},\chi_\sigma (\phi) \rangle \right|^2$, 
we obtain a numerical value for $\Delta \rho (\phi)/\rho$ 
of Eq. (\ref{AMR}) with 
Eqs. (\ref{rho_phi})$-$(\ref{tau_sd_inv}). 
The numerical values of $\Delta \rho (0)/\rho$ 
and $\Delta \rho (\pi/4)/\rho$ 
are represented by $f_0$ and $f_{\pi/4}$, respectively. 
\item[(iv)] 
When the AMR ratio is expressed as Eq. (\ref{AMR_c1}), we have
\begin{eqnarray}
\label{f_0}
&&\frac{\Delta \rho (0)}{\rho}=2 C_2 = f_0, \\
\label{f_pi/4}
&&\frac{\Delta \rho (\pi/4)}{\rho}=C_2 - 2 C_4 = f_{\pi/4}. 
\end{eqnarray}
From Eqs. (\ref{f_0}) and (\ref{f_pi/4}), 
we obtain $C_2$ and $C_4$ as
\begin{eqnarray}
\label{C_2_exact}
&&C_2=\frac{f_0}{2}, \\
\label{C_4_exact}
&&C_4=\frac{f_0}{4} - \frac{f_{\pi/4}}{2}.
\end{eqnarray}
\end{enumerate}

\section{Application to Strong Ferromagnets}
\label{appl_strong}

On the basis of 
$C_2$ of Eq. (\ref{C_2}) and $C_4$ of Eq. (\ref{C_4}), 
we obtain expressions for $C_2$ and $C_4$ for a strong ferromagnet 
with $D_{i,+}^{(d)}$=0 and $D_{i,-}^{(d)}$$\ne$0. 
The coefficients $C_2$ and $C_4$ 
are compared with those obtained by the EDM. 
In addition, from the results of the EDM 
we find that 
$C_4$ 
appears under a crystal field of tetragonal symmetry, 
whereas 
it vanishes under a crystal field of cubic symmetry. 

\subsection{Expressions for $C_2$ and $C_4$}
\label{C2C4}

Using Eqs. (\ref{rho_0_pm^(0)})$-$(\ref{rho_4_pm^(2)}), 
(\ref{C_2}), and (\ref{C_4}), 
we obtain expressions for $C_2$ and $C_4$ for a simple system with 
$D_{\xi_+,-}^{(d)}$=$D_{\xi_-,-}^{(d)}$ and 
$D_{x^2-y^2,-}^{(d)}$=$D_{3z^2-r^2,-}^{(d)}$. 
The relation 
$D_{x^2-y^2,-}^{(d)}$=$D_{3z^2-r^2,-}^{(d)}$ 
gives 
\begin{eqnarray}
\label{000}
\rho_{2,-}^{(1)}=0, 
\end{eqnarray}
where $\rho_{2,-}^{(1)}$ is given by Eq. (\ref{rho_2_pm^(1)}). 
In addition, 
in accordance with previous studies\cite{previous} 
we assume 
$n_+$=$n_-$, 
$m_+^*$=$m_-^*$, and $v_+$=$v_-$, 
where 
$v_+$=$v_-$ is satisfied by setting $k_+$=$k_-$ 
in Eqs. (\ref{v_sigma}) and (\ref{g_sigma}). 
The expressions for $C_2$ and $C_4$ 
are then written as 
\begin{eqnarray}
\label{C_2_strong}
&&C_2= 
\frac{1}
{\rho_{0,+}^{(0)} + \rho_{0,-}^{(0)}}
\left( \frac{\rho_{0,+}^{(0)}}{\rho_{0,-}^{(0)}}
\rho_{2,-}^{(2)}
+ \frac{\rho_{0,-}^{(0)}}{\rho_{0,+}^{(0)}}
\rho_{2,+}^{(2)}\right) \nonumber \\
&&\hspace*{0.5cm}=\frac{3}{8} \frac{1}{1 + r + r_\gamma}
\left[ 
\left( \frac{\lambda}{\Delta} \right)^2 
\frac{r_\gamma-r_{\varepsilon1}}{r + r_\gamma}
+\left( \frac{\lambda}{H - \Delta} \right)^2 
r_{\varepsilon 2} (r + r_\gamma) 
- 
\left( \frac{\lambda}{H + \Delta} \right)^2 
\frac{r_\gamma}{r + r_\gamma}
\right],\nonumber \\\\
\label{C_4_strong0}
&&C_4= 
\frac{\rho_{0,+}^{(0)} \rho_{4,-}^{(2)}}
{\rho_{0,-}^{(0)} ( \rho_{0,+}^{(0)} + \rho_{0,-}^{(0)} )}
+
\frac{\rho_{0,-}^{(0)} \rho_{4,+}^{(2)}}
{\rho_{0,+}^{(0)} ( \rho_{0,+}^{(0)} + \rho_{0,-}^{(0)} )} \nonumber \\
&&\hspace*{0.5cm}
=\frac{3}{32} \frac{r_{\varepsilon1} - r_{\varepsilon2}}{1+r+r_\gamma}
\left[
\left( \frac{\lambda}{\Delta} \right)^2 
\frac{1}{r+r_\gamma}
-\left(\frac{\lambda}{H-\Delta} \right)^2 (r+r_\gamma)
\right] +c_4', \\
\label{c_4'_strong}
&&
c_4'=\frac{3}{32} 
\frac{ r_\gamma}{(1 + r+r_\gamma)(r + r_\gamma)}
\left( \frac{\lambda}{\delta_\gamma} \right)^2 
\left( \frac{\lambda}{\Delta} - \frac{\lambda}{H+\Delta} \right)^2. 
\end{eqnarray}
Here, we have
\begin{eqnarray}
&&r=\frac{\rho_{s,-}}{\rho_{s,+}}, \\
&&r_{\varepsilon 1}=r_{\delta_\varepsilon,-}, \\
&&r_{\varepsilon 2}=r_{\xi_+,-}=r_{\xi_-,-}, \\
&&r_\gamma=r_{x^2-y^2,-}=r_{3z^2-r^2,-}, 
\end{eqnarray}
where 
\begin{eqnarray}
r_{i,-}= \frac{\rho_{s \to d_i,-}}{\rho_{s,+}},
\end{eqnarray}
with $i$=$\xi_+$, $\delta_\varepsilon$, $\xi_-$, $x^2-y^2$, and $3z^2-r^2$. 
The resistivity 
$\rho_{s \to d_i,-}$ is given by Eqs. (\ref{rho_i_pm}) and (\ref{1/tau_i_pm}), 
where $\sigma$ in $\rho_{s,\sigma \to d_i,-}$ is unspecified 
because 
the $\sigma$ dependences of 
$n_\sigma$, $m_\sigma^*$, and $k_\sigma$ 
are ignored as noted above. 
Furthermore, we note that 
$r_{i,-}$ satisfies the relation 
$r_{i,-} \propto D_{i,-}^{(d)}$ [see Eq. (\ref{1/tau_i_pm})]. 

\begin{figure}[ht]
\begin{center}
\includegraphics[width=.5\linewidth]{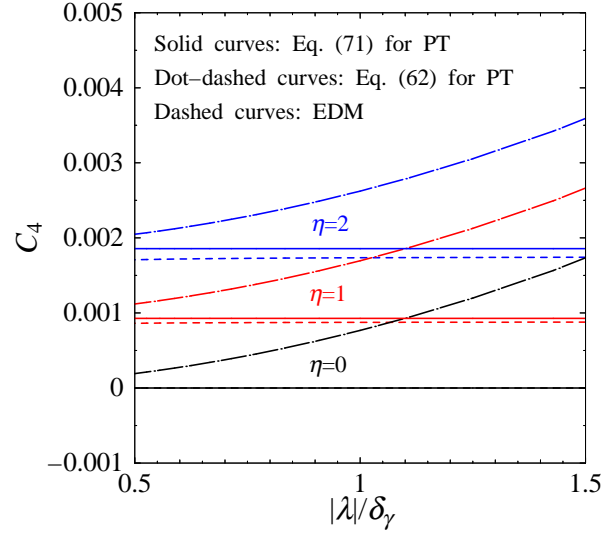} 
\caption{
(Color online) 
The quantity $|\lambda|/\delta_\gamma$ dependence of $C_4$ 
for the systems with the crystal field of tetragonal symmetry. 
We here set $H$=1 eV, $\Delta$=0.1 eV, $\lambda$=$-$0.01 eV, 
$r$=0, $r_\gamma$=0.01, $r_{\varepsilon}/r_\gamma$=1, 
and 
$\eta$=0, 1, and 2. 
The solid curves represent $C_4$ of Eq. (\ref{C_4_strong}) for PT. 
The dot-dashed curves represent $C_4$ of Eq. (\ref{C_4_strong0}) for PT. 
The dashed curves represent $C_4$ of Eq. (\ref{C_4_exact}) for the EDM. 
}
\label{c4_comp}
\end{center}
\end{figure}

On the basis of (i) and (ii) of Sec. \ref{sec_resistivity}, 
the features of $C_2 \cos 2\phi$ and $C_4 \cos 4\phi$ 
are described as follows: 
\begin{enumerate}
\item[(i)] 
The term $C_2 \cos 2\phi$ 
is related to the real part of the probability amplitudes 
of $|3z^2 -r^2, \chi_\sigma (\phi) \rangle$ and $|x^2 -y^2, \chi_\sigma (\phi) \rangle$, 
which are given by 
${\rm Re}[w_{3z^2-r^2,\sigma}^{i,\varsigma}]\cos 2\phi$ 
and 
${\rm Re}[w_{x^2-y^2,\sigma}^{i,\varsigma}]\cos 2\phi$, respectively 
[see Eqs. (\ref{X_{2phi,up}}) and (\ref{X_{2phi,dw}})]. 
Concretely, 
$C_2 \cos 2\phi$ contains 
a single $\rho_{2,\sigma}^{(2)} \cos 2\phi$ in the numerator of each term 
in $C_2 \cos 2\phi$ [see Eq. (\ref{C_2_strong})]. 
Here, $\rho_{2,\sigma}^{(2)} \cos 2\phi$ is related to 
the real part of the probability amplitudes 
of $|3z^2 -r^2, \chi_\sigma (\phi) \rangle$ and $|x^2 -y^2, \chi_\sigma (\phi) \rangle$ 
as noted in (i) of Sec. \ref{sec_resistivity}. 
\item[(ii)] 
The term $C_4 \cos 4\phi$ is related to 
the probabilities of 
$|3z^2 -r^2, \chi_\sigma (\phi) \rangle$ 
and $|x^2 -y^2, \chi_\sigma (\phi) \rangle$, 
which are given by 
$|w_{3z^2-r^2,\sigma}^{i,\varsigma}|^2 ( 1 \pm \cos 4\phi)/2$ 
and 
$|w_{x^2-y^2,\sigma}^{i,\varsigma}|^2 ( 1+\cos 4\phi)/2$, respectively 
[see Eqs. (\ref{X_{4phi,up}}) and (\ref{X_{4phi,dw}})]. 
Concretely, 
$C_4 \cos 4\phi$ contains 
a single $\rho_{4,\sigma}^{(2)} \cos 4\phi$ in the numerator of each term 
in $C_4 \cos 4\phi$ [see Eq. (\ref{C_4_strong0})]. 
Here, $\rho_{4,\sigma}^{(2)} \cos 4\phi$ is related to the probabilities 
of $|3z^2 -r^2, \chi_\sigma (\phi) \rangle$ and $|x^2 -y^2, \chi_\sigma (\phi) \rangle$ 
as noted in (ii) of Sec. \ref{sec_resistivity}. 
Also, $c_4'$ of Eq. (\ref{c_4'_strong}) 
arises from 
high-order processes of $d\gamma - d\varepsilon - d\gamma'$, 
in which 
the $d\gamma$ states are hybridized to the $d\gamma'$ states 
via the $d\varepsilon$ states. 
Such processes 
reflect the fact that there are no off-diagonal matrix elements 
in the subspace of the d$\gamma$ states (see Table \ref{matrix2}). 
\end{enumerate}
We next determine the effective value of 
the undefined parameter $\lambda/\delta_\gamma$ 
by comparing $C_4$ obtained by PT with that obtained by the EDM. 
We here put
\begin{eqnarray}
\label{r_1-}
&&r_{\varepsilon 1} = r_{\varepsilon} (1+\eta), \\
\label{r_2-}
&&r_{\varepsilon 2} = r_{\varepsilon}, 
\end{eqnarray}
where $\eta$ represents 
the difference between $r_{\varepsilon 1}/r_{\varepsilon}$ 
and $r_{\varepsilon 2}/r_{\varepsilon}$. 
Figure \ref{c4_comp} shows 
the $|\lambda|/\delta_\gamma$ dependence of $C_4$ 
of Eqs. (\ref{C_4_strong0}) and (\ref{C_4_exact}) 
for the systems with 
$H$=1 eV, $\Delta$=0.1 eV, $\lambda$=$-$0.01 eV,\cite{parameter} 
$r$=0, $r_\gamma$=0.01,\cite{com_param} 
$r_{\varepsilon}/r_\gamma$=1, and $\eta$=0, 1, and 2. 
Here, 
$r$=0 and $r_\gamma$=0.01 are set 
on the basis of those for Fe$_4$N.\cite{Kokado1} 
The range of $|\lambda|/\delta_\gamma$ is roughly assumed to be 
$0.5 \le |\lambda|/\delta_\gamma \le 1.5$ 
by consideration of 
the above parameters and $\delta_\gamma/\Delta \ll 1$. 
At each $\eta$, 
$C_4$ obtained by PT 
decreases with decreasing $|\lambda|/\delta_\gamma$ 
because of $c_4' \propto (\lambda/\delta_\gamma)^2$. 
In contrast, $C_4$ obtained by the EDM is nearly constant. 
In particular, when 
each $d\varepsilon$ state has the same PDOS at $E_{\mbox{\tiny F}}$ 
(i.e., $\eta$=0), 
$C_4$ of Eq. (\ref{C_4_strong0}) for PT becomes $C_4$=$c_4'$, 
whereas $C_4$ of Eq. (\ref{C_4_exact}) for the EDM 
is evaluated to be $C_4$$\sim$0 independently of $|\lambda|/\delta_\gamma$. 
In addition, 
the difference in $C_4$ between PT 
and the EDM decreases with decreasing $|\lambda|/\delta_\gamma$. 
From these results, 
the effective value of $|\lambda|/\delta_\gamma$ for PT 
is considered to be $|\lambda|/\delta_\gamma$$\sim$1/2. 
In other words, 
the present PT 
is unsuitable for 
application to systems with 
$|\lambda| / \delta_\gamma \gtrsim 1$. 

With regard to $C_4$, 
from now on 
we focus on the dominant term 
with $\left[ \lambda/(H \pm \Delta) \right]^2$ or $(\lambda/\Delta)^2$ 
under the condition $|\lambda|/\delta_\gamma$$\sim$1/2. 
Namely, 
we neglect 
$c_4'$ with $( \lambda/\delta_\gamma )^2 
\left[ (\lambda/\Delta) - \lambda/(H+\Delta) \right]^2$, 
which corresponds to high-order processes. 
The dominant term in $C_4$ is thus expressed as
\begin{eqnarray}
\label{C_4_strong}
&&\hspace*{-0.9cm}C_4= 
\frac{3}{32} \frac{r_{\varepsilon1}-r_{\varepsilon2}}{1+r+r_\gamma}
\left[
\left( \frac{\lambda}{\Delta} \right)^2 
\frac{1}{r+r_\gamma}
-\left( \frac{\lambda}{H-\Delta} \right)^2 (r+r_\gamma)
\right]. 
\end{eqnarray}
As seen from Fig. \ref{c4_comp}, 
$C_4$ of Eq. (\ref{C_4_strong}) agrees relatively well with 
that obtained by the EDM with $|\lambda|/\delta_\gamma$=1/2. 

Furthermore, we extract 
the dominant terms from $C_2$ of Eq. (\ref{C_2_strong}) 
and $C_4$ of Eq. (\ref{C_4_strong}) 
taking into account 
the relation of typical ferromagnets, $|\Delta/H| \ll 1$. 
The dominant terms are 
\begin{eqnarray}
\label{C_2_Fe4N_0}
&&\hspace*{-0.3cm}C_2 = \frac{3}{8} 
\left(\frac{\lambda}{\Delta} \right)^2 
\frac{r_\gamma - r_{\varepsilon1}}{(1+r+r_\gamma)(r+r_\gamma)}, \\
\label{C_4_Fe4N_0}
&&\hspace*{-0.3cm}C_4=\frac{3}{32} 
\left( \frac{\lambda}{\Delta} \right)^2 
\frac{r_{\varepsilon1}-r_{\varepsilon2}}{(1+r+r_\gamma)(r+r_\gamma)}. 
\end{eqnarray}
As a characteristic feature, 
$C_2$ of Eq. (\ref{C_2_Fe4N_0}) 
is proportional to 
$r_\gamma - r_{\varepsilon 1}$ 
($\propto D_{\gamma,-}^{(d)}-D_{\varepsilon 1,-}^{(d)}$), 
%
while $C_4$ of Eq. (\ref{C_4_Fe4N_0}) is proportional to 
$r_{\varepsilon 1}-r_{\varepsilon 2}$ 
($\propto D_{\varepsilon 1,-}^{(d)}-D_{\varepsilon 2,-}^{(d)}$). 



\subsection{
Various features 
of $C_2$ and $C_4$}
\label{c2c4_edm_r}

We investigate various features of 
$C_2$ and $C_4$ for a strong ferromagnet with 
$H$=1 eV and $\lambda$=$-$0.01 eV. 
We here use 
$C_2$ of Eq. (\ref{C_2_strong}) and $C_4$ of Eq. (\ref{C_4_strong}) 
for PT and 
$C_2$ of Eq. (\ref{C_2_exact}) and $C_4$ of Eq. (\ref{C_4_exact}) 
for the EDM, 
where $|\lambda|/\delta_\varepsilon$=$|\lambda|/\delta_\gamma$=1/2 is set 
for $C_2$ and $C_4$ for the EDM. 
We also utilize Eqs. (\ref{r_1-}) and (\ref{r_2-}). 
As a particularly important result, 
we find that 
$C_4$ appears under the crystal field of tetragonal symmetry, 
whereas 
it vanishes under the crystal field of cubic symmetry.\cite{W_FM_C4} 

\begin{figure}[ht]
\begin{center}
\includegraphics[width=.5\linewidth]{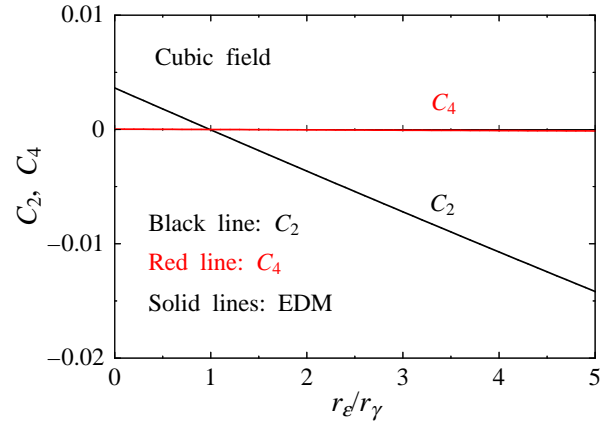} 
\caption{
(Color online) 
The quantity $r_{\varepsilon}/r_\gamma$ dependences of $C_2$ and $C_4$ 
for the system with the crystal field of cubic symmetry. 
We here set 
$H$=1 eV, $\Delta$=0.1 eV, $\lambda$=$-$0.01 eV, 
$\delta_\varepsilon$=$\delta_\gamma$=0, 
$r$=0, $r_\gamma$=0.01, and $\eta$=0. 
The solid lines represent 
$C_2$ of Eq. (\ref{C_2_exact}) and $C_4$ of Eq. (\ref{C_4_exact}) for the EDM. 
}
\label{cubic}
\end{center}
\end{figure}

\begin{figure}[ht]
\begin{center}
\includegraphics[width=.5\linewidth]{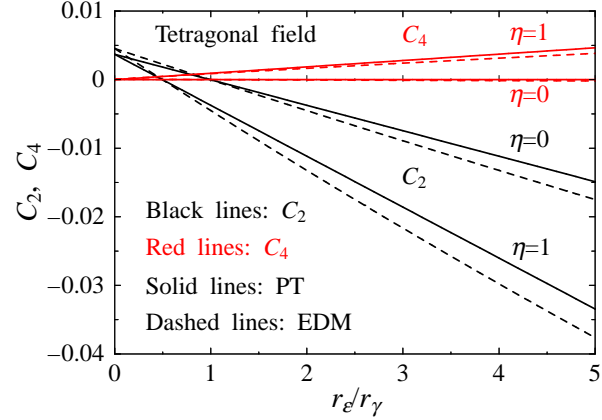} 
\caption{
(Color online) 
The quantity $r_{\varepsilon}/r_\gamma$ dependences of $C_2$ and $C_4$ 
for the system with the crystal field of tetragonal symmetry. 
We here set $H$=1 eV, $\Delta$=0.1 eV, $\lambda$=$-$0.01 eV, $r$=0, 
$r_\gamma$=0.01, and $\eta$=0 and 1. 
The solid lines represent 
$C_2$ of Eq. (\ref{C_2_strong}) 
and $C_4$ of Eq. (\ref{C_4_strong}) for PT. 
The dashed lines represent 
$C_2$ of Eq. (\ref{C_2_exact}) 
and $C_4$ of Eq. (\ref{C_4_exact}) for the EDM, 
where $|\lambda|/\delta_\varepsilon$=$|\lambda|/\delta_\gamma$=1/2. 
}
\label{tetra}
\end{center}
\end{figure}

Using the EDM, 
we obtain the $r_{\varepsilon}/r_\gamma$ dependences of 
$C_2$ and $C_4$ for a system with the crystal field of cubic symmetry, 
where 
$\Delta$=0.1 eV, $\delta_\varepsilon$=$\delta_\gamma$=0, 
$r$=0, $r_\gamma$=0.01, and $\eta$=0 (see Fig. \ref{cubic}). 
We find that 
$C_2$ can be expressed as a linear function of 
$r_{\varepsilon}/r_\gamma$. 
The sign of $C_2$ changes in the vicinity of 
$r_{\varepsilon}/r_\gamma$$\sim$1. 
Furthermore, $C_4$ takes a value of almost 0. 

\begin{figure}[ht]
\begin{center}
\includegraphics[width=0.45\linewidth]{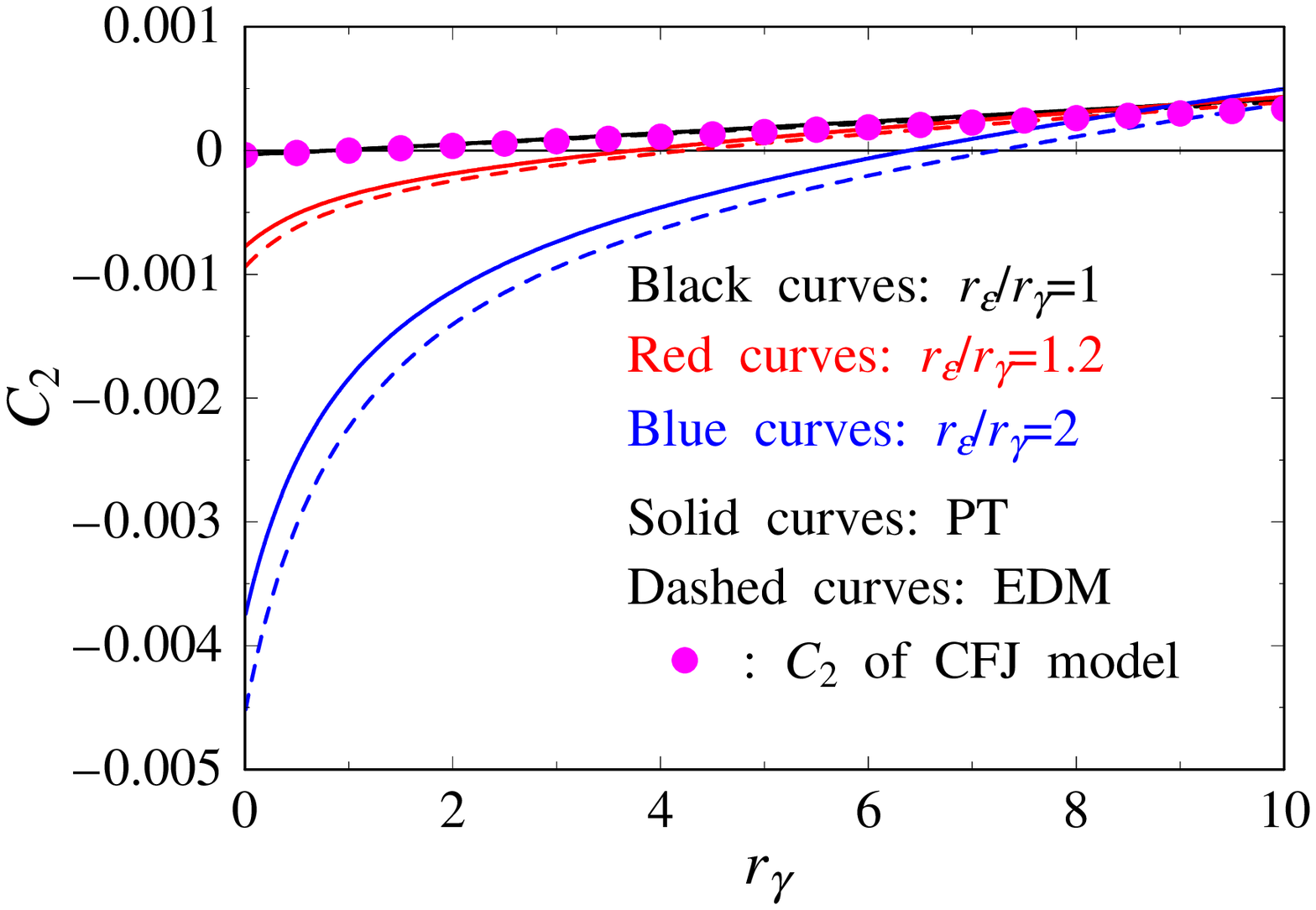}\\
\includegraphics[width=0.45\linewidth]{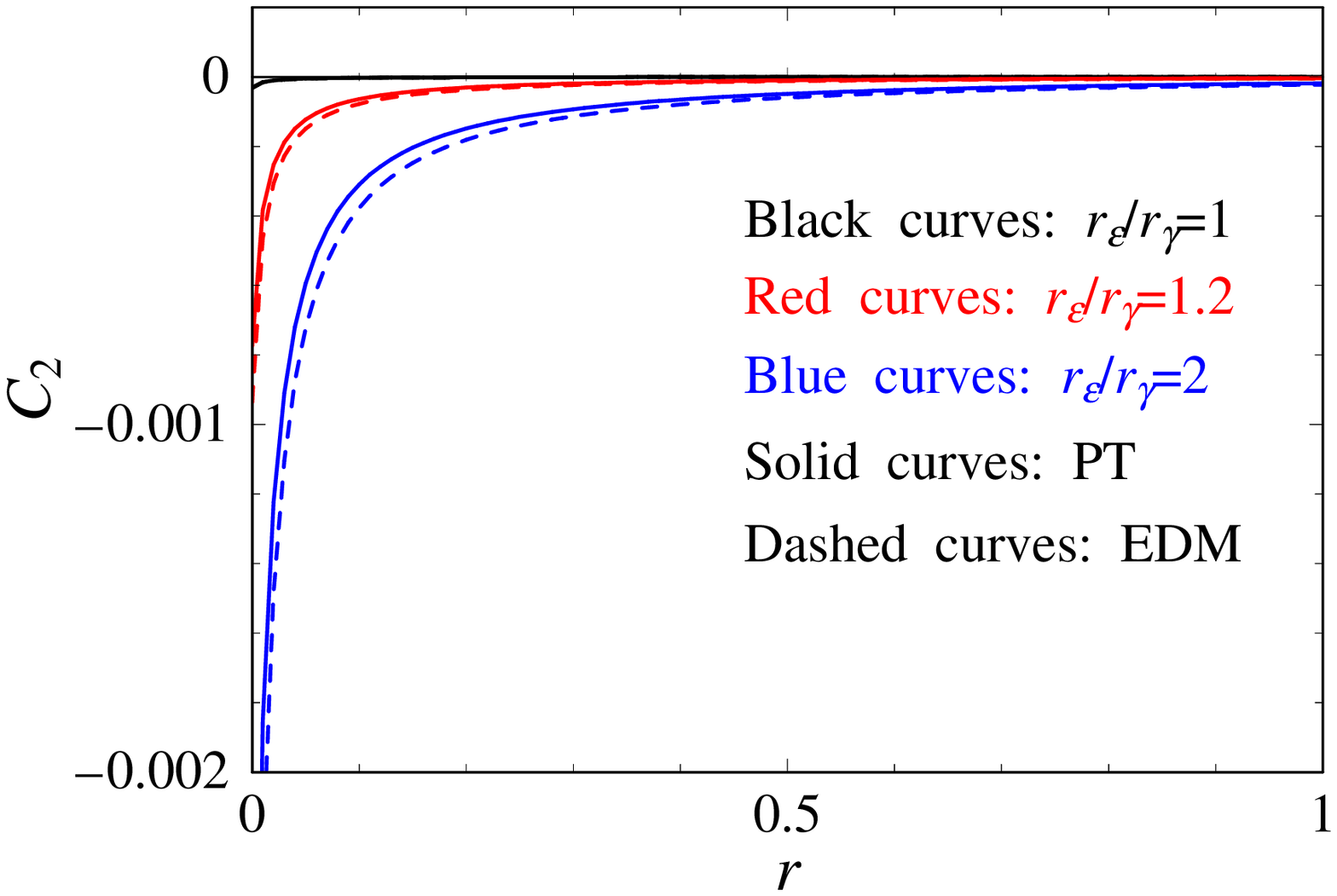}\\
\includegraphics[width=0.45\linewidth]{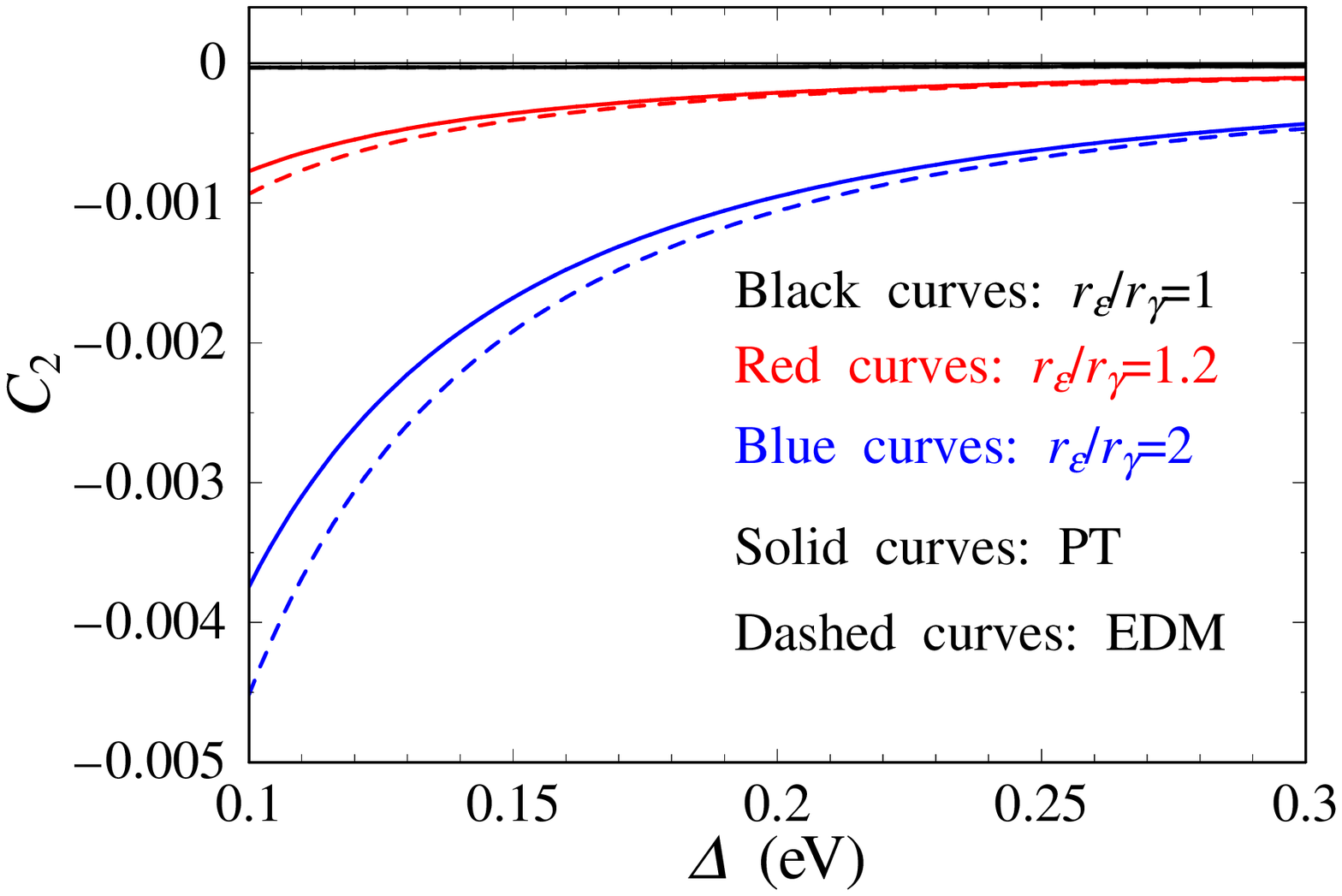}
\caption{
(Color online) 
The coefficient $C_2$ 
for strong ferromagnets 
with the crystal field of tetragonal symmetry. 
We here set $H$=1 eV, $\lambda$=$-$0.01 eV, 
$r_{\varepsilon}/r_\gamma$=1, 1.2, and 2, 
and $\eta$=0. 
Upper panel: 
The quantity 
$r_\gamma$ dependence of $C_2$ 
for the systems with $\Delta$=0.1 eV and $r$=0. 
Middle panel: 
The quantity $r$ dependence of $C_2$ 
for the systems with $\Delta$=0.1 eV and $r_\gamma$=0.01. 
Lower panel: 
The energy $\Delta$ dependence 
of $C_2$ 
for the systems with $r$=0 and $r_\gamma$=0.01. 
The solid curves represent 
$C_2$ of Eq. (\ref{C_2_strong}) for PT. 
The dashed curves represent $C_2$ of Eq. (\ref{C_2_exact}) for the EDM, 
where $|\lambda|/\delta_\varepsilon$=$|\lambda|/\delta_\gamma$=1/2. 
The dots represent $C_2$ of the CFJ model.\cite{Campbell1} 
Note that 
$C_4$ is not shown because 
$C_4$ for PT is 0 and $C_4$ for the EDM is much smaller than $|C_2|$. 
}
\label{amr_same}
\end{center}
\end{figure}

\begin{figure}[ht]
\begin{center}
\includegraphics[width=0.45\linewidth]{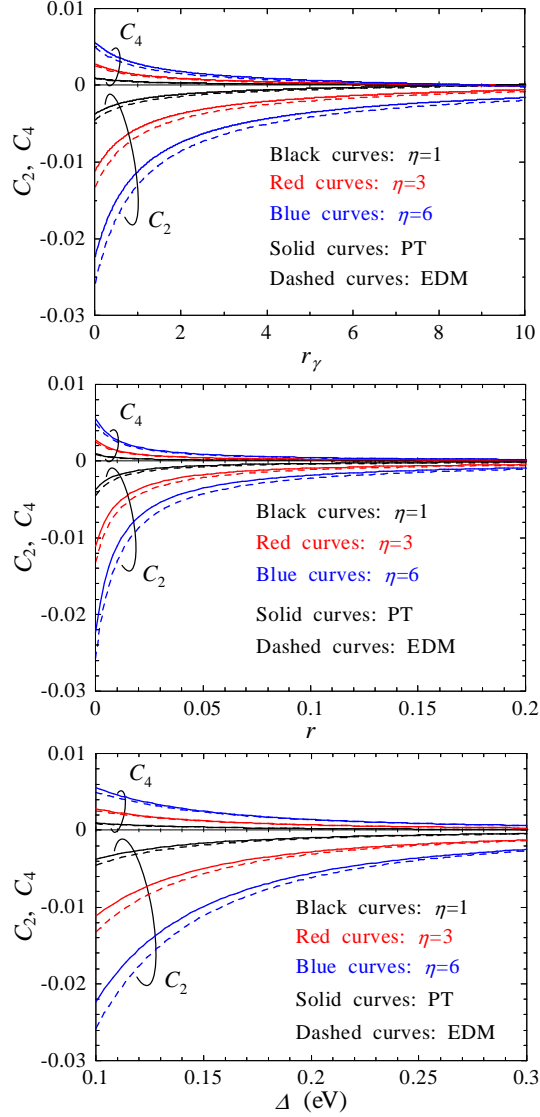}
\caption{
(Color online) 
The coefficients $C_2$ and $C_4$ 
for strong ferromagnets with the crystal field of tetragonal symmetry. 
We here set $H$=1 eV, $\lambda$=$-$0.01 eV, 
$r_{\varepsilon}/r_\gamma$=1, and 
$\eta$=1, 3, and 6. 
Upper panel: 
The quantity 
$r_\gamma$ dependences of $C_2$ and $C_4$ 
for the systems with $\Delta$=0.1 eV and $r$=0. 
Middle panel: 
The quantity $r$ dependences of $C_2$ and $C_4$ 
for the systems with $\Delta$=0.1 eV and $r_\gamma$=0.01. 
Lower panel: 
The energy $\Delta$ dependences of $C_2$ and $C_4$ 
for the systems with $r$=0 and $r_\gamma$=0.01. 
The solid curves show 
$C_2$ of Eq. (\ref{C_2_strong}) and $C_4$ of 
Eq. (\ref{C_4_strong}) for PT. 
The dashed curves represent $C_2$ of Eq. (\ref{C_2_exact}) 
and $C_4$ of Eq. (\ref{C_4_exact}) for the EDM, 
where $|\lambda|/\delta_\gamma$=1/2 and $|\lambda|/\delta_\varepsilon$=1/2. 
}
\label{amr_diff}
\end{center}
\end{figure}

Figure \ref{tetra} shows 
the $r_{\varepsilon}/r_\gamma$ or $\eta$ dependences of 
$C_2$ and $C_4$ for a system 
with the crystal field of tetragonal symmetry, 
where 
$\Delta$=0.1 eV, $r$=0, and $\eta$=0 and 1. 
From the results of PT, 
we find 
$C_2$$\sim$0 for the system 
with $r_{\varepsilon 1}/r_\gamma$=$r_{\varepsilon}(1+\eta)/r_\gamma$=1 
and $C_2$$\ne$0 for that with 
$r_{\varepsilon 1}/r_\gamma$=$r_{\varepsilon}(1+\eta)/r_\gamma$$\ne$1. 
This feature mainly reflects Eq. (\ref{C_2_Fe4N_0}). 
We also obtain $C_4$=0 for the system with $\eta$=0 
and $C_4$$\ne$0 for that with $\eta$$\ne$0 
because of 
$C_4 \propto r_{\varepsilon1}-r_{\varepsilon2}$ $(= r_{\varepsilon} \eta)$. 
The coefficients $C_2$ 
and $C_4$ 
obtained by PT 
qualitatively agree well with 
those obtained by the EDM.




In Fig. \ref{amr_same}, we show 
$C_2$ 
for systems with 
the crystal field of tetragonal symmetry, 
where 
$r_\varepsilon/r_\gamma$=1, 1.2, and 2 and $\eta$=0. 
Here, $C_4$ for PT takes a value of 0 because of $\eta$=0, 
and $|C_4|$ for the EDM is much smaller than $|C_2|$. 
The upper panel shows the $r_\gamma$ dependence of $C_2$ 
for systems with $\Delta$=0.1 eV and $r$=0. 
The middle panel shows the $r$ dependence of 
$C_2$ for systems with $\Delta$=0.1 eV and $r_\gamma$=0.01. 
The lower panel 
shows the $\Delta$ dependence of $C_2$ 
for systems with $r$=0 and $r_\gamma$=0.01. 
In the upper panel, 
when $r_\varepsilon/r_\gamma$=1, 
$C_2$ for PT is close to that for the CFJ model, i.e., 
$C_2$=$(1/2) \Delta \rho(0)/\rho$=$(3/8)(\lambda/H)^2 (\alpha -1)$, 
with $\alpha$=$r_\varepsilon$=$r_\gamma$ (see Appendix \ref{CFJ}). 
In the middle and lower panels, 
$C_2$ for PT takes a value of almost 0 
in the case of $r_\varepsilon/r_\gamma$=1. 
The sign of $C_2$ for PT is negative 
in the case of $r_\varepsilon/r_\gamma$=1.2 or 2. 
In addition, 
$|C_2|$ for PT 
increases 
with decreasing $r$ or $\Delta$ and 
with increasing $r_\varepsilon/r_\gamma$. 
These features mainly reflect Eq. (\ref{C_2_Fe4N_0}). 
In all panels, 
$C_2$ 
for PT 
qualitatively agrees well with that for the EDM. 

Figure \ref{amr_diff} shows 
$C_2$ and $C_4$ for systems 
with the crystal field of tetragonal symmetry, 
where $r_\varepsilon/r_\gamma$=1 and $\eta$=1, 3, and 6. 
The upper panel shows the $r_\gamma$ dependences of $C_2$ and $C_4$ 
for systems with $\Delta$=0.1 eV and $r$=0. 
The middle panel shows the $r$ dependences of 
$C_2$ and $C_4$ 
for systems with $\Delta$=0.1 eV and $r_\gamma$=0.01. 
The lower panel 
shows the $\Delta$ dependences of 
$C_2$ and $C_4$ 
for systems with $r$=0 and $r_\gamma$=0.01. 
In all panels, 
the sign of $C_2$ for PT is negative, 
while 
that of $C_4$ for PT is positive. 
In addition, 
$|C_2|$ and $C_4$ for PT 
increase 
with decreasing $r_\gamma$, $r$, or $\Delta$ 
and with increasing $\eta$. 
Such features 
are mainly due to 
Eqs. (\ref{C_2_Fe4N_0}) and (\ref{C_4_Fe4N_0}). 
The coefficients $C_2$ and $C_4$ 
for PT 
qualitatively agree well with 
those 
for the EDM. 

\section{Simple Analysis of $C_2$ and $C_4$ for F\MakeLowercase{e}$_4$N}
\label{appl_fe4n}

\begin{figure}[ht]
\begin{center}
\includegraphics[width=.43\linewidth]{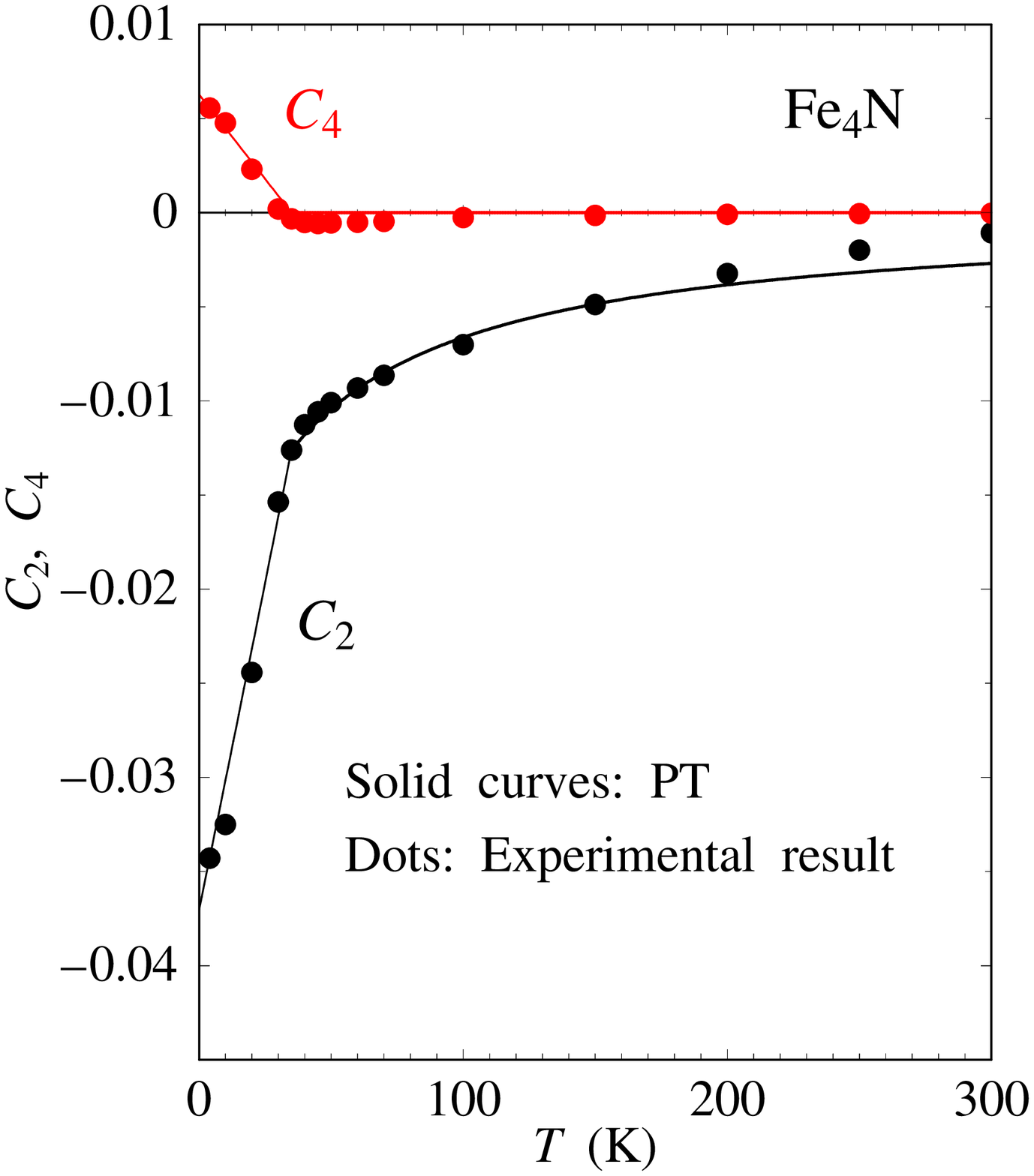} \\
\hspace*{0.3cm}\includegraphics[width=.41\linewidth]{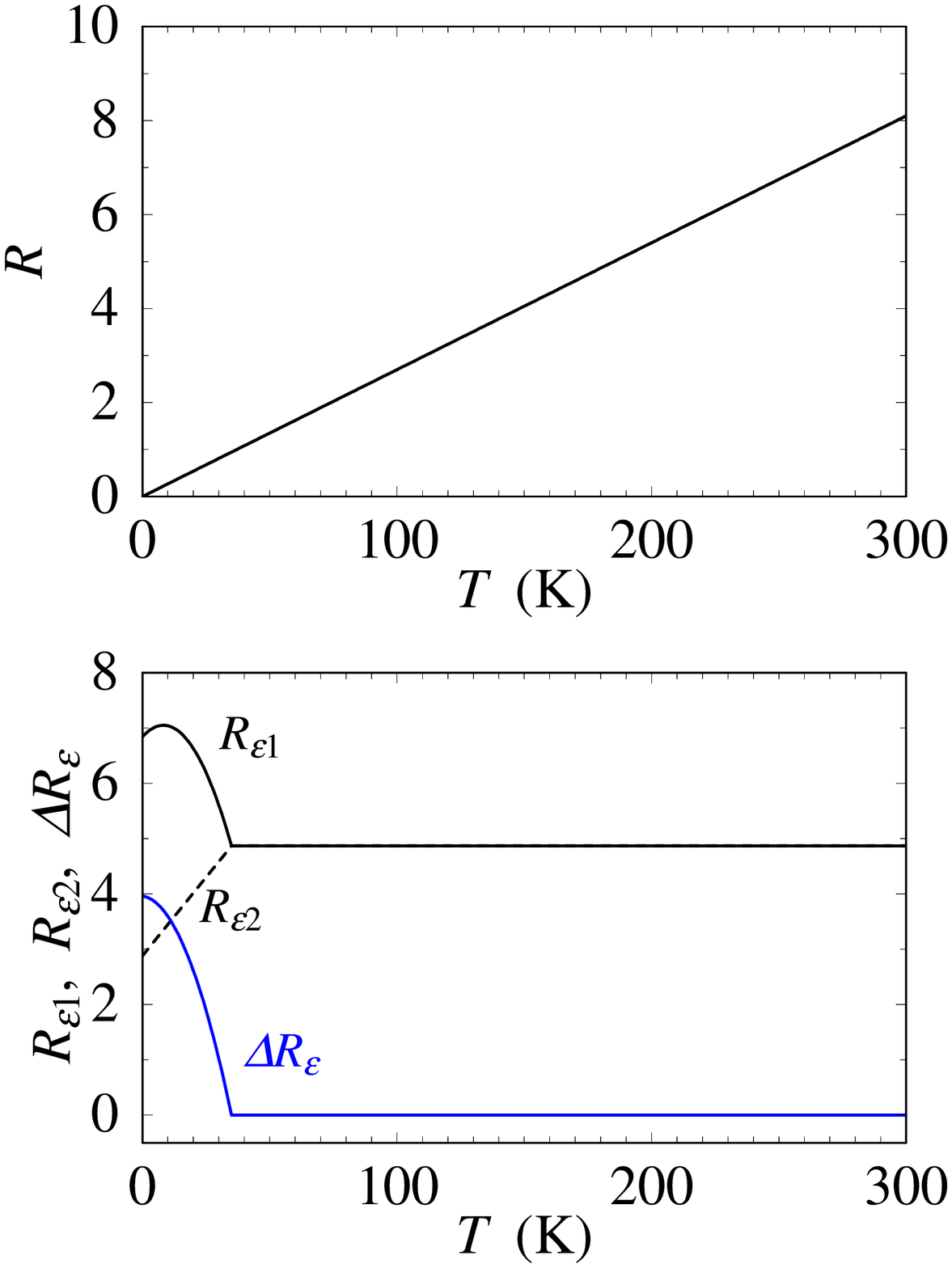} 
\caption{
(Color online) 
Upper panel: 
The $T$ dependences of $C_2$ and $C_4$ for Fe$_4$N. 
The solid curves represent Eqs. (\ref{C_2_Fe4N_1}) 
and (\ref{C_4_Fe4N_1}) for PT. 
The dots represent the experimental values 
in the temperature range from 4 to 300 K 
for the case of 
${\mbox{\boldmath $I$}}//$Fe$_4$N [100].\cite{Tsunoda2} 
Middle panel: The $T$ dependence of $R$ in Eq. (\ref{param}). 
The expression for $R$ is given by Eqs. (\ref{R_high}) and (\ref{R_low}). 
Lower panel: The $T$ dependences of 
$R_{\varepsilon 1}$, $R_{\varepsilon 2}$, and $\Delta R_\varepsilon$ 
in Eq. (\ref{param}). 
The expressions for 
$R_{\varepsilon 1}$, $R_{\varepsilon 2}$, and $\Delta R_\varepsilon$ 
are given by Eqs. (\ref{R2_high}) and (\ref{R1_low}), 
(\ref{R2_high}) and (\ref{R2_low}), 
and (\ref{del_R_high}) and (\ref{del_R_low}), respectively. 
The black solid curve (black dashed curve) represents 
$R_{\varepsilon 1}$ ($R_{\varepsilon 2}$). 
The blue solid curve represents $\Delta R_\varepsilon$. 
}
\label{best}
\end{center}
\end{figure}

Utilizing the above results, 
we perform a simple analysis of the experimental results\cite{Tsunoda2} 
for the $T$ dependences of $C_2$ and $C_4$ 
for an Fe$_4$N\cite{Jack,Sakuma,Kokado3} 
film on a MgO(001) substrate, 
where 
${\mbox{\boldmath $I$}}$ flows along Fe$_4$N [100]. 
The experimental results clearly show 
the difference in the behaviors between 
the low-temperature range of 4 K $\le T \le 35$ K 
and 
the high-temperature range of 35 K $< T \le$ 300 K 
(see circles in Fig. \ref{best}). 
Here, we regard Fe$_4$N as 
a strong ferromagnet with $D_{i,+}^{(d)}$$\sim$0.\cite{Kokado1} 
In addition, we mainly focus on 
the effect of the PDOSs of the $d \varepsilon$ states on $C_4$. 
Note that 
we do not take into account 
the realistic crystal structure of Fe$_4$N 
(i.e., a perovskite-type structure\cite{Jack}) 
for simplicity.\cite{com_model1} 

%
From Eqs. (\ref{C_2_Fe4N_0}) and (\ref{C_4_Fe4N_0}), 
we first obtain simple expressions for $C_2$ and $C_4$ for Fe$_4$N. 
By taking into account 
the relation for Fe$_4$N, i.e., $r \ll 1$ 
and $r_\gamma \ll 1$,~\cite{Kokado1} 
$C_2$ and $C_4$ are given by
\begin{eqnarray}
\label{C_2_Fe4N_1}
&&C_2 = \frac{3}{8} 
\left(\frac{\lambda}{\Delta} \right)^2 
\frac{r_\gamma - r_{\varepsilon1}}{r+r_\gamma}
= \frac{\kappa (1-R_{\varepsilon 1})}{1+R} 
=
\frac{\kappa (1-R_{\varepsilon 2}- \Delta R_\varepsilon)}{1+R}, \\
\label{C_4_Fe4N_1}
&&C_4=\frac{3}{32} 
\left( \frac{\lambda}{\Delta} \right)^2 
\frac{r_{\varepsilon1}-r_{\varepsilon2}}{r+r_\gamma} 
=\frac{\displaystyle{\frac{\kappa}{4}}(R_{\varepsilon 1}-R_{\varepsilon 2})}{1+R} 
=\frac{ \displaystyle{\frac{\kappa}{4}\Delta R_\varepsilon}}{1+R}, 
\end{eqnarray}
where 
\begin{eqnarray}
\label{param}
&&\kappa = \frac{3}{8} \left( \frac{\lambda}{\Delta} \right)^2, 
R_{\varepsilon 1}= \frac{r_{\varepsilon 1}}{r_\gamma}, ~
R_{\varepsilon 2}= \frac{r_{\varepsilon 2}}{r_\gamma}, 
~\Delta R_\varepsilon=R_{\varepsilon 1}-R_{\varepsilon 2}, \nonumber \\
\label{RR}
&&R = \frac{r}{r_\gamma}
=\frac{\rho_{s,-}/\rho_{s,+}}{\rho_{s \to d_\gamma,-}/\rho_{s,+}}=\frac{\rho_{s,-}}{\rho_{s \to d_\gamma,-}}, 
\end{eqnarray}
with 
$\rho_{s \to d_\gamma,-}$=$\rho_{s \to d_{x^2-y^2},-}$=$\rho_{s \to d_{3z^2-r^2},-}$. 
Here, 
$\Delta R_\varepsilon$ is proportional to 
$D_{\varepsilon 1,-}^{(d)} - D_{\varepsilon 2,-}^{(d)}$, 
which is 
the difference in the PDOSs at $E_{\mbox{\tiny F}}$ 
among the d$\varepsilon$ states. 

We next determine parameter sets 
for $\lambda$, $\Delta$, 
$R$, $R_{\varepsilon 1}$, $R_{\varepsilon 2}$, and $\Delta R_\varepsilon$ 
that can reproduce the experimental result 
for the $T$ dependences of $C_2$ and $C_4$. 
The quantity $\lambda$ is set to $\lambda$=$-$0.013 eV for Fe.\cite{Yosida} 
The quantity $\Delta$ is assumed to be $\Delta$=0.1 eV.\cite{parameter} 
Here, the $T$ dependence of $\Delta$ is considered 
to be negligibly small, 
because 
the decrease in the lattice constant 
due to a decrease in $T$ is less than 0.5\%,\cite{Tsunoda2} 
where 
$\Delta$ of 0.1 eV is due to the Coulomb interaction 
between a magnetic ion and the surrounding ions. 
We accordingly adopt the $T$ dependences of 
$R$, $R_{\varepsilon 1}$, $R_{\varepsilon 2}$, and $\Delta R_\varepsilon$. 
The $T$ dependence of $R$ 
($R_{\varepsilon 1}$, $R_{\varepsilon 2}$, and $\Delta R_\varepsilon$) 
is shown in the middle (lower) panel of Fig. \ref{best}. 
Details of the parameter sets are given below.


We express $R$ of Eq. (\ref{RR}) as
\begin{eqnarray}
\label{RRR}
&&\hspace*{-0.4cm}R = \frac{\rho_{s,-}}{\rho_{s \to d_\gamma,-}}=
 \frac{\rho_{s,-}^{\rm imp}+\rho_{s,-}^{\rm ph}}{\rho_{s \to d_\gamma,-}^{\rm imp}}, 
\end{eqnarray}
where 
$\rho_{s,-}$ and $\rho_{s \to d_\gamma,-}$ 
are assumed to be 
$\rho_{s,-}$=$\rho_{s,-}^{\rm imp} + 
\rho_{s,-}^{\rm ph}$ and 
$\rho_{s \to d_\gamma,-}$=$\rho_{s \to d_\gamma,-}^{\rm imp}$ 
(see Sec. \ref{sec_resistivity}). 
The quantity $\rho_{s,-}^{\rm imp}$ 
is the $s$--$s$ resistivity due to impurities 
and 
$\rho_{s \to d_\gamma,-}^{\rm imp}$ 
is the $s$--$d$ resistivity due to impurities, 
where $d_\gamma,-$ represents the $d\gamma$ states of the down spin. 
Here, 
$\rho_{s,-}^{\rm imp}/\rho_{s \to d_\gamma,-}^{\rm imp}$ 
is set so that
\begin{eqnarray}
\label{assumption}
\frac{\rho_{s,-}^{\rm imp}}{\rho_{s \to d_\gamma,-}^{\rm imp}} 
\ll 1, 
\end{eqnarray}
by considering that 
$\rho_{s \to d_\gamma,-}^{\rm imp}$ 
($\rho_{s,-}^{\rm imp}$) satisfies the relation 
$\rho_{s \to d_\gamma,-}^{\rm imp} \propto D_{\gamma,-}^{(d)}$ 
($\rho_{s,-}^{\rm imp} \propto D_{-}^{(s)}$\cite{com_Ds}) 
and also 
Fe$_4$N satisfies 
$D_{\gamma,-}^{(d)} \gg D_{-}^{(s)}$. 
On the other hand, 
$\rho_{s,-}^{\rm ph}$ is the $s$--$s$ resistivity 
due to the phonons. 
This $\rho_{s,-}^{\rm ph}$ depends on $T$ 
through the influence of the number of phonons, which depends on $T$. 

The parameter sets 
in the high- and low-temperature ranges are noted below. 
\begin{enumerate}
\item[(i)] In the high-temperature range of $T >$ 35 K, we have
\begin{eqnarray}
\label{R_high}
&&R=\Theta T, \\
\label{del_R_high}
&&\Delta R_\varepsilon=0, \\
\label{R2_high}
&&R_{\varepsilon 2}=R_{\varepsilon 1} =1 - \frac{C_2^{(T^*)}}{\kappa}
(1+R^*), 
\end{eqnarray}
with 
$\Theta$=0.0270, 
$T^*$=35 K, 
$R^*$=$\Theta T^*$, 
and $C_2^{(T^*)}$=$-$0.0126, 
where 
$C_2^{(T^*)}$ 
is the experimental value of $C_2$ at $T$=$T^*$.\cite{Tsunoda2} 


The procedure for determining this parameter set is as follows: 
First, $C_4$ is experimentally observed to be almost 0. 
Since $C_4 \propto \Delta R_\varepsilon$ [see Eq. (\ref{C_4_Fe4N_1})], 
we assume $\Delta R_\varepsilon$=0 
(or $R_{\varepsilon 1}$=$R_{\varepsilon 2}$); 
that is, 
the PDOSs of the $d\varepsilon$ states at $E_{\mbox{\tiny F}}$ 
take the same value. 
From the viewpoint of the crystal structure of Fe$_4$N, 
this assumption may imply that 
the crystal exhibits cubic symmetry.\cite{com_model1} 
Next, 
since $|C_2|$ gradually decreases with increasing $T$, 
we straightforwardly take into consideration 
the $T$ dependence of $R$ of Eq. (\ref{RRR}), 
where $R$ is included in the denominator 
of $C_2$ of Eq. (\ref{C_2_Fe4N_1}). 
The denominator $1+R$ 
is expressed as
\begin{eqnarray}
1+R \approx 1 + \frac{\rho_{s,-}^{\rm ph}}{\rho_{s \to d_\gamma,-}^{\rm imp}}, 
\end{eqnarray}
by using Eqs. (\ref{RRR}) and (\ref{assumption}). 
Here, 
$\rho_{s,-}^{\rm ph}$ is assumed to be proportional to $T$ 
on the basis of the experimental result for 
the $T$ dependence of the total resistivity.\cite{Tsunoda_exp} 
Thereby, $R$ ($\approx \rho_{s,-}^{\rm ph}/\rho_{s \to d_\gamma,-}^{\rm imp}$) 
is given by
\begin{eqnarray}
\label{R=ThT}
R
=\theta T,
\end{eqnarray}
where $\theta$ is a constant number. 
On the other hand, 
$R_{\varepsilon 2}$ 
is 
determined so that 
Eq. (\ref{C_2_Fe4N_1}) satisfies the condition 
$(T, C_2)$=$(T^*, C_2^{(T^*)})$. 
Namely, $R_{\varepsilon 2}$ is expressed as 
$R_{\varepsilon 2}$=$1-(C_2^{(T^*)}/\kappa)(1+\theta T^*)$. 
Substituting this $R_{\varepsilon 2}$ 
and $\Delta R_\varepsilon$=0 into Eq. (\ref{C_2_Fe4N_1}), 
we obtain 
\begin{eqnarray}
\label{fitting}
C_2=\frac{C_2^{(T^*)}(1+\theta T^*)}{1+\theta T}. 
\end{eqnarray}
From the least-square fitting of Eq. (\ref{fitting}) 
to the experimental result for $C_2$, 
we determine $\theta$ to be $\Theta$ (=0.0270),\cite{validity} 
where the fitting is done for $T$=35$-$150 K 
by paying attention to the relatively low temperature side. 
Using $\Theta$=0.0270, 
we can also evaluate $R_{\varepsilon 2}$ of Eq. (\ref{R2_high}). 

\item[(ii)] In the low-temperature range of $T \le$ 35 K, we have
\begin{eqnarray}
\label{R_low}
&&R=\Theta T, \\
\label{del_R_low}
&&\Delta R_\varepsilon
=\frac{T-T^*}{T_l - T^*}\left( \frac{1+R}{\kappa/4} C_4^{(T_l)} \right), \\
\label{R2_low}
&&R_{\varepsilon 2}=
\frac{R_{\varepsilon 2}^* - R_{\varepsilon 2}^{(l)}}{T^* - T_l} T
+ \frac{R_{\varepsilon 2}^{(l)}T^* -R_{\varepsilon 2}^* T_l}{T^* - T_l}, \\
\label{R1_low}
&&R_{\varepsilon 1}=R_{\varepsilon 2} + \Delta R_\varepsilon,
\end{eqnarray}
with $T_l$=4 K, 
$R_{\varepsilon 2}^*$=$1-(1+R^*)C_2^{(T^*)}/\kappa$, 
$R_{\varepsilon 2}^{(l)}$=$1-(1+R_l)(C_2^{(T_l)}+4C_4^{(T_l)})/\kappa$, 
$R_l$=$\Theta T_l$, 
$C_2^{(T_l)}$=$-$0.0343, and $C_4^{(T_l)}$=0.00556, 
where $C_2^{(T_l)}$ ($C_4^{(T_l)}$) 
is the experimental value of $C_2$ at $T$=$T_l$ 
($C_4$ at $T$=$T_l$).\cite{Tsunoda2}

The procedure for determining this parameter set is as follows: 
We first adopt $R$=$\Theta T$, 
which is the same as Eq. (\ref{R_high}) in the high-temperature range, 
on the basis of the expeimental result 
of the $T$ dependence of the total resistivity.\cite{Tsunoda_exp} 
Second, since 
$C_4$ was experimentally observed to be 
a linear function of $T$, 
we assume 
$C_4$ to be 
$C_4$=$p T + q$, where $p$ and $q$ are constants. 
The constants $p$ and $q$ are determined 
so that Eq. (\ref{C_4_Fe4N_1}) 
satisfies the condition 
$(T, C_4)$=$(T_l, C_4^{(T_l)})$, $(T^*, 0)$. 
As a result, $C_4$ is expressed as 
$C_4$=$(T-T^*)C_4^{(T_l)}/(T_l-T^*)$. 
From this $C_4$ and Eq. (\ref{C_4_Fe4N_1}), 
we obtain $\Delta R_\epsilon$ of Eq. (\ref{del_R_low}). 
The obtained $\Delta R_\epsilon$ may indicate the following two properties: 
One is that the crystal has tetragonal symmetry, 
which generates $\Delta R_\varepsilon$$\ne$0 
due to the difference of the PDOS at $E_{\mbox{\tiny F}}$ 
among the d$\varepsilon$ states. 
The other is that 
the tetragonal distortion increases with decreasing $T$. 
Third, $R_{\varepsilon 2}$ 
is assumed to be 
$R_{\varepsilon 2}$=$p'T+q'$ as a simple form, 
where $p'$ and $q'$ are constants. 
The constants $p'$ and $q'$ are determined 
so that 
Eq. (\ref{C_2_Fe4N_1}) satisfies the condition 
$(T, C_2)$=$(T_l, C_2^{(T_l)})$, $(T^*, C_2^{(T^*)})$. 
\end{enumerate}



Substituting the above-mentioned $\lambda$ and $\Delta$, 
Eqs. (\ref{R_high})$-$(\ref{R2_high}), 
and Eqs. (\ref{R_low})$-$(\ref{R2_low}) 
into Eqs. (\ref{C_2_Fe4N_1}) and (\ref{C_4_Fe4N_1}), 
we obtain $C_2$ and $C_4$ for Fe$_4$N, 
where 
$C_4$ in the low-temperature range was described above. 
In the upper panel of Fig. \ref{best}, 
we show the $T$ dependences of $C_2$ and $C_4$. 
We find that 
$C_2$ and $C_4$ for PT 
successfully reproduce 
the experimental results. 
In particular, 
the experimental results 
in the range 4 K $\le T \le$ 35 K, 
in which 
the change of $|C_2|$ is about four times as large as 
that of $|C_4|$, 
can be explained by 
the ratio of the coefficients of $\Delta R_\varepsilon$ 
between Eqs. (\ref{C_2_Fe4N_1}) and (\ref{C_4_Fe4N_1}). 
Finally, 
we comment on 
the above-mentioned $T$ dependence of $\Delta R_\varepsilon$, i.e., 
the difference in the PDOSs at $E_{\mbox{\tiny F}}$ 
among the d$\varepsilon$ states. 
The $T$ dependence of $\Delta R_\varepsilon$ 
has been assumed to arise from 
the increase of the tetragonal distortion due to a decrease in $T$. 
The tetragonal distortion may originate from 
the anisotropic thermal compression of the lattice. 
This compression 
is considered to be 
due to the adhesion between the Fe$_4$N film and the MgO substrate. 
We expect that 
such an assumption 
will be verified experimentally in the future. 


\section{Conclusions}
\label{conclusion}
We theoretically studied 
the twofold and fourfold symmetric AMR effects of ferromagnets. 
In particular, 
we obtained the coefficients of 
the twofold symmetric term ($\cos 2\phi$ term) 
and the fourfold symmetric term ($\cos 4\phi$ term) in the AMR ratio, 
denoted as $C_2$ and $C_4$, respectively. 
We used the two-current model for 
the system consisting of the conduction state and localized d states. 
The localized d states were obtained from 
the Hamiltonian with the spin--orbit interaction, 
the exchange field, and the crystal field. 
Details 
are given as follows: 
\begin{enumerate}
\item[(i)] 
We performed the numerical calculation of $C_2$ and $C_4$ 
for a strong ferromagnet 
using d states, which were obtained by applying 
the EDM to the Hamiltonian. 
The result revealed that 
$C_4$ appears under the crystal field of tetragonal symmetry, 
whereas 
it vanishes under the crystal field of cubic symmetry. 
\item[(ii)] 
We derived general expressions for the resistivity, $C_2$, and $C_4$ 
for ferromagnets with the tetragonal field 
using the d states, which were obtained 
by applying first- and second-order PT to 
the Hamiltonian. 
From the expressions, 
we obtained expressions for $C_2$ and $C_4$ 
for the strong ferromagnet with the tetragonal field. 
The result showed that 
$C_2 \cos 2\phi$ is related to the real part of the probability amplitudes 
of the specific hybridized states 
$|3z^2-r^2, \chi_\sigma (\phi) \rangle$ 
and $|x^2-y^2, \chi_\sigma (\phi) \rangle$ 
and $C_4 \cos 4\phi$ is related to the probabilities of 
$|3z^2-r^2, \chi_\sigma (\phi) \rangle$ 
and $|x^2-y^2, \chi_\sigma (\phi) \rangle$. 
In addition, we investigated 
various features of $C_2$ and $C_4$ obtained by PT 
and found that they 
qualitatively agreed well 
with those obtained by the EDM. 
\item[(iii)] 
We analyzed the experimental results of 
the $T$ dependences of $C_2$ and $C_4$ for 
an Fe$_4$N film on a MgO substrate 
using the dominant terms in $C_2$ and $C_4$ obtained by PT. 
The dominant term in $C_2$ was proportional to 
the difference in the PDOSs at $E_{\mbox{\tiny F}}$ 
between the $d\varepsilon$ and $d\gamma$ states, 
and 
that in $C_4$ was proportional to 
the difference in the PDOSs at $E_{\mbox{\tiny F}}$ 
among the $d\varepsilon$ states. 
The experimental results in the high-temperature range (35 K $< T \le$ 300 K) 
were well reproduced by taking into account 
the $T$ dependence of the $s$--$s$ resistivity 
and 
by assuming that 
the PDOSs of the $d\varepsilon$ states at $E_{\mbox{\tiny F}}$ 
took the same value. 
This assumption might imply that 
the crystal structure of Fe$_4$N exhibits cubic symmetry. 
Also, 
the experimental results in the low-temperature range (4 K $\le T \le$ 35 K) 
were successfully reproduced 
by assuming that 
the difference in 
the PDOSs at $E_{\mbox{\tiny F}}$ among the $d\varepsilon$ states 
increased with decreasing $T$. 
This assumption suggested 
that the tetragonal distortion increases with decreasing $T$. 
Here, the tetragonal distortion was considered to 
originate from 
the anisotropic thermal compression of the lattice 
due to 
the adhesion between the MgO substrate and Fe$_4$N film. 
\end{enumerate}

\begin{acknowledgments}
We would like to thank Prof. M. Shirai of Tohoku University 
for the useful discussion. 
We acknowledge the stimulating discussion 
in the meeting of the Cooperative Research Project (H26/A04) 
of the Research Institute of Electrical Communication, Tohoku University. 
This work has been supported by 
Grants-in-Aid for Scientific Research (C) (Nos. 25390055 and 25410092) 
and 
(A) (No. 26249037) 
from the Japan Society for the Promotion of Science. 
\end{acknowledgments}

\appendix

\section{Matrix Representation of ${\cal H}$}
\label{H_matrix}

We construct the matrix of ${\cal H}$ of Eq. (\ref{Hamiltonian}) 
as shown in Table \ref{matrix1}.

In the construction, 
we perform, for example, the following operations: 
\begin{eqnarray}
\label{LySy}
\lambda L_yS_y |xz,\chi_- (\phi) \rangle &=&
\lambda L_y |xz \rangle S_y |\chi_- (\phi) \rangle \nonumber \\
&=&
i \lambda ( - |x^2-y^2 \rangle + \sqrt{3}|3z^2-r^2 \rangle ) 
\frac{1}{2} 
( i \cos \phi |\chi_+ (\phi) \rangle + \sin \phi | \chi_- (\phi) \rangle ), \nonumber \\\\
\label{LxSx}
\lambda L_xS_x |yz,\chi_- (\phi) \rangle &=&
\lambda L_x |yz \rangle S_x |\chi_- (\phi) \rangle \nonumber \\
&=& 
i \lambda ( |x^2-y^2 \rangle + \sqrt{3}|3z^2-r^2 \rangle ) 
\frac{1}{2} 
( i \sin \phi |\chi_+ (\phi) \rangle - \cos \phi | \chi_- (\phi) \rangle ). \nonumber \\
\end{eqnarray}
Equations (\ref{LySy}) and (\ref{LxSx}) 
play an important role in $C_2$ and $C_4$, 
as described 
when we discuss the $\phi$ dependence of the wave functions 
(see Sec. \ref{wf}). 

\begin{table}
\caption{
Matrix representation of ${\cal H}$ of Eq. (\ref{Hamiltonian}) 
in the basis set 
$|xy,\chi_+ (\phi) \rangle$, 
$|yz,\chi_+ (\phi) \rangle$, 
$|xz,\chi_+ (\phi) \rangle$, 
$|xy,\chi_- (\phi) \rangle$, 
$|yz,\chi_- (\phi) \rangle$, 
$|xz,\chi_- (\phi) \rangle$, 
$|x^2-y^2, \chi_+ (\phi) \rangle$, 
$|3z^2-r^2, \chi_+ (\phi) \rangle$, 
$|x^2-y^2, \chi_- (\phi) \rangle$, and 
$|3z^2-r^2, \chi_- (\phi) \rangle$. 
In this table, 
$(\phi)$ in $\chi_\sigma (\phi)$ is omitted due to limited space. 
}
{\tiny 
\hspace*{-2cm}
\begin{tabularx}{19.5cm}{XXXXXXXXXXX}
\hline 
 & $|xy,\chi_+  \rangle$ & $|yz,\chi_+  \rangle$ & $|xz,\chi_+  \rangle$ & $|xy,\chi_-  \rangle$ & $|yz,\chi_-  \rangle$ & $|xz,\chi_-  \rangle$ & $|x^2-y^2$, $\chi_+  \rangle$ & $|3z^2-r^2$, $\chi_+ \rangle$ & $|x^2-y^2$, $\chi_-  \rangle$ & $|3z^2-r^2$, $\chi_- \rangle$ \\
\hline 
$\langle xy,\chi_+  |$ & $-\frac{H}{2}$ & $i\frac{\lambda}{2} \sin\phi$ & $-i\frac{\lambda}{2}\cos\phi$ & 0 & $\frac{\lambda}{2} \cos \phi$ & $\frac{\lambda}{2}\sin \phi$ & 0 & 0 & $-i\lambda$ & 0 \\ \\
$\langle yz,\chi_+  |$ & $-i\frac{\lambda}{2} \sin\phi$ &  $- \frac{H}{2}+\delta_\varepsilon$ & 0 & $- \frac{\lambda}{2} \cos \phi$ & 0 & $-i \frac{\lambda}{2}$ & $-i \frac{\lambda}{2} \cos \phi$ & $-i \frac{\sqrt{3}\lambda}{2} \cos \phi$ & $\frac{\lambda}{2} \sin \phi$ & $\frac{\sqrt{3}\lambda}{2} \sin \phi$ \\ \\
$\langle xz,\chi_+  |$ & $i\frac{\lambda}{2}\cos\phi$ & 0 & $- \frac{H}{2}+\delta_\varepsilon$ & $- \frac{\lambda}{2} \sin \phi$ & $i \frac{\lambda}{2}$ & 0 & $-i \frac{\lambda}{2} \sin \phi$ & $i \frac{\sqrt{3} \lambda}{2} \sin \phi$ & $-\frac{\lambda}{2} \cos \phi$ & $\frac{\sqrt{3} \lambda}{2} \cos \phi$ \\ \\
$\langle xy,\chi_-  |$ & 0 & $- \frac{\lambda}{2} \cos \phi$ & $- \frac{\lambda}{2} \sin \phi$ & $\frac{H}{2}$ & $-i \frac{\lambda}{2} \sin \phi$ & $i \frac{\lambda}{2} \cos \phi$ & $-i \lambda$ & 0 & 0 & 0 \\ \\
$\langle yz,\chi_-  |$ & $\frac{\lambda}{2} \cos \phi$ & 0 & $-i \frac{\lambda}{2}$ & $i \frac{\lambda}{2} \sin \phi$ & $\frac{H}{2}+\delta_\varepsilon$ & 0 & $-\frac{\lambda}{2} \sin \phi$ & $-\frac{\sqrt{3} \lambda}{2}\sin \phi$ & $i \frac{\lambda}{2} \cos \phi$ & $i \frac{\sqrt{3}\lambda}{2} \cos \phi$ \\ \\
$\langle xz,\chi_- |$ & $\frac{\lambda}{2}\sin \phi$ & $i \frac{\lambda}{2}$ & 0 & $-i \frac{\lambda}{2} \cos \phi$ & 0 & $\frac{H}{2}+\delta_\varepsilon$ & $\frac{\lambda}{2} \cos \phi$ & $- \frac{\sqrt{3} \lambda}{2} \cos \phi$ & $i \frac{\lambda}{2} \sin \phi$ & $- i \frac{\sqrt{3} \lambda}{2} \sin \phi$ \\ \\
$\langle x^2-y^2$, $\chi_+  |$ & 0 & $i \frac{\lambda}{2} \cos \phi$ & $i \frac{\lambda}{2} \sin \phi$ & $i \lambda$ & $-\frac{\lambda}{2} \sin \phi$ & $\frac{\lambda}{2} \cos \phi$ & $- \frac{H}{2}+\Delta$ & 0 & 0 & 0 \\ \\
$\langle 3z^2-r^2$, $\chi_+  |$ & 0 & $i \frac{\sqrt{3}\lambda}{2} \cos \phi$ & $-i \frac{\sqrt{3} \lambda}{2} \sin \phi$ & 0 & $-\frac{\sqrt{3} \lambda}{2}\sin \phi$ & $- \frac{\sqrt{3} \lambda}{2} \cos \phi$ & 0 & $- \frac{H}{2} + \Delta + \delta_\gamma$ & 0 & 0 \\ \\
$\langle x^2-y^2$, $\chi_- |$ & $i \lambda$ & $\frac{\lambda}{2} \sin \phi$ & $-\frac{\lambda}{2} \cos \phi$ & 0 & $-i \frac{\lambda}{2} \cos \phi$ & $-i \frac{\lambda}{2} \sin \phi$ & 0 & 0 & $\frac{H}{2} + \Delta$ & 0 \\ \\
$\langle 3z^2-r^2$, $\chi_- |$ & 0 & $\frac{\sqrt{3}\lambda}{2} \sin \phi$ & $\frac{\sqrt{3} \lambda}{2} \cos \phi$ & 0 & $-i \frac{\sqrt{3}\lambda}{2} \cos \phi$ & $i \frac{\sqrt{3} \lambda}{2} \sin \phi$ & 0 & 0 & 0 & $\frac{H}{2} + \Delta + \delta_\gamma$ \\
\hline 
\end{tabularx}
}
\label{matrix1}
\end{table} 

\section{Zero-Order States}
\label{zero-order}

Performing the unitary transformation for $V$ of Eq. (\ref{V}), 
we obtain the zero-order states.\cite{Sakurai}

Table \ref{V_+-} shows the matrix representation of $V$ in the subspace 
with the basis set 
$|xy,\chi_\sigma (\phi) \rangle$, 
$|yz,\chi_\sigma (\phi) \rangle$, and 
$|xz,\chi_\sigma (\phi) \rangle$, 
with $\sigma$=$+$ or $-$. 
The eigenvalues of $V$ are obtained as 
$\xi_+$, $\delta_\varepsilon$, and $\xi_-$, 
with 
\begin{eqnarray}
\label{xi_pm}
\xi_\pm=
\frac{\delta_\varepsilon \pm \sqrt{\delta_\varepsilon^2 + \lambda^2}}{2}. 
\end{eqnarray}
In the case of $\sigma$=$+$, 
the eigenstates for 
$\xi_+$, $\delta_\varepsilon$, and $\xi_-$ 
are respectively given by 
$|\xi_+,\chi_+(\phi) \rangle$ of Eq. (\ref{+,+}), 
$|\delta_\varepsilon,\chi_+(\phi) \rangle$ of Eq. (\ref{d,+}), 
and 
$|\xi_-,\chi_-(\phi) \rangle$ of Eq. (\ref{-,+}). 
In the case of $\sigma$=$-$, 
the eigenstates for 
$\xi_+$, $\delta_\varepsilon$, and $\xi_-$ 
are 
$|\xi_+,\chi_-(\phi) \rangle$ of Eq. (\ref{+,-}), 
$|\delta_\varepsilon,\chi_-(\phi) \rangle$ of Eq. (\ref{d,-}), 
and 
$|\xi_-,\chi_-(\phi) \rangle$ of Eq. (\ref{-,-}), respectively. 
These states correspond to 
the zero-order states in PT. 

\begin{table}[h]
\caption{
Matrix representation of $V$ of Eq. (\ref{V}) 
in the subspace with the basis set 
$|xy,\chi_\pm (\phi) \rangle$, 
$|yz,\chi_\pm (\phi) \rangle$, and 
$|xz,\chi_\pm (\phi) \rangle$. 
}
\begin{center}
\begin{tabular}{lccc}
\hline 
 & $|xy,\chi_\pm (\phi) \rangle$ & $|yz,\chi_\pm (\phi) \rangle$ & $|xz,\chi_\pm (\phi) \rangle$ \\
\hline 
$\langle xy,\chi_\pm (\phi)|$ & 0  & $\pm i \frac{\lambda}{2} \sin \phi$ & $\mp i \frac{\lambda}{2} \cos \phi$ \\
$\langle yz,\chi_\pm (\phi)|$ & $\mp i \frac{\lambda}{2} \sin \phi$ & $\delta_\varepsilon$ & 0 \\
$\langle xz,\chi_\pm (\phi)|$ & $\pm i \frac{\lambda}{2} \cos \phi$ & 0 & $\delta_\varepsilon$ \\
\hline 
\end{tabular}
\end{center}
\label{V_+-}
\end{table} 

\section{Overlap Integral of $s$--$d$ Scattering Rate}
\label{selection}

We briefly discuss 
$|(i,\chi_\varsigma (\phi)|e^{ik_\sigma x},\chi_\sigma (\phi) \rangle |^2$ 
in Eq. (\ref{tau_sd_inv}), 
where $|i,\chi_\varsigma (\phi))$ is represented by 
a linear combination of 
$|xy,\chi_\sigma (\phi) \rangle$, 
$|yz,\chi_\sigma (\phi) \rangle$, 
$|xz,\chi_\sigma (\phi) \rangle$, 
$|x^2 -y^2,\chi_\sigma (\phi) \rangle$, 
and $|3z^2 -r^2,\chi_\sigma (\phi) \rangle$. 

On the basis of a previous study,\cite{Potter} 
we first give the following overlap integral: 
\begin{eqnarray}
\label{integral}
&&\hspace*{-2.5cm}
\langle \mu \nu,\chi_{\sigma'} (\phi) |e^{i{\mbox{\boldmath $k$}}_\sigma \cdot {\mbox{\boldmath $r$}}},\chi_\sigma (\phi) \rangle 
=
\int_{-\infty}^\infty  \int_{-\infty}^\infty  \int_{-\infty}^\infty 
f(r) 
\mu \nu 
\frac{1}{\sqrt{\Omega}}
e^{i{\mbox{\boldmath $k$}}_\sigma \cdot {\mbox{\boldmath $r$}}}
dx dy dz \delta_{\sigma,\sigma'}\nonumber \\
&&\hspace*{2.cm}=
\frac{32 \pi \Gamma \zeta 
}{\sqrt{\Omega}(k_\sigma^2 + \zeta^2)^3}
\left( \delta_{\mu, \nu} 
- \frac{ 6 k_{\mu,\sigma} k_{\nu,\sigma}}{k_\sigma^2 + \zeta^2} \right)\delta_{\sigma,\sigma'}. 
\end{eqnarray}
Here, we have 
$|e^{i{\mbox{\boldmath $k$}}_\sigma 
\cdot {\mbox{\boldmath $r$}}},\chi_\sigma (\phi) \rangle$=
$(1/\sqrt{\Omega}) e^{i{\mbox{\boldmath $k$}}_\sigma 
\cdot {\mbox{\boldmath $r$}}}\chi_\sigma (\phi)$ 
(see Sec. \ref{sec_resistivity}), 
${\mbox{\boldmath $k$}}_\sigma$
=$(k_{x,\sigma}, k_{y,\sigma}, k_{z,\sigma})$, 
$k_\sigma$=$|{\mbox{\boldmath $k$}}_\sigma |$, and 
$|\mu \nu,\chi_{\sigma'} (\phi) \rangle$=$f(r) \mu \nu \chi_{\sigma'}$, 
with $\mu$=$x$, $y$, $z$, $\nu$=$x$, $y$, $z$, 
$\sigma$=$+$, $-$, and $\sigma'$=$+$, $-$, 
where 
$f(r)$ is the radial part of the 3d orbital expressed by 
$f(r)$=$\Gamma e^{-\zeta r}$, 
and $\Gamma$ and $\zeta$ are constants. 
The state 
$|\mu \nu,\chi_{\sigma'}(\phi) \rangle$ denotes the $\mu \nu$ orbital 
with $\sigma'$ spin. 

Using Eq. (\ref{integral}), we can calculate 
the overlap integrals for realistic orbitals. 
In the case of ${\mbox{\boldmath $k$}}_\sigma$=$(k_\sigma, 0, 0)$, 
corresponding to ${\mbox{\boldmath $I$}}//x$ (see Sec. \ref{sec_resistivity}), 
we have 
\begin{eqnarray}
\label{selection_x2-y2}
&&\langle x^2 -y^2,\chi_{\sigma'} (\phi)|e^{ik_\sigma x},\chi_\sigma (\phi) \rangle=
\frac{1}{2} g_\sigma \delta_{\sigma,\sigma'}, \\
\label{selection_3z2-r2}
&&\langle 3z^2 -r^2,\chi_{\sigma'} (\phi)|e^{ik_\sigma x},\chi_\sigma (\phi) \rangle=
-\frac{1}{2\sqrt{3}} g_\sigma \delta_{\sigma,\sigma'}, \\
\label{selection_xy}
&&\langle xy,\chi_{\sigma'} (\phi) |e^{ik_\sigma x},\chi_\sigma (\phi) \rangle=\langle yz,\chi_{\sigma'} (\phi) |e^{ik_\sigma x},\chi_\sigma (\phi) \rangle 
=\langle xz,\chi_{\sigma'} (\phi) |e^{ik_\sigma x},\chi_\sigma (\phi) \rangle=0, \nonumber \\
\end{eqnarray}
with 
\begin{eqnarray}
\label{g_sigma}
&&g_\sigma = 
- \frac{192 \pi \Gamma \zeta k_\sigma^2 
}{\sqrt{\Omega}(k_\sigma^2 + \zeta^2)^4}. 
\end{eqnarray}
Equations (\ref{selection_x2-y2})$-$(\ref{selection_xy}) 
mean that 
only $|3z^2-r^2,\chi_\sigma (\phi) \rangle$ and 
$|x^2-y^2,\chi_\sigma (\phi) \rangle$ 
contribute to the transport 
of ${\mbox{\boldmath $I$}}//x$. 



Next, using Eqs. (\ref{selection_x2-y2})$-$(\ref{g_sigma}), 
we obtain 
$\left| (i,\chi_\varsigma (\phi)|e^{ik_\sigma x},\chi_\sigma (\phi) \rangle \right|^2$ in Eq. (\ref{tau_sd_inv}). 
Here, $|i,\chi_\varsigma (\phi) )$ is given simply by 
$|i,\chi_\varsigma (\phi) )$=$\sum_j \sum_\sigma 
a_{j,\sigma}^{i,\varsigma}(\phi) |j,\chi_\sigma (\phi) \rangle$, 
where $a_{j,\sigma}^{i,\varsigma}(\phi)$ is the coefficient 
of $|j,\chi_\sigma (\phi) \rangle$. 
In the case of $j$=$x^2-y^2$ or $3z^2-r^2$, 
$a_{j,\sigma}^{i,\varsigma}(\phi)$ 
corresponds to 
$c_{i,\varsigma}$, 
$c_{i,\varsigma}w_{j,\sigma}^{i,\varsigma} \cos 2 \phi$, 
or 
$c_{i,\varsigma}w_{j,\sigma}^{i,\varsigma} \sin 2 \phi$, 
as seen from Eqs. (\ref{|+,+)})$-$(\ref{|3z2,-)}). 
As a result, 
$\left| (i,\chi_\varsigma (\phi)|e^{ik_\sigma x},\chi_\sigma (\phi) \rangle \right|^2$ is expressed as
\begin{eqnarray}
\label{selection1}
&& \hspace*{-1cm}
\left| (i,\chi_\varsigma (\phi)|e^{ik_\sigma x},\chi_\sigma (\phi) \rangle \right|^2 
= 
\left| \frac{1}{2} a_{x^2-y^2,\sigma}^{i,\varsigma} (\phi) g_\sigma
- \frac{1}{2\sqrt{3}} a_{3z^2-r^2,\sigma}^{i,\varsigma}(\phi) g_\sigma 
\right|^2. 
\end{eqnarray}
In addition, 
$\left| (i,\chi_\varsigma (\phi)|e^{ik_\sigma x},\chi_\sigma (\phi) \rangle \right|^2$ 
leads to the following two types of expressions 
(see Table \ref{tab1}). 
Type 1 is written as 
\begin{eqnarray}
\label{type1}
&&\left| \ell_0 + \ell_{\rm c} \cos 2\phi \right|^2 
=
|\ell_0 |^2 + \frac{1}{2}|\ell_{\rm c}|^2 + 
(\ell_0^* \ell_{\rm c} + \ell_0 \ell_{\rm c}^*) \cos 2\phi 
+ \frac{1}{2} |\ell_{\rm c}|^2 \cos 4\phi, 
\end{eqnarray}
where $\ell_0$ ($\ell_{\rm c}$) is the coefficient of the constant term 
($\cos 2\phi$ term). 
Type 2 is 
\begin{eqnarray}
\label{type2}
&&\left|\ell_{\rm s} \sin 2\phi \right|^2 
= \frac{1}{2} |\ell_{\rm s}|^2 - \frac{1}{2} |\ell_{\rm s}|^2 \cos 4\phi, 
\end{eqnarray}
where $\ell_{\rm s}$ is the coefficient of the $\sin 2\phi$ term. 
Type 1 generates the twofold and fourfold symmetric terms 
and type 2 generates the fourfold symmetric term.

\begin{table*}
\caption{
Expressions for 
$\left| (i,\chi_\varsigma (\phi)|e^{ik_\sigma x},\chi_\sigma (\phi) \rangle \right|^2$. 
Types 1 and 2 are given by 
Eqs. (\ref{type1}) and (\ref{type2}), respectively. 
Here, Eqs. (\ref{selection_x2-y2})$-$(\ref{g_sigma}) are used. 
}
\begin{center}
\begin{tabular}{llllc}
\hline 
$i$ & $\varsigma$ & $\sigma$ & 
$\left| (i,\chi_\varsigma (\phi)|e^{ik_\sigma x},\chi_\sigma (\phi) \rangle \right|^2$ & 
type \\
\hline 
$\xi_+$ & $+$ & $+$ & 
$g_+^2 \left|
\frac{1}{2\sqrt{3}} c_{\xi_+,+} w_{3z^2-r^2,+}^{\xi_+,+} \sin 2\phi
\right|^2$ & 2 \\
$\delta_\varepsilon$ & $+$ & $+$ 
& 
$g_+^2 \left| 
c_{\delta_\varepsilon,+}
\left( \frac{1}{2} - \frac{1}{2\sqrt{3}}w_{3z^2-r^2,+}^{\delta_\varepsilon,+} \cos 2\phi
\right)
\right|^2$ & 1 \\
$\xi_-$ & $+$ & $+$ &
$g_+^2 \left|
\frac{1}{2\sqrt{3}} c_{\xi_-,+} w_{3z^2-r^2,+}^{\xi_-,+} \sin 2\phi
\right|^2$ & 2 \\
$x^2-y^2$ & $+$ & $+$
&
$g_+^2 \left| 
c_{x^2-y^2,+}
\left( \frac{1}{2} - \frac{1}{2\sqrt{3}}w_{3z^2-r^2,+}^{x^2-y^2,+} \cos 2\phi
\right)
\right|^2$ & 1 \\
$3z^2-r^2$ & $+$ & $+$ 
& $g_+^2 \left| 
c_{3z^2-r^2,+}
\left( - \frac{1}{2\sqrt{3}} +
\frac{1}{2} w_{x^2-y^2,+}^{3z^2-r^2,+} \cos 2\phi
\right)
\right|^2$ & 1 \\
$\xi_+$ & $+$ & $-$ 
& 
$g_-^2 \left| 
c_{\xi_+,+}\left( \frac{1}{2} - \frac{1}{2\sqrt{3}}w_{3z^2-r^2,-}^{\xi_+,+} \cos 2\phi
\right)
\right|^2$ & 1 \\
$\delta_\varepsilon$ & $+$ & $-$ 
& 
$g_-^2 \left|
\frac{1}{2\sqrt{3}} c_{\delta_\varepsilon,+} w_{3z^2-r^2,-}^{\delta_\varepsilon,+} \sin 2\phi
\right|^2$ & 2 \\
$\xi_-$ & $+$ & $-$ 
&
$g_-^2 \left| 
c_{\xi_-,+}
\left( \frac{1}{2} - \frac{1}{2\sqrt{3}}w_{3z^2-r^2,-}^{\xi_-,+} \cos 2\phi
\right)
\right|^2$ & 1 \\
$\xi_+$ & $-$ & $-$ & 
$g_-^2 \left|
\frac{1}{2\sqrt{3}} c_{\xi_+,-} w_{3z^2-r^2,-}^{\xi_+,-} \sin 2\phi
\right|^2$ & 2 \\
$\delta_\varepsilon$ & $-$ & $-$ 
& 
$g_-^2 \left| 
c_{\delta_\varepsilon,-}
\left( \frac{1}{2} - \frac{1}{2\sqrt{3}}w_{3z^2-r^2,+}^{\delta_\varepsilon,-} \cos 2\phi
\right)
\right|^2$ & 1 \\
$\xi_-$ & $-$ & $-$
&
$g_-^2 \left|
\frac{1}{2\sqrt{3}} c_{\xi_-,-} w_{3z^2-r^2,-}^{\xi_-,-} \sin 2\phi
\right|^2$ & 2 \\
$x^2-y^2$ & $-$ & $-$ 
&
$g_-^2 \left| 
c_{x^2-y^2,-}
\left( \frac{1}{2} - \frac{1}{2\sqrt{3}}w_{3z^2-r^2,-}^{x^2-y^2,-} \cos 2\phi
\right)
\right|^2$ & 1 \\
$3z^2-r^2$ & $-$ & $-$ 
& $g_-^2 \left| 
c_{3z^2-r^2,-}
\left( - \frac{1}{2\sqrt{3}} 
+ \frac{1}{2} w_{x^2-y^2,-}^{3z^2-r^2,-} \cos 2\phi
\right)
\right|^2$ & 1 \\
$\xi_+$ & $-$ & $+$ 
& 
$g_+^2 \left| 
c_{\xi_+,-}\left( \frac{1}{2} - \frac{1}{2\sqrt{3}}w_{3z^2-r^2,+}^{\xi_+,-} \cos 2\phi
\right)
\right|^2$ & 1 \\
$\delta_\varepsilon$ & $-$ & $+$ 
& 
$g_+^2 \left|
\frac{1}{2\sqrt{3}} c_{\delta_\varepsilon,-} w_{3z^2-r^2,+}^{\delta_\varepsilon,-} \sin 2\phi
\right|^2$ & 2 \\
$\xi_-$ & $-$ & $+$ 
&
$g_+^2 \left| 
c_{\xi_-,-}
\left( \frac{1}{2} - \frac{1}{2\sqrt{3}}w_{3z^2-r^2,+}^{\xi_-,-} \cos 2\phi
\right)
\right|^2$ & 1 \\
\hline 
\end{tabular}
\end{center}
\label{tab1}
\end{table*}

\section{$s$--$d$ Scattering Rate}
\label{sd_scatter}
Using Eqs. (\ref{tau_sd_inv}), 
(\ref{|+,+)})$-$(\ref{|3z2,-)}), 
and (\ref{selection_x2-y2})$-$(\ref{type2}), 
we obtain the sum of the $s$--$d$ scattering rates 
of the second term in the right-hand side of Eq. (\ref{tau_inv}), 
i.e., 
$\sum_{i}\sum_{\varsigma}  1/\tau_{s,\sigma \to d_i,\varsigma} (\phi)$. 

We can express 
$\sum_{i}\sum_{\varsigma}  1/\tau_{s,\sigma \to d_i,\varsigma} (\phi)$ 
as
\begin{eqnarray}
&&\hspace*{-1cm}
\sum_{i} \sum_{\varsigma} 
\frac{1}{\tau_{s,\sigma \to d_i,\varsigma} (\phi)} 
=\frac{2\pi}{\hbar} n_{\rm imp}N_{\rm n} {V_{\rm imp}(R_{\rm n})}^2
( X_{0,\sigma} + X_{2\phi,\sigma} + X_{4\phi,\sigma} ). 
\end{eqnarray}
Here, 
$X_{0,\sigma}$ is the constant term, which is independent of $\phi$, 
$X_{2\phi,\sigma}$ is proportional to $\cos 2\phi$, 
and $X_{4\phi,\sigma}$ is proportional to $\cos 4\phi$. 
The terms of $\sigma$=$+$ 
are as follows: 
\begin{eqnarray}
&&X_{0,+} = 
\frac{1}{2}\left|c_{\xi_+,+}\right|^2 o_2^2 \left|w_{3z^2-r^2,+}^{\xi_+,+}\right|^2 
D_{\xi_+,+}^{(d)} \nonumber \\
&&\hspace*{1.2cm}+
\left|c_{\delta_\varepsilon,+}\right|^2 \left( o_1^2 + \frac{1}{2}o_2^2 \left|w_{3z^2-r^2,+}^{\delta_\varepsilon,+} \right|^2 \right) 
D_{\delta_\varepsilon,+}^{(d)} \nonumber \\
&&\hspace*{1.2cm}
+\frac{1}{2}\left|c_{\xi_-,+}\right|^2 o_2^2 \left|w_{3z^2-r^2,+}^{\xi_-,+}\right|^2 
D_{\xi_-,+}^{(d)} \nonumber \\
&&\hspace*{1.2cm}+
\left|c_{x^2-y^2,+}\right|^2 \left( o_1^2 + \frac{1}{2}o_2^2 \left|w_{3z^2-r^2,+}^{x^2-y^2,+} \right|^2 \right) 
D_{x^2-y^2,+}^{(d)} \nonumber \\
&&\hspace*{1.2cm}
+\left|c_{3z^2-r^2,+}\right|^2 \left( o_2^2 + \frac{1}{2}o_1^2 \left|w_{x^2-y^2,+}^{3z^2-r^2,+} \right|^2 \right) 
D_{3z^2-r^2,+}^{(d)} \nonumber \\
&&\hspace*{1.2cm}
+ \left|c_{\xi_+,-}\right|^2 \left( o_1^2 + \frac{1}{2}o_2^2 \left|w_{3z^2-r^2,+}^{\xi_+,-} \right|^2 \right) 
D_{\xi_+,-}^{(d)} \nonumber \\
&&\hspace*{1.2cm}
+\frac{1}{2}\left|c_{\delta_\varepsilon,-}\right|^2 o_2^2 \left|w_{3z^2-r^2,+}^{\delta_\varepsilon,-}\right|^2 
D_{\delta_\varepsilon,-}^{(d)} \nonumber \\
&&\hspace*{1.2cm}
+\left|c_{\xi_-,-}\right|^2 \left( o_1^2 + \frac{1}{2}o_2^2 \left|w_{3z^2-r^2,+}^{\xi_-,-} \right|^2 \right) 
D_{\xi_-,-}^{(d)}, 
\end{eqnarray}
\begin{eqnarray}
\label{X_{2phi,up}}
&&X_{2\phi,+}=
2o_1 o_2 \Bigg[ 
\left|c_{\delta_\varepsilon,+}\right|^2 {\rm Re}\left[ w_{3z^2-r^2,+}^{\delta_\varepsilon,+} \right] D_{\delta_\varepsilon,+}^{(d)} \nonumber \\
&&\hspace*{1.4cm}+\left|c_{x^2-y^2,+}\right|^2 {\rm Re}\left[ w_{3z^2-r^2,+}^{x^2-y^2,+} \right] D_{x^2-y^2,+}^{(d)} \nonumber \\
&&\hspace*{1.4cm}+\left|c_{3z^2-r^2,+}\right|^2 {\rm Re}\left[ w_{x^2-y^2,+}^{3z^2-r^2,+} \right] D_{3z^2-r^2,+}^{(d)} \nonumber \\
&&\hspace*{1.4cm}+
\left|c_{\xi_+,-}\right|^2 {\rm Re}\left[ w_{3z^2-r^2,+}^{\xi_+,-} \right] D_{\xi_+,-}^{(d)} \nonumber \\
&&\hspace*{1.4cm}+\left|c_{\xi_-,-}\right|^2 {\rm Re}\left[ w_{3z^2-r^2,+}^{\xi_-,-} \right]D_{\xi_-,-}^{(d)} \Bigg]\cos 2\phi, \\
\label{X_{4phi,up}}
&&X_{4\phi,+}=
\frac{1}{2} \Bigg[ - \left|c_{\xi_+,+}\right|^2 o_2^2 \left|w_{3z^2-r^2,+}^{\xi_+,+}\right|^2 D_{\xi_+,+}^{(d)} \nonumber \\
&&\hspace*{1.4cm}
+ \left|c_{\delta_\varepsilon,+}\right|^2 o_2^2 \left|w_{3z^2-r^2,+}^{\delta_\varepsilon,+}\right|^2 D_{\delta_\varepsilon,+}^{(d)} \nonumber \\
&&\hspace*{1.4cm}
- \left|c_{\xi_-,+}\right|^2 o_2^2 \left|w_{3z^2-r^2,+}^{\xi_-,+}\right|^2 D_{\xi_-,+}^{(d)} \nonumber \\
&&\hspace*{1.4cm}
+ \left|c_{x^2-y^2,+}\right|^2 o_2^2 \left|w_{3z^2-r^2,+}^{x^2-y^2,+}\right|^2 D_{x^2-y^2,+}^{(d)} \nonumber \\
&&\hspace*{1.4cm}
+ \left|c_{3z^2-r^2,+}\right|^2 o_1^2 \left|w_{x^2-y^2,+}^{3z^2-r^2,+}\right|^2 D_{3z^2-r^2,+}^{(d)} \nonumber \\
&&\hspace*{1.4cm}
+ \left|c_{\xi_+,-}\right|^2 o_2^2 \left|w_{3z^2-r^2,+}^{\xi_+,-}\right|^2 D_{\xi_+,-}^{(d)} \nonumber \\
&&\hspace*{1.4cm}-\left|c_{\delta_\varepsilon,-}\right|^2 o_2^2 \left|w_{3z^2-r^2,+}^{\delta_\varepsilon,-}\right|^2 D_{\delta_\varepsilon,-}^{(d)} \nonumber \\
&& \hspace*{1.4cm}+\left|c_{\xi_-,-}\right|^2 o_2^2 \left|w_{3z^2-r^2,+}^{\xi_-,-}\right|^2 D_{\xi_-,-}^{(d)} \Bigg]\cos 4\phi, 
\end{eqnarray}
with 
$o_1$=$\langle x^2-y^2, \chi_\sigma (\phi) |e^{ik_\sigma x}, \chi_\sigma (\phi) \rangle$ 
and 
$o_2$=$\langle 3z^2-r^2, \chi_\sigma (\phi) |e^{ik_\sigma x}, \chi_\sigma (\phi) \rangle$, 
where $o_1$ and $o_2$ are calculated in 
Eqs. (\ref{selection_x2-y2}) and (\ref{selection_3z2-r2}), respectively. 
The terms of $\sigma$=$-$ 
are as follows: 
\begin{eqnarray}
&&X_{0,-} = 
\left|c_{\xi_+,+}\right|^2 
\left( o_1^2 + \frac{1}{2}o_2^2 
\left|w_{3z^2-r^2,-}^{\xi_+,+} \right|^2 \right) 
D_{\xi_+,+}^{(d)} \nonumber \\
&&\hspace*{1.2cm}
+\frac{1}{2}\left|c_{\delta_\varepsilon,+}\right|^2 o_2^2 \left|w_{3z^2-r^2,-}^{\delta_\varepsilon,+}\right|^2 D_{\delta_\varepsilon,+}^{(d)} \nonumber \\
&&\hspace*{1.2cm}+ \left|c_{\xi_-,+}\right|^2 
\left( o_1^2 + \frac{1}{2}o_2^2 
\left|w_{3z^2-r^2,-}^{\xi_-,+} \right|^2 \right) 
D_{\xi_-,+}^{(d)} \nonumber \\
&&\hspace*{1.2cm}+ 
\frac{1}{2} \left|c_{\xi_+,-}\right|^2 o_2^2 \left|w_{3z^2-r^2,-}^{\xi_+,-} \right|^2 D_{\xi_+,-}^{(d)} \nonumber \\
&&\hspace*{1.2cm}+ \left|c_{\delta_\varepsilon,-}\right|^2 
\left( o_1^2 + \frac{1}{2}o_2^2 
\left|w_{3z^2-r^2,-}^{\delta_\varepsilon,-} \right|^2 \right) 
D_{\delta_\varepsilon,-}^{(d)} \nonumber \\
&&\hspace*{1.2cm}
+\frac{1}{2}\left|c_{\xi_-,-}\right|^2 o_2^2 \left|w_{3z^2-r^2,-}^{\xi_-,-}\right|^2 
D_{\xi_-,-}^{(d)} \nonumber \\
&&\hspace*{1.2cm}
+
\left|c_{x^2-y^2,-}\right|^2
\left( o_1^2 + \frac{1}{2}o_2^2 \left|w_{3z^2-r^2,-}^{x^2-y^2,-} \right|^2 \right) D_{x^2-y^2,-}^{(d)} \nonumber \\
&&\hspace*{1.2cm}+\left|c_{3z^2-r^2,-}\right|^2 \left( o_2^2 + \frac{1}{2} o_1^2 \left|w_{x^2-y^2,-}^{3z^2-r^2,-}\right|^2 \right) D_{3z^2-r^2,-}^{(d)}, 
\end{eqnarray}
\begin{eqnarray}
\label{X_{2phi,dw}}
&&X_{2\phi,-}=
2o_1 o_2 \Bigg[ 
\left|c_{\xi_+,+}\right|^2 
{\rm Re}\left[ w_{3z^2-r^2,-}^{\xi_+,+} \right] 
D_{\xi_+,+}^{(d)} \nonumber \\
&&\hspace*{1.4cm}+\left|c_{\xi_-,+}\right|^2 
{\rm Re}\left[ w_{3z^2-r^2,-}^{\xi_-,+} \right] 
D_{\xi_-,+}^{(d)} \nonumber \\
&&\hspace*{1.4cm}
+ \left|c_{\delta_\varepsilon,-}\right|^2 {\rm Re}\left[ w_{3z^2-r^2,-}^{\delta_\varepsilon,-} \right] D_{\delta_\varepsilon,-}^{(d)} 
\nonumber \\
&&\hspace*{1.4cm}+\left|c_{x^2-y^2,-}\right|^2 
{\rm Re}\left[ w_{3z^2-r^2,-}^{x^2-y^2,-} \right] 
D_{x^2-y^2,-}^{(d)} \nonumber \\
&&\hspace*{1.4cm}
+\left|c_{3z^2-r^2,-}\right|^2{\rm Re}
\left[ w_{x^2-y^2,-}^{3z^2-r^2,-} \right]
D_{3z^2-r^2,-}^{(d)} \Bigg] \cos 2\phi, \\
\label{X_{4phi,dw}}
&&X_{4\phi,-}=
\frac{1}{2} \Bigg[ 
|c_{\xi_+,+}|^2 o_2^2\left|w_{3z^2-r^2,-}^{\xi_+,+}\right|^2 D_{\xi_+,+}^{(d)} \nonumber \\
&&\hspace*{1.4cm} - |c_{\delta_\varepsilon,+}|^2 o_2^2\left|w_{3z^2-r^2,-}^{\delta_\varepsilon,+}\right|^2 D_{\delta_\varepsilon,+}^{(d)} \nonumber \\
&&\hspace*{1.4cm} + |c_{\xi_-,+}|^2 o_2^2\left|w_{3z^2-r^2,-}^{-,+}\right|^2 D_{\xi_-,+}^{(d)} \nonumber \\
&&\hspace*{1.4cm} 
-\left|c_{\xi_+,-}\right|^2 o_2^2\left|w_{3z^2-r^2,-}^{\xi_+,-}\right|^2 D_{\xi_+,-}^{(d)} \nonumber \\
&&\hspace*{1.4cm}+\left|c_{\delta_\varepsilon,-}\right|^2 o_2^2\left|w_{3z^2-r^2,-}^{\delta_\varepsilon,-}\right|^2 D_{\delta_\varepsilon,-}^{(d)}\nonumber \\
&&\hspace*{1.4cm} - |c_{\xi_-,-}|^2 o_2^2\left|w_{3z^2-r^2,-}^{\xi_-,-}\right|^2 D_{\xi_-,-}^{(d)} \nonumber \\
&&\hspace*{1.4cm} + 
\left|c_{x^2-y^2,-}\right|^2 o_2^2\left|w_{3z^2-r^2,-}^{x^2-y^2,-}\right|^2 D_{x^2-y^2,-}^{(d)} \nonumber \\
&&\hspace*{1.4cm} + 
\left|c_{3z^2-r^2,-}\right|^2 o_1^2\left|w_{x^2-y^2,-}^{3z^2-r^2,-}\right|^2 D_{3z^2-r^2,-}^{(d)} \Bigg] \cos 4\phi. 
\end{eqnarray}

\section{Relation between the Present Model and Previous Models}
\label{relation}
%


\subsection{Correspondence to our previous model}
\label{previous}
We show that 
$\Delta \rho(0)/\rho$ (=2$C_2$) of the present model 
coincides with 
that of 
our previous model\cite{Kokado1} 
under the condition $D_{i,\sigma}^{(d)}$=$D_\sigma^{(d)}$, 
which indicates that 
the orbital $i$ dependence of the PDOS is ignored. 
Under this condition, 
$\rho_{s,\sigma \to d_i,\varsigma}$ is replaced by 
$\rho_{s,\sigma \to d,\varsigma}$ 
[see Eqs. (\ref{rho_i_pm}) and (\ref{1/tau_i_pm})]. 
This replacement leads to $\rho_{2,\pm}^{(1)}$=0 
[see Eq. (\ref{rho_2_pm^(1)})]. 


On the basis of Eq. (7) in Ref. \citen{Kokado1}, 
we first give an expression for the resistivity 
with spin-flip scattering $\rho_\sigma (\phi)$, i.e., 
\begin{eqnarray}
\label{rho with sf}
&&
\hspace*{-0.3cm}
\rho (\phi) = 
\frac{\rho_{+}(\phi) \rho_{-} (\phi)
+ \rho_{+}(\phi) \rho_{-,+} (\phi)
+\rho_{-}(\phi) \rho_{+,-}(\phi)}
{\rho_{+} (\phi)
+ \rho_{-} (\phi)
+ (1 +a)\rho_{+,-}(\phi)
+ (1 + a^{-1})
\rho_{-,+}(\phi)}, 
\end{eqnarray}
with 
$a$=$m_-^* n_+/(m_+^* n_-)$,\cite{Kokado1} 
where 
$\rho_{+,-}(\phi)$ [$\rho_{-,+}(\phi)$] is the resistivity 
of the spin-flip scattering from the up spin to the down spin 
(from the down spin to the up spin).

Using Eqs. (\ref{AMR}), (\ref{rho with sf}), 
and (\ref{rho_s_phi})$-$(\ref{rho_4_pm^(2)}), 
we can obtain an expression for $\Delta \rho (\phi)/\rho$. 
The coefficient $C_2$ is finally obtained as
\begin{eqnarray}
\label{C2_JPSJ}
&&C_2=
\frac{1}{X}( 
\tilde{\rho}_{2,+}^{(2)}Y_1 +
\tilde{\rho}_{2,-}^{(2)} Y_2 ),
\end{eqnarray}
with
\begin{eqnarray}
&&X=
[ (\rho_{s,+} + \rho_{s,+ \to d,+}) 
(\rho_{s,-} + \rho_{s,- \to d,-} + \rho_{-,+}) 
+(\rho_{s,-} + \rho_{s,- \to d,-}) \rho_{+,-} 
] \nonumber \\
&&\hspace*{0.8cm}\times 
[
\rho_{s,+} + \rho_{s,+ \to d,+} 
+ \rho_{s,-} + \rho_{s,- \to d,-} 
+ (1+a)\rho_{+,-} + (1+a^{-1}) \rho_{-,+} 
],
\\
&&Y_1=
(\rho_{s,-} + \rho_{s,- \to d,-}) 
(\rho_{s,-} + \rho_{s,- \to d,-} + 
\rho_{-,+} - \rho_{+,-}) \nonumber \\
&&\hspace*{0.8cm}+(\rho_{s,-} + \rho_{s,- \to d,-} + \rho_{-,+}) 
[(1+a)\rho_{+,-} + (1+a^{-1}) \rho_{-,+} 
],
\\
&&Y_2=
(\rho_{s,+} + \rho_{s,+ \to d,+}) 
(\rho_{s,+} + \rho_{s,+ \to d,+} + 
\rho_{+,-} - \rho_{-,+}) \nonumber \\
&&\hspace*{0.8cm}+(\rho_{s,+} + \rho_{s,+ \to d,+} + \rho_{+,-}) 
[(1+a)\rho_{+,-} + (1+a^{-1}) \rho_{-,+} 
],
\\
&&\tilde{\rho}_{2,+}^{(2)}=\frac{3}{8} \left( \frac{\lambda}{H - \Delta} \right)^2 ( \rho_{s,+ \to d,-} - \rho_{s,+ \to d,+}),
\\
&&\tilde{\rho}_{2,-}^{(2)}=\frac{3}{8} \left( \frac{\lambda}{H + \Delta} \right)^2 ( \rho_{s,- \to d,+} - \rho_{s,- \to d,-}). 
\end{eqnarray}
Here, we have ignored the $\phi$ dependences 
of $\rho_{+,-}(\phi)$ and $\rho_{-,+}(\phi)$ 
in the same manner as in Ref. \citen{Kokado1}; 
that is, 
$\rho_{+,-}(\phi)$$\equiv$$\rho_{+,-}$ 
and 
$\rho_{-,+}(\phi)$$\equiv$$\rho_{-,+}$ have been used. 

On the assumption of $H \gg \Delta$, 
we express 
$C_2$ of Eq. (\ref{C2_JPSJ}) as
\begin{eqnarray}
C_2 = \frac{3}{8}\left( \frac{\lambda}{H} \right)^2 \frac{1}{X}
[ (\rho_{s,+ \to d,-} - \rho_{s,+ \to d,+}) Y_1
+ (\rho_{s,- \to d,+} - \rho_{s,- \to d,-}) Y_2 ], 
\end{eqnarray}
where 
$\left[ \lambda/(H \pm \Delta) \right]^2$$\simeq$$(\lambda/H)^2$ has been used. 
As a result, 
$\Delta \rho(0)/\rho$ (=$2C_2$) becomes 
\begin{eqnarray}
\label{JPSJ_eq_28}
\frac{\Delta \rho(0)}{\rho}
= \frac{3}{4}\left( \frac{\lambda}{H} \right)^2 \frac{1}{X}
[ (\rho_{s,+ \to d,-} - \rho_{s,+ \to d,+}) Y_1
+ (\rho_{s,- \to d,+} - \rho_{s,- \to d,-}) Y_2 ]. \nonumber \\
\end{eqnarray}
Equation (\ref{JPSJ_eq_28}) corresponds to 
Eq. (28) in Ref. \citen{Kokado1}. 


\subsection{Correspondence to CFJ model}
\label{CFJ}

We show that 
$\Delta \rho(0)/\rho$ (=2$C_2$) of the present model 
coincides with 
that of the CFJ model\cite{Campbell1} 
under the condition of the CFJ model, i.e., 
$D_{i,+}^{(d)}$=0, 
$r_{\varepsilon1}$=$r_{\varepsilon2}$=$r_\gamma$=$\alpha$, 
$r \ll \alpha$, and $r \ll 1$.\cite{Kokado1} 
Here, $C_2$ is given by Eq. (\ref{C_2_strong}). 



Under the above condition and the assumption of $H \gg \Delta$, 
we express 
$C_2$ of Eq. (\ref{C_2_strong}) as
\begin{eqnarray}
\label{C_2_CFJ}
C_2 = \frac{3}{8} \left(\frac{\lambda}{H} \right)^2  (\alpha -1), 
\end{eqnarray}
where 
$\left[ \lambda/(H \pm \Delta) \right]^2$$\simeq$$(\lambda/H)^2$. 
As a result, 
$\Delta \rho(0)/\rho$ (=$2C_2$) becomes 
\begin{eqnarray}
\label{CFJ_model}
\frac{\Delta \rho(0)}{\rho}= 
\frac{3}{4} \left(\frac{\lambda}{H} \right)^2(\alpha -1).
\end{eqnarray}
Equation (\ref{CFJ_model}) is 
$\Delta \rho(0)/\rho$ of the CFJ model in Ref. \citen{Campbell1}. 


\begin{thebibliography}{10}
\bibitem{half-metal}
The half-metallic ferromagnet is defined as 
having a finite density 
of states (DOS) at the Fermi energy ($E_{\mbox{\tiny F}}$) 
in one spin channel 
and a zero DOS at $E_{\mbox{\tiny F}}$ in the other spin channel.
\bibitem{comment_H}
When the impurities are randomly located in a crystal, 
we obtain 
the $s$--$d$ scattering rate of Eq. (\ref{tau_sd_inv}) 
with the following features: 
(i) The final states $|i,\chi_\varsigma (\phi))$ are 
the d states of the single atom, 
which are obtained from the Hamiltonian of the single atom. 
(ii) The scattering rate includes $D_{i,\varsigma}^{(d)}$, i.e., 
the PDOS 
of the wave function of the tight-binding model 
for the d state of the $i$ orbital and $\varsigma$ spin 
at $E_{\mbox{\tiny F}}$. 
Details were described in Appendix B and 
Eqs. (B$\cdot$4) and (B$\cdot$16)$-$(B$\cdot$18) 
in Ref. \citen{Kokado1}.
\bibitem{tetra}
In the case of the crystal field of cubic symmetry 
with $\delta_\varepsilon$=$\delta_\gamma$=0, 
the coefficients could not be analytically derived 
within the framework of second-order PT.
\bibitem{parameter}
We roughly estimate 
$\Delta/H$ and $|\lambda|/\Delta$ 
to be $\Delta/H$$\sim$0.1 and $|\lambda|/\Delta$$\sim$0.013. 
We here use 
$\lambda$=$-$0.013 eV for Fe$^{2+}$,\cite{Yosida} 
$\Delta$$\sim$0.1 eV for fcc-Fe in the ferromagnetic state, 
and 
$H$$\sim$1 eV for typical ferromagnets. 
Here, fcc-Fe in the ferromagnetic state 
is regarded as a simple system 
that is similar to Fe$_4$N. 
In addition, $\Delta$ is evaluated 
by fitting the dispersion curves obtained by the tight-binding model to 
those obtained by the first-principles calculation. 
This calculation method was described in Ref. \citen{Kokado3}. 
Regarding $H$$\sim$1 eV, 
see Sec. 13.1 
in Ref. \citen{Yosida}.
\bibitem{previous}
For example, see Ref. \citen{Malozemoff1} 
and Sec. 3 in Ref. \citen{Kokado1}. 
As shown in Sec. 3 in Ref. \citen{Kokado1}, 
experimental results were satisfactorily analyzed 
by using a model based on such an assumption 
in spite of the rough assumption.
\bibitem{com_param}
As noted in Table I in Ref. \citen{Kokado1}, 
$r$ (i.e., $\rho_{s \downarrow}/\rho_{s \uparrow}$) 
was evaluated to be $r$=$1.6 \times 10^{-3} \ll 1$. 
In addition, 
as shown in Sec. 3 in Ref. \citen{Kokado1}, 
$r_\gamma$ (i.e., $\rho_{s \to d \downarrow}/\rho_{s\uparrow}$) 
may be considered to be $\sim$0.01.
\bibitem{W_FM_C4}
Using the EDM of Sec. \ref{edm}, 
we can obtain $C_2$ and $C_4$ for weak ferromagnets 
with $D_+$$\ne$0 and $D_-$$\ne$0. 
The weak ferromagnets 
exhibit $C_4$$\sim$0 at $\eta$=0, similarly to strong ferromagnets.
\bibitem{com_model1}
We report a basic feature of Fe$_4$N. 
The crystal structure of Fe$_4$N is a perovskite-type structure, 
in which N is located at the body center position of fcc-Fe.\cite{Jack} 
The unit cell with a cubic shape 
consists of a corner site and three face-center sites. 
Here, 
the $d \varepsilon$ states for each spin 
at the corner site are considered to be degenerate. 
The three face-center sites are in the $xy$, $yz$, and $xz$ planes 
and are denoted $F_{xy}$, $F_{yz}$, and $F_{xz}$, respectively. 
We can then specify 
the ground and excited states 
of the $d \varepsilon$ orbitals at $F_{xy}$, $F_{yz}$, and $F_{xz}$ 
taking into account the effect of N at the body center site. 
At $F_{xy}$, 
the ground state is the $xy$ orbital, 
while the excited states are the $yz$ and $xz$ orbitals. 
At $F_{yz}$, 
the ground state is the $yz$ orbital, 
while the excited states are the $xy$ and $xz$ orbitals. 
At $F_{xz}$, 
the ground state is the $xz$ orbital, 
while the excited states are the $xy$ and $yz$ orbitals. 
In this system, 
the PDOSs of the $xy$, $yz$, and $xz$ states for each spin 
at $E_{\mbox{\tiny F}}$ 
take the same value. 
In contrast, 
when 
the system has tetragonal distortion in the $z$ direction, 
the PDOS of the $xy$ state at $E_{\mbox{\tiny F}}$ 
is different from that of the $yz$ or $xz$ state at $E_{\mbox{\tiny F}}$.
\bibitem{Tsunoda_exp}
From Fig. 2 in Ref. \citen{Tsunoda2}, 
we roughly evaluate 
the total resistivity $\rho_{\rm total}$ to be 
$\rho_{\rm total}$$\simeq$$T/3 + 14$ $\mu \Omega \cdot {\rm cm}$. 
This $\rho_{\rm total}$ is expressed as 
$\rho_{\rm total}$=
$\rho_+ \rho_-/(\rho_+ + \rho_-) 
\simeq \rho_-$ owing to the relation for Fe$_4$N, i.e., 
$\rho_-/\rho_+ \ll 1$.\cite{Kokado1}  
Here, $\rho_-$ is 
simply given by 
$\rho_- = \rho_{s,-} + \rho_{s \to d,-}$=
$\rho_{s,-}^{\rm ph} 
+ \rho_{s,-}^{\rm imp} + \rho_{s \to d,-}^{\rm imp}$. 
Since $\rho_{s,-}^{\rm ph}$ depends on $T$, 
we assume 
$\rho_{s,-}^{\rm ph}$=$T/3$ and 
$\rho_{s,-}^{\rm imp} + \rho_{s \to d,-}^{\rm imp}$=14. 
Namely, $\rho_{s,-}^{\rm ph}$ is proportional to $T$.
\bibitem{validity}
In this study, 
we 
only consider the relation $R \propto T$ [see Eq. (\ref{R=ThT})] 
on the basis of Ref. \citen{Tsunoda_exp}. 
We here do not judge the validity of $\Theta$=0.0270. 
The present model, which consists of the d states of a single atom, 
does not take into account 
the d states in the unit cell of the realistic crystal structure 
(i.e., perovskite-type stucture). 
In such a model, 
it is inconsequential 
to judge the validity of the numerical value of $\Theta$.
\end{thebibliography}
\end{document}